%% file: main.tex
\documentclass[10pt,journal,compsoc]{IEEEtran}

\usepackage{amsmath}
\usepackage{enumitem}
\usepackage{booktabs}
\usepackage{graphicx}
\usepackage{environ}
\usepackage{amsthm}
\usepackage{algorithm}
\usepackage[noend]{algpseudocode}
\algrenewcommand\algorithmicdo{}
\usepackage{tikz}
\usepackage{tabularx}
\usepackage{tabulary}
\usepackage{multirow}
\usepackage{todonotes}
\usepackage{setspace}
\usepackage[singlelinecheck=false]{caption}
\usepackage{cleveref}
\usepackage{anyfontsize}
\usepackage[compact]{titlesec}
\usepackage{makecell}

\usepackage[style=ieee,backend=bibtex,url=false,doi=false,isbn=false,eprint=false]{biblatex}
\renewbibmacro*{bbx:savehash}{}
\AtEveryBibitem{\clearfield{issn}}
\AtEveryCitekey{\clearfield{issn}}

\AtEveryBibitem{%
\ifentrytype{inproceedings}{
    \clearfield{month}%
    \clearfield{volume}
}{}
\ifentrytype{article}{
    \clearfield{month}
}{}
\ifentrytype{book}{
    \clearfield{url}%
    \clearfield{urldate}%
    \clearfield{review}%
    \clearfield{series}
}{}
\ifentrytype{collection}{
    \clearfield{url}%
    \clearfield{urldate}%
    \clearfield{review}%
    \clearfield{series}
}{}
\ifentrytype{incollection}{
    \clearfield{url}%
    \clearfield{urldate}%
    \clearfield{review}%
    \clearfield{series}
}{}
}

\theoremstyle{definition}

\usetikzlibrary{backgrounds}
\tikzset{font={\fontsize{9pt}{10}\selectfont}}
\input{notation.tex}
\input{appendices/counts.tex}

\input{custom_graphics_commands.tex}

\newcommand{\graphicscale}{0.75\linewidth}

\begin{document}
\input{custom_tikz_styles.tex}
\title{A Taxonomy and Review of Algorithms for \\ Modeling and Predicting Human Driver Behavior}

\author{Raunak P. Bhattacharyya, 
        Kyle Brown,
        Juanran Wang,
        Katherine Driggs-Campbell, and
		Mykel J. Kochenderfer 
    \thanks{R.P. Bhattacharyya is with the Yardi School of Artificial Intelligence at the Indian Institute of Technology Delhi.
	E-mail: raunakbh@iitd.ac.in}
	\thanks{K. Brown and M.J. Kochenderfer are with the Stanford Intelligent Systems Laboratory in the Department of Aeronautics and Astronautics at Stanford University. 
    E-mail: \{kjbrown7,mykel\} @stanford.edu}
  \thanks{K. Driggs-Campbell is with the Department of Electrical and Computer Engineering at the University of Illinois at Urbana-Champaign. 
    E-mail: krdc@illinois.edu}
    \thanks{J. Wang is with the Department of Computer Science at Stanford University.
    E-mail: jun2026@stanford.edu}
}

\markboth{}%
{ Brown \MakeLowercase{\textit{et al.}}: Modeling and Prediction of Human Driver Behavior: A Survey}

\IEEEtitleabstractindextext{%
\begin{abstract}
An open problem in autonomous driving research is modeling human driving behavior, which is needed for the planning component of the autonomy stack, safety validation through traffic simulation, and causal inference for generating explanations for autonomous driving. Modeling human driving behavior is challenging because it is stochastic, high-dimensional, and involves interaction between multiple agents. This problem has been studied in various communities with a vast body of literature. Existing reviews have generally focused on one aspect: motion prediction. In this article, we present a unification of the literature that covers intent estimation, trait estimation, and motion prediction. This unification is enabled by modeling multi-agent driving as a partially observable stochastic game, which allows us to cast driver modeling tasks as inference problems. We classify driver models into a taxonomy based on the specific tasks they address and the key attributes of their approach. Finally, we identify open research opportunities in the field of driver modeling.
\end{abstract}

}

\maketitle
\IEEEdisplaynontitleabstractindextext
\IEEEpeerreviewmaketitle

\input{Sections/1_Introduction.tex}
\input{Sections/2_ProblemFormulation.tex}

\input{Sections/3_Taxonomy.tex}
\input{Sections/2_canonical_tasks_taxonomy.tex}
\input{Sections/5_Conclusion.tex}
\ifCLASSOPTIONcompsoc
    \section*{Acknowledgments}
\else
    \section*{Acknowledgment}
\fi
This work was supported by Qualcomm and by the National Science Foundation under grant no. DGE – 1656518. The authors would like to thank Ahmed Sadek, Mohammad Naghshvar, and the team at Qualcomm Corporate Research for their insightful feedback. RB would like to acknowledge support from ANRF via the Early Career Research Grant under file no. ECRG/2024/006446.


\newcommand{\bioheight}{1.25in} 
\newcommand{\biospace}{-0.7cm} 
\newcommand{\biowidth}{1in} 

\vspace{\biospace}

\begin{IEEEbiography}
  [{\includegraphics[width=\biowidth,height=\bioheight,clip,keepaspectratio]{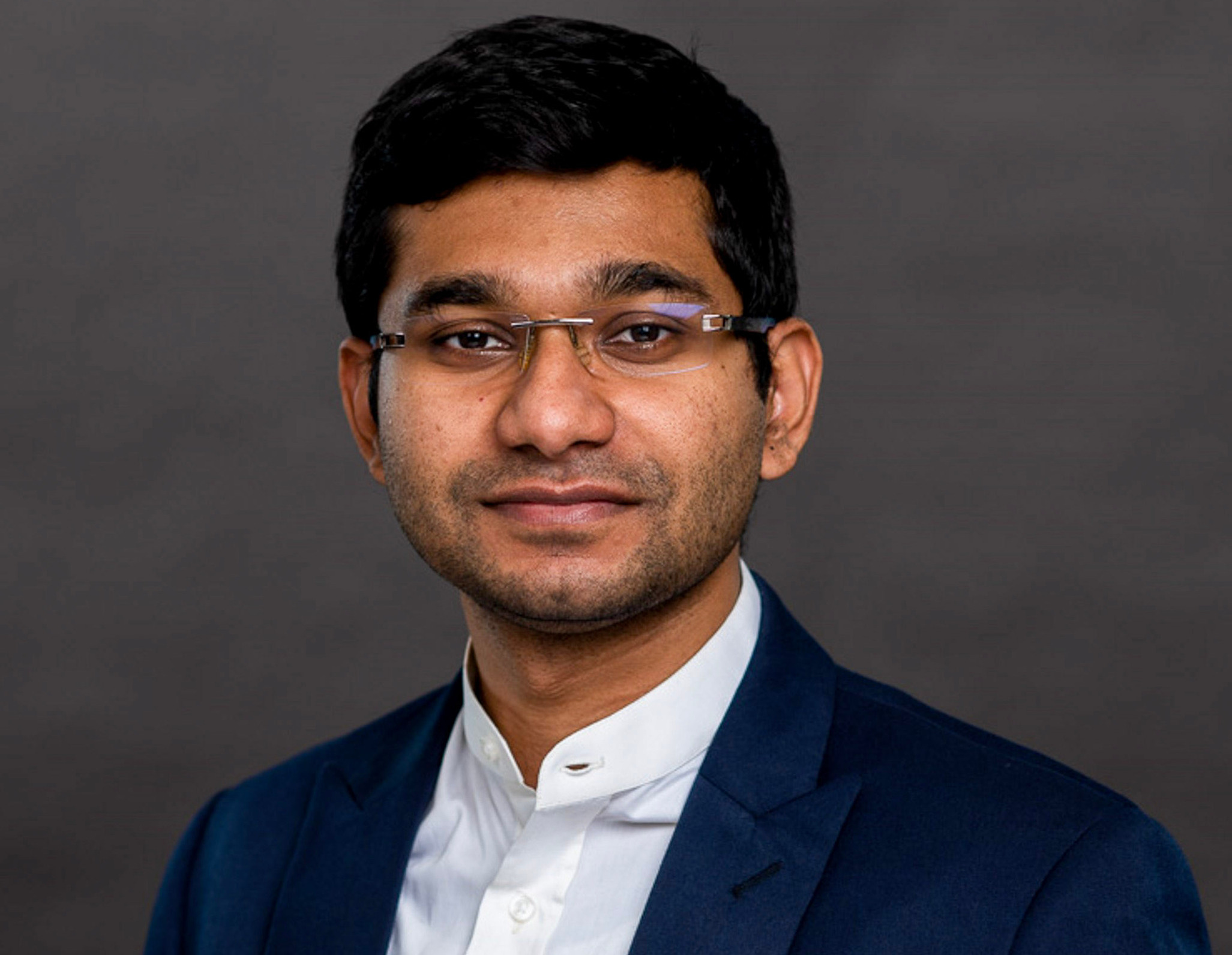}}]{Raunak P. Bhattacharyya}
  received the B.Tech. degree from the Indian Institute of Technology, Bombay in 2013, the M.S. degree from the Georgia Institute of Technology in 2016, and the Ph.D. degree from Stanford University in 2021. He is currently an Assistant Professor in the School of Artificial Intelligence at the Indian Institute of Technology, Delhi. His research focuses on safe and effective decision-making for autonomous systems in uncertain environments, with interests in imitation learning, human-AI collaboration, mission planning, and reinforcement learning.
\end{IEEEbiography}

\vspace{\biospace}

\begin{IEEEbiography}
  [{\includegraphics[width=\biowidth,height=\bioheight,clip,keepaspectratio]{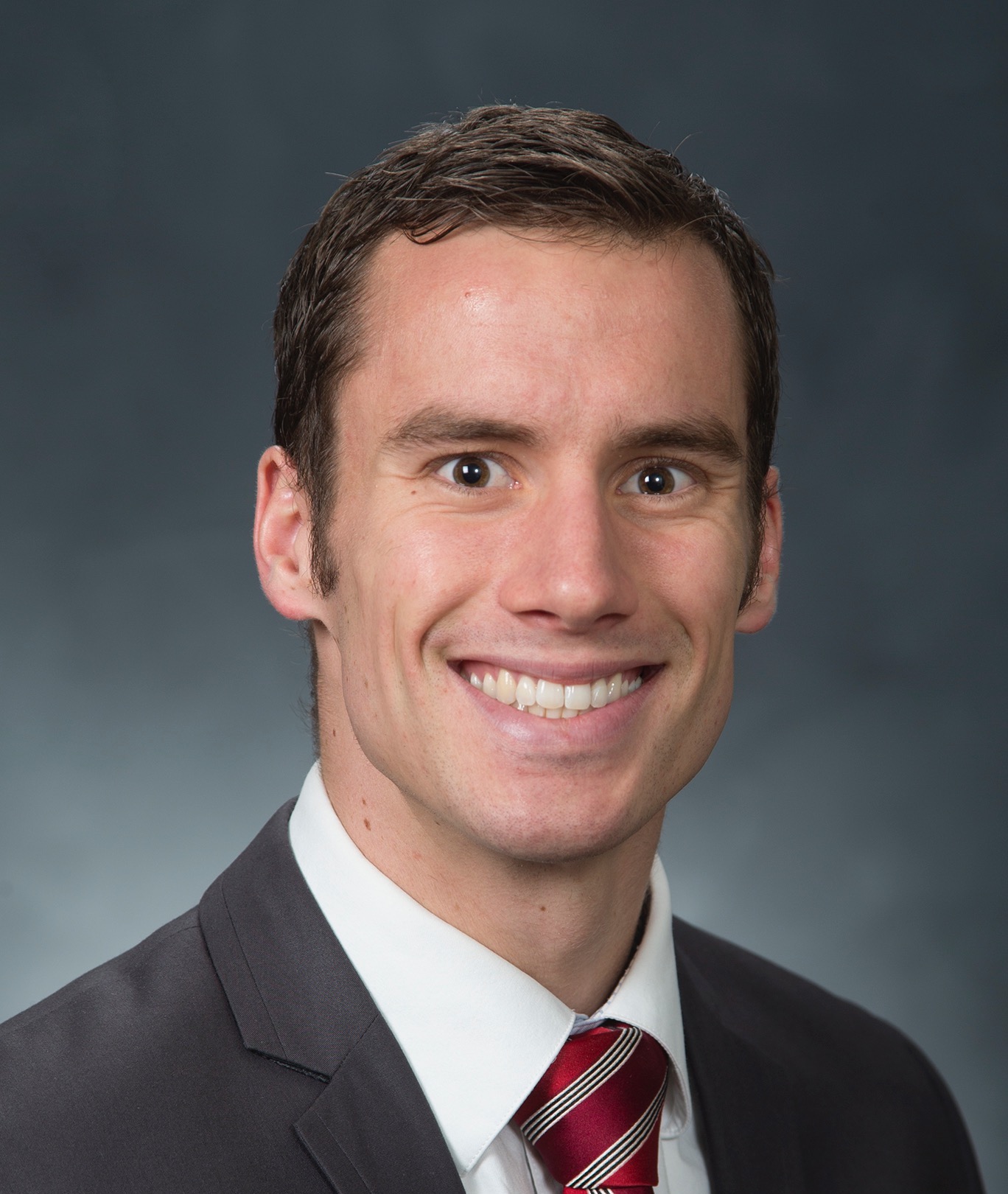}}]{Kyle J. Brown}
  received the B.S. degree in Mechanical Engineering from Brigham Young University in 2016 and the M.S. and Ph.D. degrees from Stanford University in 2021. He is currently with Anduril Industries.
\end{IEEEbiography}

\vspace{\biospace}

\begin{IEEEbiography}
  [{\includegraphics[width=\biowidth,height=\bioheight,clip,keepaspectratio]{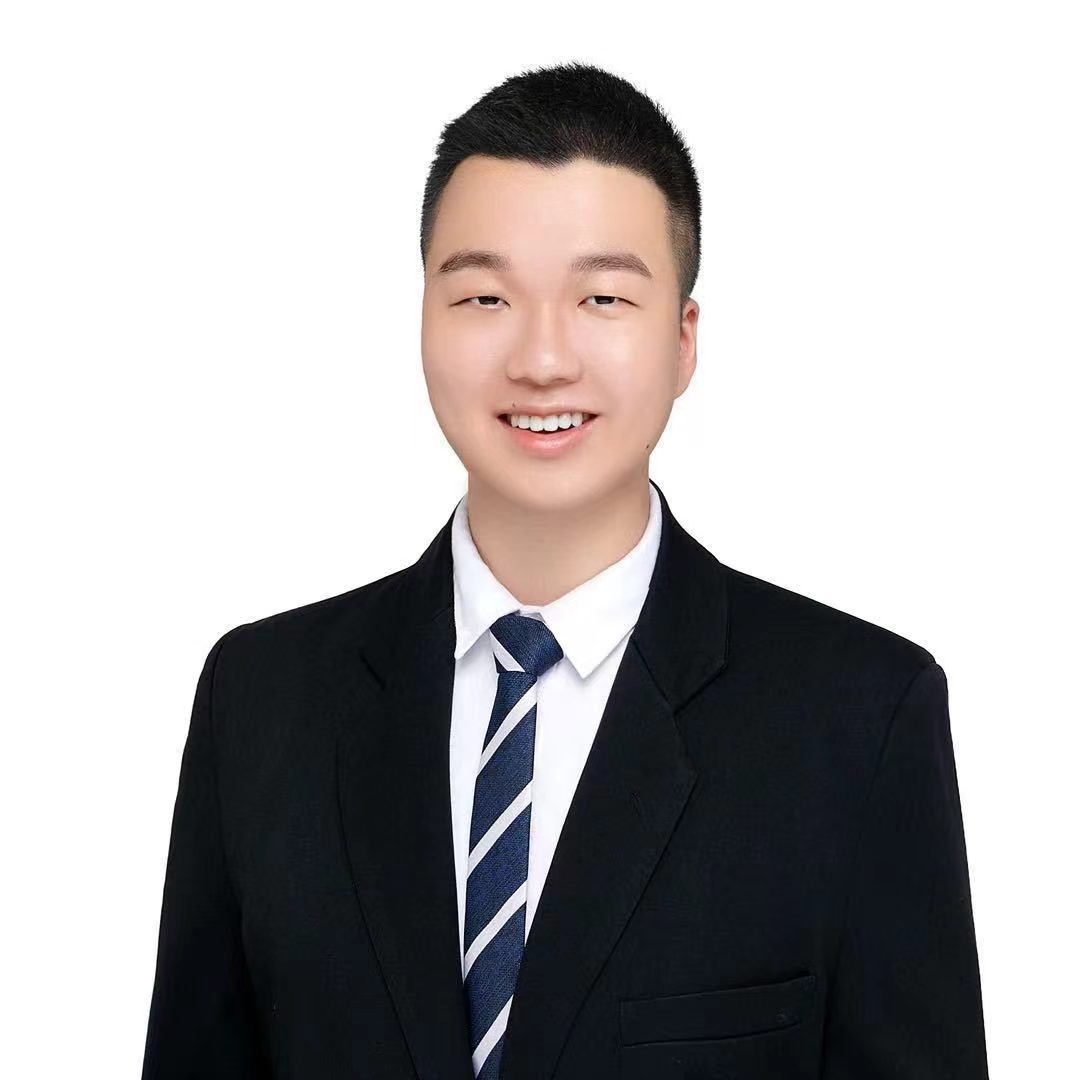}}]{Juanran Wang}
  is a B.S. and M.S. candidate in Computer Science at Stanford University, Stanford, CA, USA. He has served as an undergraduate research assistant at the Stanford Intelligent Systems Laboratory (SISL), participating in research projects on autonomous driving, decision making under uncertainty, and generative modeling.
\end{IEEEbiography}

\vspace{\biospace}

\begin{IEEEbiography}
  [{\includegraphics[width=\biowidth,height=\bioheight,clip,keepaspectratio]{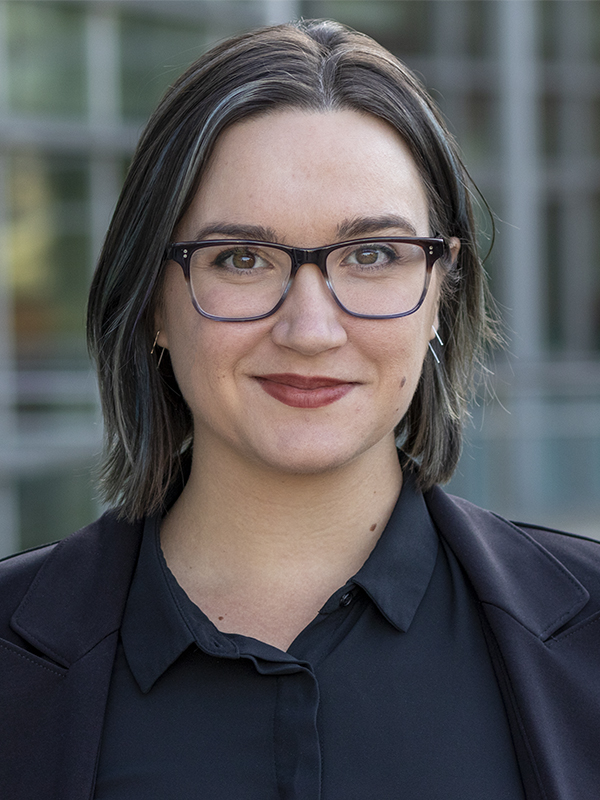}}]{Katherine Driggs-Campbell}
  received her BSE in Electrical Engineering from Arizona State University and her MS and PhD in Electrical Engineering and Computer Science from the University of California, Berkeley. She is currently an Assistant Professor in the ECE Department at the University of Illinois at Urbana-Champaign. Her research focuses on exploring and uncovering structure in complex human-robot systems to create more intelligent, interactive autonomy. She draws from the fields of optimization, learning \& AI, and control theory, applied to human-robot interaction and autonomous vehicles.
\end{IEEEbiography}

\vspace{\biospace}

\begin{IEEEbiography}
  [{\includegraphics[width=\biowidth,height=\bioheight,clip,keepaspectratio]{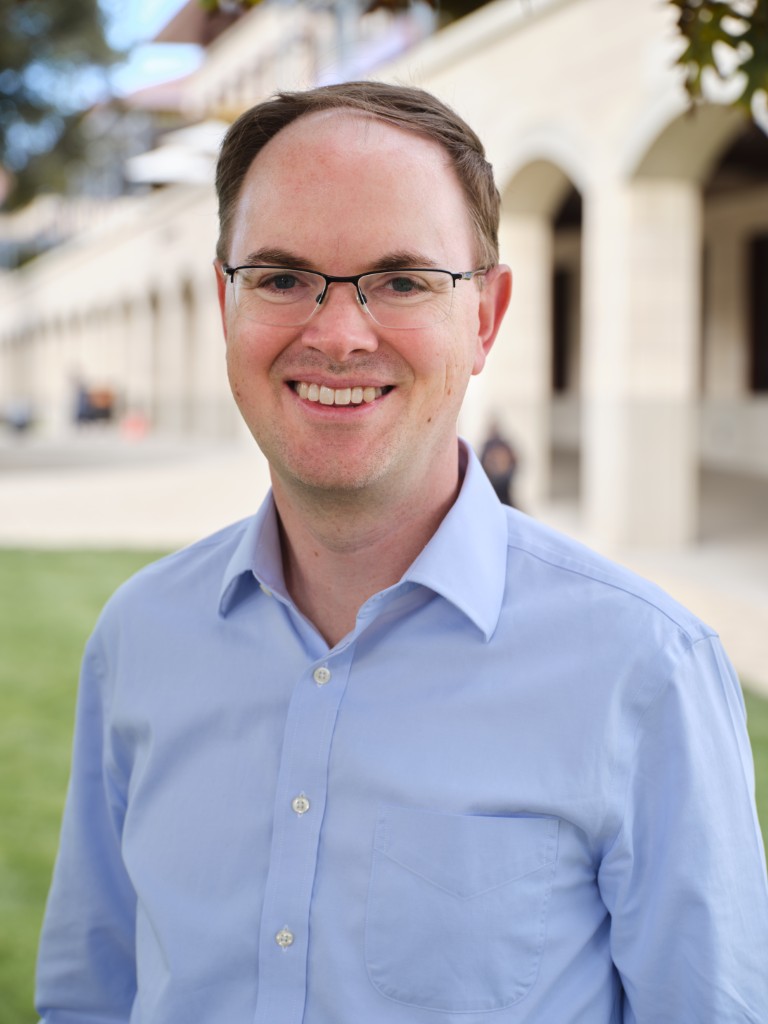}}]{Mykel J. Kochenderfer}
  received the B.S. and M.S. degrees in computer science from Stanford University, Stanford, CA, USA, in 2003, and the Ph.D. degree from The University of Edinburgh in 2006. He is currently an Associate Professor of aeronautics and astronautics with Stanford University. He is also the Director of the Stanford Intelligent Systems Laboratory (SISL), conducting research on advanced algorithms and analytical methods for the design of robust decision making systems. He is an author of the textbooks Decision Making under Uncertainty: Theory and Application (MIT Press, 2015), Algorithms for Optimization (MIT Press, 2019), and Algorithms for Decision Making (MIT Press, 2022).
\end{IEEEbiography}

\vfill

\renewcommand*{\bibfont}{\scriptsize}
\printbibliography

\newpage
\appendices
\input{Sections/6_Appendices.tex}

\end{document}

%% file: notation.tex
\newcommand{\checkempty}[3]{
	\def\temp{#2}\ifx\temp\empty
		\def\tempb{#3}\ifx\tempb\empty
			{#1}%
		\else
			{#1}^{(#3)}%
		\fi
	\else
		\def\tempb{#3}\ifx\tempb\empty
		{#1}_{#2}%
		\else
		{#1}^{(#3)}_{#2}%
		\fi
	\fi
}
\newcommand{\state}[2]{\checkempty{\mathbf{x}}{#1}{#2}} 
\newcommand{\ctrlaction}[2]{\checkempty{\mathbf{u}}{#1}{#2}} 
\newcommand{\ctrlactionspace}[1]{\checkempty{\mathcal{U}}{#1}{}} 
\newcommand{\obs}[2]{\checkempty{\mathbf{z}}{#1}{#2}} 
\newcommand{\statespace}[1]{\checkempty{\mathcal{X}}{#1}{}} 
\newcommand{\observationspace}[1]{\checkempty{\mathcal{Z}}{#1}{}} 
\newcommand{\internalstate}[2]{\checkempty{\mathbf{b}}{#1}{#2}} 
\newcommand{\internalstatespace}[1]{\checkempty{\mathcal{B}}{#1}{}} 
\newcommand{\transitionmodel}[1]{\checkempty{\boldsymbol{F}}{#1}{}} 
\newcommand{\obsmodel}[1]{\checkempty{\boldsymbol{G}}{#1}{}} 
\newcommand{\policymodel}[1]{\checkempty{\mathbf{\pi}}{#1}{}}  
\newcommand{\internaltransitionmodel}[1]{\checkempty{\boldsymbol{H}}{#1}{}} 

\newcommand{\egoindex}{r}

\newcommand{\initialtime}{0}
\newcommand{\currenttime}{t}
\newcommand{\finaltime}{t_f}

\newcommand{\numcars}{n}



%% file: appendices/counts.tex
\newcommand{\numpapers}{200}
\newcommand{\numstateestimation}{30}
\newcommand{\numintentionestimation}{107}
\newcommand{\numtraitestimation}{45}
\newcommand{\nummotionprediction}{137}

%% file: custom_graphics_commands.tex
\newcommand{\vertsep}{48mm}
\newcommand{\horzsep}{30mm}
\newcommand{\tilehorzsep}{.13*\horzsep}
\newcommand{\tilevertsep}{.04*\vertsep}

\newcommand{\nodesize}{10mm}
\newcommand{\timeidx}{t}
\newcommand{\positionB}{(       0.0,                    0.0                 )}
\newcommand{\positionA}{(       {0.4*\horzsep},         {-0.4*\vertsep}      )}
\newcommand{\positionO}{(       0.0,                    {-0.6*\vertsep}  )}
\newcommand{\positionS}{(       0.0,                    {-\vertsep}            )}

\newcommand{\positionTHi}{(     0.0,                    {-1.25*\vertsep}  )}

\newcommand{\positionT}{(       0.0,                    {-1.6*\vertsep}  )}
\newcommand{\saturation}{1.0}

\newcommand{\FadeDynamicModelOne}       {\tikzstyle{DynamicModelOne}        +=[style=FadedEdge]}
\newcommand{\FadeDynamicModelTwo}       {\tikzstyle{DynamicModelTwo}        +=[style=FadedEdge]}
\newcommand{\FadeObservationModel}      {\tikzstyle{ObservationModel}       +=[style=FadedEdge]}
\newcommand{\FadeEgoObservationModel}   {\tikzstyle{EgoObservationModel}    +=[style=FadedEdge]}
\newcommand{\FadeBeliefUpdateModelOne}  {\tikzstyle{BeliefUpdateModelOne}   +=[style=FadedEdge]}
\newcommand{\FadeBeliefUpdateModelTwo}  {\tikzstyle{BeliefUpdateModelTwo}   +=[style=FadedEdge]}
\newcommand{\FadePolicyModel}           {\tikzstyle{PolicyModel}            +=[style=FadedEdge]}
\newcommand{\FadeDynamicModel}{\FadeDynamicModelOne \FadeDynamicModelTwo}
\newcommand{\FadeBeliefUpdateModel}{\FadeBeliefUpdateModelOne \FadeBeliefUpdateModelTwo}

\newcommand{\FadeStateNode}   {\tikzstyle{StateNode}   +=[style=FadedNode]}
\newcommand{\FadeActionNode}  {\tikzstyle{ActionNode}  +=[style=FadedNode]}
\newcommand{\FadeObsNode}     {\tikzstyle{ObsNode}     +=[style=FadedNode]}
\newcommand{\FadeEgoObsNode}  {\tikzstyle{EgoObsNode}  +=[style=FadedObservedNode]}
\newcommand{\FadeBeliefNode}  {\tikzstyle{BeliefNode}  +=[style=FadedNode]}

\newcommand{\UnLabelDynamicModelOne}      {\tikzstyle{DynamicModelOne}      +=[text opacity=0.0]}
\newcommand{\UnLabelDynamicModelTwo}      {\tikzstyle{DynamicModelTwo}      +=[text opacity=0.0]}
\newcommand{\UnLabelObservationModel}     {\tikzstyle{ObservationModel}     +=[text opacity=0.0]}
\newcommand{\UnLabelEgoObservationModel}  {\tikzstyle{EgoObservationModel}  +=[text opacity=0.0]}
\newcommand{\UnLabelBeliefUpdateModelOne} {\tikzstyle{BeliefUpdateModelOne} +=[text opacity=0.0]}
\newcommand{\UnLabelBeliefUpdateModelTwo} {\tikzstyle{BeliefUpdateModelTwo} +=[text opacity=0.0]}
\newcommand{\UnLabelPolicyModel}          {\tikzstyle{PolicyModel}          +=[text opacity=0.0]}
\newcommand{\UnLabelDynamicModel}         {\UnLabelDynamicModelOne \UnLabelDynamicModelTwo}
\newcommand{\UnLabelBeliefUpdateModel}    {\UnLabelBeliefUpdateModelOne \UnLabelBeliefUpdateModelTwo}

\newcommand{\LabelHTwo}       {\tikzstyle{BeliefUpdateModelTwo} +=[text opacity=1.0]}

\newcommand{\LabelH}          {\LabelH \LabelHTwo}

\newcommand{\DrawO}[1][]{\node[ObsNode]          (o)     at \positionO {$\obs{#1}{}$};}
\newcommand{\DrawB}[1][]{\node[BeliefNode]       (b)     at \positionB {$\internalstate{#1}{}$};}
\newcommand{\DrawA}[1][]{\node[ActionNode]       (a)     at \positionA {$\ctrlaction{#1}{}$};}
\newcommand{\DrawS}[1][]{\node[StateNode]        (s)     at \positionS {$\state{#1}{}$};}
\newcommand{\DrawTime}[1][]{\node[TimeNode]      (t)     at \positionT {\ifthenelse{\equal{#1}{}}{$\timeidx$}{#1}};}

\newcommand{\FadeAllNodes}{
    \FadeStateNode
    \FadeActionNode
    \FadeObsNode
    \FadeEgoObsNode
    \FadeBeliefNode
    }
\newcommand{\FadeAllEdges}{
    \FadeDynamicModel
    \FadeObservationModel
    \FadeEgoObservationModel
    \FadeBeliefUpdateModel
    \FadePolicyModel
    }

\newcommand{\UnLabelEdges}{
    \UnLabelDynamicModel
    \UnLabelObservationModel
    \UnLabelEgoObservationModel
    \UnLabelBeliefUpdateModel
    \UnLabelPolicyModel
}

\newcommand{\ShiftRight}[2]{
        \pgftransformshift{\pgfpoint{\horzsep*#1}{0.0}}
        \renewcommand{\timeidx}{#2}
}
\newcommand{\ShiftLeft}[2]{
        \pgftransformshift{\pgfpoint{-{\horzsep*#1}}{0.0}}
        \renewcommand{\timeidx}{#2}
}
\newcommand{\ShiftBack}[2]{
    \pgftransformshift{\pgfpoint{{#1*\tilehorzsep}}{#2*\tilevertsep}}
    }

\makeatletter
\newsavebox{\measure@tikzpicture}
\NewEnviron{scaletikzpicturetowidth}[1]{%
    \def\tikz@width{#1}%
    \def\tikzscale{1}\begin{lrbox}{\measure@tikzpicture}%
    \BODY
    \end{lrbox}%
    \pgfmathparse{#1/\wd\measure@tikzpicture}%
    \edef\tikzscale{\pgfmathresult}%
    \BODY
}
\makeatother

\newcommand{\POSGsnapshot}[4]{
    \POMDPsnapshot[#1]
    \begin{scope}[opacity=\saturation]
        \DrawObsEdges{#2}{#3}{#4}
    \end{scope}
}

\newcommand{\POMDPsnapshot}[1][]{
    \DrawB[#1]
    \DrawA[#1]
    \DrawO[#1]
    \DrawS[#1]
    
    \begin{scope}[shift={(\horzsep,0.0)}]
        \node[EmptyNode]   (bplus) at \positionB {};
        \node[EmptyNode]   (splus) at \positionS {};
    \end{scope}
    \begin{scope}[opacity=\saturation]
        \draw [style=ObservationModel,->]       (s) -- node[left=1pt]   {$\obsmodel{#1}$} (o);
        \draw [style=DynamicModelOne,->]        (s) -- node[above=1pt]  {$\transitionmodel{#1}$} (splus);
        \draw [style=DynamicModelTwo,->]        (a) -- node[left=1pt]   {$\transitionmodel{#1}$} (splus);
        \draw [style=BeliefUpdateModelOne,->]   (o) -- node[left=1pt]   {$\internaltransitionmodel{#1}$} (b);
        \draw [style=BeliefUpdateModelTwo,->]   (b) -- node[above=1pt]  {$\internaltransitionmodel{#1}$} (bplus);
        \draw [style=PolicyModel,->]            (b) -- node[right=1pt]  {\Large $\policymodel{#1}$} (a);
    \end{scope}
}

\newcommand{\POMDPghostsnapshot}[1][]{
    {
        \node[BeliefNode]       (b)     at \positionB {$\cdots$};
        \node[ActionNode]       (a)     at \positionA {$\ctrlaction{#1}{}$};
        \node[StateNode]        (s)     at \positionS {$\cdots$};
    }
    
    \begin{scope}[shift={(\horzsep,0.0)}]
        \node[EmptyNode]   (bplus) at \positionB {};
        \node[EmptyNode]   (splus) at \positionS {};
    \end{scope}
    \begin{scope}[opacity=\saturation]
        \draw [style=DynamicModelOne,->]        (s) -- node[above=1pt]  {$\transitionmodel{#1}$} (splus);
        \draw [style=DynamicModelTwo,->]        (a) -- node[left=1pt]   {$\transitionmodel{#1}$} (splus);
        \draw [style=BeliefUpdateModelTwo,->]   (b) -- node[above=1pt]  {$\internaltransitionmodel{#1}$} (bplus);
        \draw [style=PolicyModel,->]            (b) -- node[right=1pt]  {\Large $\policymodel{#1}$} (a);
    \end{scope}
}

\newcommand{\POMDPend}[1][]{
    {
        \ShiftRight{1}{1}
        \node[BeliefNode]   (bplus) at \positionB {$\internalstate{#1}{}$};
        \node[StateNode]    (splus) at \positionS {$\state{#1}{}$};
    }
}
\newcommand{\DrawObsEdges}[3]{
    \begin{scope}[shift={({#1*\tilehorzsep},{#1*\tilevertsep})}]
        \node[InvisibleNode]    (on)     at \positionO {};
    \end{scope}
    \foreach \i in {#2,...,#3}
    {
        \begin{scope}[shift={({\i*\tilehorzsep},{\i*\tilevertsep})}]
            \node[InvisibleNode]    (sn)     at \positionS {};
        \end{scope}
        \begin{scope}[opacity=\saturation]
            \draw [style=ObservationModel,->]       (sn) -- node[left=1pt]   {$\obsmodel{#1}$} (on);
        \end{scope}
    }
}

%% file: custom_tikz_styles.tex
\newcommand{\FadedBorder}{black!30}
\newcommand{\FadedFill}{black!10}
\newcommand{\FadedEdge}{black!30}
\newcommand{\FadedText}{black!40}
\tikzstyle{GraphNode}=[shape=circle,minimum size=\nodesize, inner sep = 0.5mm, text=black!, opacity=1]
\tikzstyle{FadedNode}=[style=GraphNode,draw=\FadedBorder, fill=\FadedFill, text=\FadedText, opacity=1]
\tikzstyle{FadedObservedNode}=[style=GraphNode,draw=\FadedBorder, text=\FadedText, opacity=1]
\tikzstyle{HighlightNode}=[style=GraphNode,text=red!, draw=red!, opacity=1]
\tikzstyle{HiddenNode}=[style=GraphNode,draw=black!,fill=black!20, text=black!,opacity=1]
\tikzstyle{ObservedNode}=[style=GraphNode,draw=black!,fill=white!]
\tikzstyle{EmptyNode}=[style=GraphNode,opacity=0]
\tikzstyle{InvisibleNode}=[style=EmptyNode,opacity=1]

\tikzstyle{InputHighlight}=[draw=black!,line width=0.9pt,text=cardinal]
\tikzstyle{OutputHighlight}=[draw=black!,line width=0.9pt,text=bluish]
\tikzstyle{ConditionalHighlight}=[draw=black!,line width=0.9pt,text=greenish]

\tikzstyle{StateNode}=[style=HiddenNode]
\tikzstyle{ActionNode}=[style=HiddenNode]
\tikzstyle{BeliefNode}=[style=HiddenNode]
\tikzstyle{ObsNode}=[style=HiddenNode]
\tikzstyle{EgoObsNode}=[style=ObservedNode]
\tikzstyle{TimeNode}=[style=InvisibleNode]

\tikzstyle{GraphEdge}=[draw=black!, shorten > = 0pt, > = latex, line width = 0.5mm, text=black!,opacity=1]
\tikzstyle{ObservationModel}=[style=GraphEdge]
\tikzstyle{EgoObservationModel}=[style=GraphEdge]
\tikzstyle{BeliefUpdateModelOne}=[style=GraphEdge]
\tikzstyle{BeliefUpdateModelTwo}=[style=GraphEdge]
\tikzstyle{PolicyModel}=[style=GraphEdge]
\tikzstyle{DynamicModelOne}=[style=GraphEdge]
\tikzstyle{DynamicModelTwo}=[style=GraphEdge]
\tikzstyle{HighlightEdge}=[text=red!, draw=red!, > = latex,red!]
\tikzstyle{FadedEdge}=[draw=black!30, > = latex, black!30, text=black!50,opacity=1]
\tikzstyle{InvisibleEdge}=[opacity=0]

%% file: Sections/1_Introduction.tex

\section{Introduction}\label{sec:introduction}

\IEEEPARstart{A}{utomated} vehicles will need to operate in close proximity to human driven vehicles.
This task is challenging because human behavior can be difficult to predict.
The cognitive processes that govern human decision-making are inherently unobservable.
Skills, preferences, and driving ``style'' vary widely among drivers.
Moreover, complex interactions between drivers are typical on the road.
The task of modeling human driver behavior, though challenging, must be addressed to enable safe and efficient automated driving systems.

The existing body of driver modeling literature includes a wide variety of problem formulations, model assumptions, and evaluation criteria.
Several reviews of existing driver behavior models have been published recently. Notable examples include the 2011 review of tactical behavior prediction models by \citeauthor{Doshi2011}~\cite{Doshi2011}, the 2014 review of motion prediction and risk estimation models by \citeauthor{Lefevre2014}~\cite{Lefevre2014}, the 2016 review of human factors both in and around automated vehicles by \citeauthor{Ohn-Bar2016}~\cite{Ohn-Bar2016}, the 2020 review of deep learning based behavior prediction methods by \citeauthor{mozaffari2020deep}~\cite{mozaffari2020deep}, and the 2022 review of motion prediction in terms of physics-based, pattern-based and planning-based models by \citeauthor{karle2022scenario}~\cite{karle2022scenario}. Earlier reviews include the 1985 critical review by \citeauthor{Michon}~\cite{Michon}, the 1994 survey of cognitive driver models by \citeauthor{Ranney1994}~\cite{Ranney1994}, and the 1999 review of car-following models by \citeauthor{Brackstone}~\cite{Brackstone}. 
    Each survey focuses on a different subset of driver behavior modeling tasks. 
    Some touch relatively lightly on driver modeling as a corollary to their discussion of a different topic~\cite{Bengler2014,Carvalho2015b,Zhan2018,Riedmaier2020}.
    
While previous reviews generally focus on motion prediction, this review also includes intent estimation and trait estimation, tasks upstream of motion prediction aimed at understanding the driver’s immediate intent, and the factors that govern how that intent may be achieved. We unify state estimation, intent estimation, trait estimation, and motion prediction models into a single framework, the Partially Observable Stochastic Game. The driver modeling tasks are cast as inference problems within this framework.

Beginning researchers will find an overview of the driver modeling ``research landscape,'' and experienced researchers will find tools for identifying meaningful connections between existing models. 
  Our analysis focuses on fundamental attributes of proposed models as described in the publications that introduce them.
    When appropriate, we point out advantages and disadvantages of specific techniques.
      We emphasize, however, that our purpose is to facilitate understanding rather than to recommend any particular algorithm.
    We avoid quantitative comparison of existing models, as such analysis is limited by the results reported in the relevant publications.

Our main contributions are (1) the introduction of a common mathematical framework for modeling driver behavior in arbitrary multi-agent traffic scenarios, (2) the construction of a taxonomy that classifies existing models based on their approaches to a set of algorithmic tasks that fall under the driver modeling umbrella, and (3) the placement of the contributions of over \numpapers{} papers into our taxonomy. 

%% file: Sections/2_ProblemFormulation.tex

\begin{figure*}[!hpt]
  \centering
  \input{utils/graphics/POSG_chain_paper.tex}
  \caption{The evolution of an n-agent Partially Observable Stochastic Game (POSG) visualized as a graphical model. Each layer (into the page) of the graphical model corresponds to a different agent. Time increases from left to right. Edges represent the direction of information flow.}
  \label{fig:posg_graphical_model}
\end{figure*}




\section{Mathematical Formulation}\label{sec:problem_formulation}

We begin by presenting a general mathematical framework for describing the microscopic dynamics of traffic---dynamics that arise from the behavior of multiple interacting, decision-making agents operating in a complex, partially observable environment. 
This framework is the discrete-time multi-agent \emph{partially observable stochastic game} (POSG)~\cite{Kuhn1953}. The POSG formulation is introduced below and summarized in \cref{fig:posg_graphical_model} and \cref{tab:posg_notation}. 


Consider an arbitrary traffic scene wherein an autonomous ego vehicle operates alongside $\numcars$ surrounding human-driven vehicles in a static environment. While both the ego vehicle and the surrounding vehicles can be considered agents participating in our traffic scene, our discussion centers upon the process wherein the ego vehicle models the behavior of the surrounding vehicles. Therefore, the term \textit{agents} in our survey refers to the $\numcars$ surrounding vehicles unless specified otherwise.\footnote{Although the framework introduced here is general, and could include any type of traffic participant (e.g., pedestrian, cyclist, animal, etc.), we focus our discussion on vehicular traffic involving human (and potentially robotic) drivers.}
  Let $\state{i}{\currenttime}$ and $\internalstate{i}{\currenttime}$ represent the \emph{physical state} and the \emph{internal state}, respectively, of agent $i$ at time $t$. 
    The physical state describes attributes such as the agent's position, orientation, and velocity, whereas the internal state encompasses attributes such as the agent's navigational goals, behavioral traits, and ``mental model'' of the surrounding environment. 

Let $\obs{i}{t} \sim \obsmodel{i}(\state{1}{t},\ldots,\state{\numcars}{t})$ represent the process by which agent $i$ observes its surroundings (including the physical states of other agents and, implicitly, the static environment) at time $t$.
  We call $\obsmodel{i}$ the \emph{observation function}, and $\obs{i}{t}$ the \emph{observation}. In general, the physical environment is only partially observed (i.e., $\obs{i}{t}$ is lossy).
    As agent $i$ processes the information in each new observation, its internal state evolves over time.
      We describe this process by $\internalstate{i}{t} \sim \internaltransitionmodel{i}(\internalstate{i}{t-1},\obs{i}{t})$, where $\internaltransitionmodel{i}$ is the \emph{internal state update function}.

At each time step, agent $i$ selects a \emph{control action} $\ctrlaction{i}{t}$ according to $\ctrlaction{i}{t} \sim \policymodel{i}(\internalstate{i}{t})$, where $\policymodel{i}$ is called the \emph{policy function} and its argument reflects the fact that the decisions originate from the agent's internal state. 
We say that the agents \textit{interact} with each other in our POSG framework because the state of each agent is observed by the other agents, thereby influencing the internal state updates and the subsequent actions of the other agents.
  Agent $i$'s physical state evolves over time according to $\state{i}{t+1} \sim \transitionmodel{i}(\state{i}{t},\ctrlaction{i}{t})$, where $\transitionmodel{i}$ is a discrete-time stochastic \emph{state-transition function}\footnote{In the context of driver modeling, the state-transition function $\transitionmodel{}$ is a matter of vehicle dynamics, which can often be modeled accurately~\cite{Plochl}.} that maps the current state $\state{i}{t}$ and control action $\ctrlaction{i}{t}$ to a distribution over next states at time $t+1$.

\begin{table}[t]
    \caption{Notation associated with the multi-agent Partially Observable Stochastic Game (POSG) framework. Subscripts denote the agent index, and superscripts denote the time step.}
    {\small
    \renewcommand{\arraystretch}{1.5}
    \begin{tabular*}{\linewidth}{@{}l @{\extracolsep{\fill}}l}
        \toprule
        \multicolumn{2}{c}{\textbf{POSG Notation}} \\
        \midrule
        $\state{i}{t} \in \statespace{}_i$ & \emph{physical state} (of agent $i$ at time $t$) \\
        $\internalstate{i}{t} \in \internalstatespace{}_i$ & \emph{internal state} \\
        $\ctrlaction{i}{t} \in \ctrlactionspace{}_i$ & \emph{control action} \\
        $\obs{i}{t} \in \observationspace{}_i$ & \emph{observation} \\
        $\obs{i}{t} \sim \obsmodel{i}(\state{1}{t},\ldots,\state{\numcars}{t})$ & \emph{observation function}\\
        $ \internalstate{i}{t+1} \sim \internaltransitionmodel{i} (\internalstate{i}{t}, \obs{i}{t})$ & \emph{internal state update function} \\
        $\ctrlaction{i}{t} \sim \policymodel{i}(\internalstate{i}{t})$ & \emph{policy function} \\
        $\state{i}{t+1} \sim \transitionmodel{i}(\state{i}{t}, \ctrlaction{i}{t})$ & \emph{state transition function} \\
        \bottomrule
        \hspace{1pt}
    \end{tabular*}
    }
    \label{tab:posg_notation}
    \vspace{-0.9cm}
\end{table}

Finally, let the special subscript $\egoindex$ (i.e., $\obs{\egoindex}{}$, $\state{\egoindex}{}$, etc.) refer to the ego vehicle that observes---and potentially participates in---the traffic scene. 
  The presence of such an ego vehicle is by no means required in the POSG formulation, but we include it for convenience, as our review focuses primarily on driver modeling tasks that might be addressed in the planning and control stack of an automated vehicle.
  For the sake of our discussion, we assume that the information available to the ego vehicle at time $t$ consists of the history of observations $\obs{\egoindex}{\initialtime:t}$ received by the ego vehicle up to and including that time.

For convenience and simplicity, we occasionally omit the subscript or the superscript from the POSG variables.
  We also use colon syntax, (e.g., $\state{1:\numcars}{}$ denotes the states of all agents, $\state{}{\initialtime:\currenttime}$ denotes a time series of states from $\initialtime$ to $\currenttime$, etc.).

Compared with existing \textit{stochastic game} (SG) models, our POSG framework is more effective in modeling the interactions among the agents in the traffic scene. Since the policy function in the SG model operates on state observations only, the SG model does not explicitly represent how each agent reasons about the intentions and traits of other agents during decision-making. In our POSG framework, each agent uses the current observation to update its internal state, which may contain the agent's hypotheses about the possible traits, intentions, and future behavior of the surrounding agents. The policy function, which operates on the internal state, processes these hypotheses about the surrounding agents, which explicitly reflects how real-world drivers consider the possible future behavior of surrounding vehicles when making decisions.

Furthermore, our POSG framework more effectively represents how human drivers change their intentions over time and adjust their policies accordingly. While agents in a stochastic game typically map state observations to actions through a fixed policy, in our POSG framework, each agent may adjust its intention by updating its internal state. Such change in intention is then reflected as the policy function operates on the updated internal state. Finally, a third advantage of our POSG framework is that, by formulating an observation function, it is more flexible in representing different degrees and types of partial observability, enabling it to more accurately capture real-world driving conditions involving various categories of sensor errors.


\section{Driver Behavior Models}\label{sec:driver_models}

\begin{table}[t]
    \caption{Target variables for each of the core driver modeling tasks.}
    {\small
    \renewcommand{\arraystretch}{1.5}
    \begin{tabular*}{\linewidth}{@{}l @{\extracolsep{\fill}}l}
        \toprule
        \textbf{Core Driver Modeling Task} & \textbf{Inference Target} \\
        \midrule
        \emph{State Estimation} & $\state{1:\numcars}{t}$ \\ 
        \emph{Intention and Trait Estimation} & $\internalstate{1:\numcars}{t}$ \\ 
        \emph{Motion Prediction} & $\state{1:\numcars}{t+1:\finaltime}$ \\ 
        \bottomrule
        \hspace{1pt}
    \end{tabular*}
    }
    \label{tab:core_modeling_tasks}
    \vspace{-0.9cm}
\end{table}

\begin{figure}[t]
  \centering
  \includegraphics[width=\linewidth]{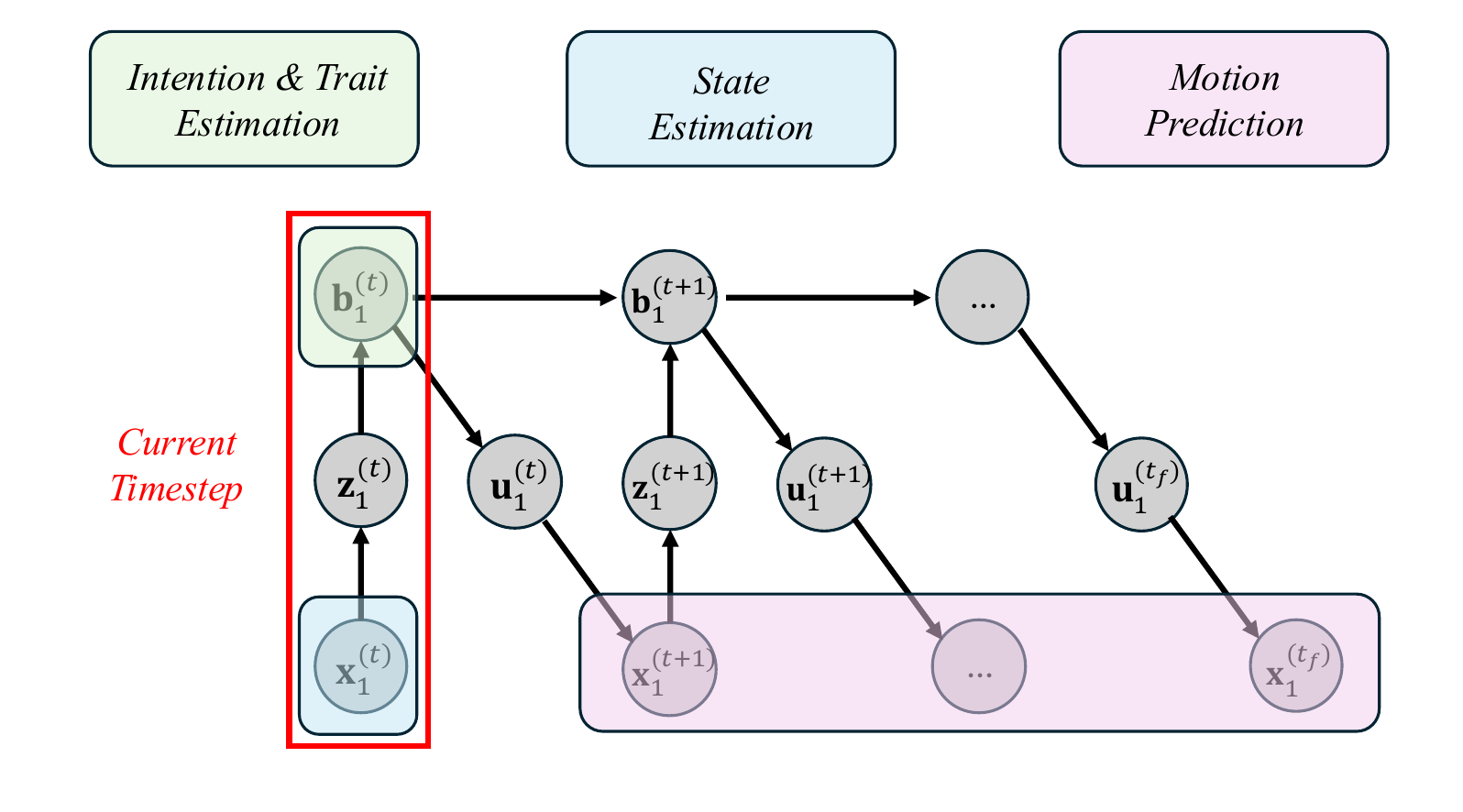}
  \caption{The methods we review address four main tasks of driver behavior modeling: state estimation, intention estimation, trait estimation, and motion prediction. Consider, for instance, the first surrounding vehicle at timestep $t$. The red box indicates variables in our POSG graphical model that describe the traffic situation at the current timestep. In the context of our POSG framework, state estimation aims to infer the current physical state of the vehicle, $\mathbf{x_1}^{(t)}$. The goal of intention estimation and trait estimation is to gain information about the current internal state of the driver of the vehicle, $\mathbf{b_1}^{(t)}$. The goal of motion prediction is to forecast the future physical state trajectory of the vehicle, $\mathbf{x_1}^{(t+1:t_f)}$.}
  \label{fig:inference_target_illustration}
\end{figure}

In the context of the POSG formulation, a driver behavior model is a collection of (not necessarily explicit) assumptions about $\obsmodel{}$, $\internaltransitionmodel{}$, $\policymodel{}$, and $\transitionmodel{}$\footnote{The state-transition function $\transitionmodel{}$ plays an important role in driver-modeling applications, although it technically has more to do with the vehicle than the driver.}. 
The implications of those assumptions depend on the model's applications.
Hence, the first layer of our taxonomy highlights specific algorithmic tasks addressed by each reviewed model.
    Some driver models address multiple tasks, while others concentrate on only one.
      All of the considered tasks fall under the driver modeling umbrella, and each is likely to play a role in the planning and control stack of an automated vehicle.

Each task involves reasoning about the present and/or future values of $\state{}{}$ and/or $\internalstate{}{}$ based on the information encoded in a history of the ego vehicle's observations $\obs{\egoindex}{\initialtime:\currenttime}$.
\emph{State estimation} is the task of inferring the current physical states $\state{1:\numcars}{t}$ of the surrounding vehicles. 
  \emph{Intention estimation} involves inferring---at a high level (e.g., turn left, change lane, etc.)---what a driver might intend to do in the immediate future.
  \emph{Trait estimation} equates to selecting the values of model parameters that can represent e.g., a driver's skills, preferences, and ``style,'' as well as properties like fatigue, distractedness, etc.
    Both ``intentions'' and ``traits'' can be considered part of a driver's current internal state $\internalstate{}{\currenttime}$.\footnote{Traits can also be interpreted as ``parameters'' of a driver's policy function $\policymodel{}$}
\emph{Motion prediction} is the task of predicting the future physical states $\state{1:\numcars}{t+1:\finaltime}$ of other vehicles. 
The respective inference goals of these four tasks in driver behavior modeling are illustrated in \cref{fig:inference_target_illustration}.

Most reviewed models address at least one of these four tasks, and we refer to them as the \emph{core} driver modeling tasks. The inference targets (i.e., the variables whose values are to be inferred) for each core task are summarized in \cref{tab:core_modeling_tasks}. 
  Before proceeding to a more thorough analysis of the core tasks, we briefly identify five additional tasks that are quite relevant to driver modeling, but for which a detailed analysis is beyond the scope of this article. \Cref{tab:tasks} also identifies existing models that address these auxiliary tasks.

\emph{Risk estimation} is the task of quantifying how ``unsafe'' one or more drivers' future motion is expected to be. 
  For examples, an advanced driver assistance system (ADAS) must quantify risk in order to decide if, when, and how to intervene in a given scenario.
  An excellent review of risk-estimation models is provided by \citeauthor{Lefevre2014}~\cite{Lefevre2014}.
The objective of \emph{Anomaly detection} is to recognize when the behavior of one or more traffic participants defies expectations.
  Such information might be crucial in activating safety features that make the ego vehicle's behavior more cautious in certain situations.
\emph{Behavior imitation} denotes the goal of making automated driving more ``human-like''. Imitating humans can be desirable, for example, if the goal is to produce a familiar-feeling ride for passengers or a familiar interaction experience for other drivers.
\emph{Microscopic traffic simulation} is related to motion prediction.
For the purposes of this survey, we make the following distinction: \emph{Motion prediction} is online and discriminative---a tool for forecasting the development of a given situation (presumably, the situation in which the ego vehicle currently finds itself). \emph{Traffic simulation} is offline and generative---a tool for exploring a wide range of potential situations, often with the goal of probing the ego vehicle's planning and control stack for weaknesses that can be addressed by system developers.
Finally, the rightmost column of \cref{tab:tasks} is used to identify models that are described in the context of a \emph{behavior planning} algorithm (i.e., an algorithm for planning the actions of an automated vehicle).

%% file: utils/graphics/POSG_chain_paper.tex

{
    \renewcommand{\vertsep}{48mm}
    \renewcommand{\horzsep}{30mm}
    \renewcommand{\tilehorzsep}{.18*\horzsep}
    \renewcommand{\tilevertsep}{.04*\vertsep}
    \renewcommand{\positionA}{({0.50*\horzsep},{-0.5*\vertsep})}
    \renewcommand{\positionO}{(0.0,{-0.5*\vertsep})}
    \renewcommand{\nodesize}{8mm}
    \renewcommand{\positionT}{\positionTHi}

\begin{scaletikzpicturetowidth}{\graphicscale}
    \pgfdeclarelayer{L2}
    \pgfdeclarelayer{L3}
    \pgfsetlayers{L3,L2,main}
    \begin{tikzpicture}[scale=\tikzscale]
    {

        \UnLabelEdges

        {
        {
            \ShiftLeft{3}{3}
            \POSGsnapshot{1}{0}{0}{2}
            \begin{pgfonlayer}{L3} 
            \end{pgfonlayer}
            \DrawTime[$\initialtime$]
        }
        {
            \ShiftLeft{2}{2}
            \POMDPghostsnapshot[1]
            \DrawTime[$\cdots$]
        }
        {
            \ShiftLeft{1}{1}
            \POSGsnapshot{1}{0}{0}{2}
            \begin{pgfonlayer}{L3} 
            \end{pgfonlayer}
            \DrawTime[$\currenttime - 1$]
        }
        {
            \POSGsnapshot{1}{0}{0}{2}
            \begin{pgfonlayer}{L3} 
            \end{pgfonlayer}
            \DrawTime[$\currenttime$]
        }
        {
            \ShiftRight{1}{1}
            \POSGsnapshot{1}{0}{0}{2}
            \begin{pgfonlayer}{L3} 
            \end{pgfonlayer}
            \DrawTime[$\currenttime+1$]
        }
        {
            \ShiftRight{2}{2}
            {
            \tikzstyle{BeliefUpdateModelTwo}+=[style=InvisibleEdge]
            \POMDPghostsnapshot[1]
            \DrawTime[$\cdots$]
            }
            {
            \tikzstyle{BeliefNode}+=[style=EmptyNode]
            \POMDPend[1]
            }
        }
        {
            \ShiftRight{3}{3}
            \begin{pgfonlayer}{L3} 
            \end{pgfonlayer}
            \DrawTime[$\finaltime$]
        }
        }
        {
        \FadeAllEdges
        \FadeAllNodes
        \begin{pgfonlayer}{L2}
            {\ShiftBack{1}{1}

            {
                \ShiftLeft{3}{3}
                \POSGsnapshot{2}{0}{-1}{1}
            }
            {
                \ShiftLeft{2}{2}
                \POMDPghostsnapshot[2]
            }
            {
                \ShiftLeft{1}{1}
                \POSGsnapshot{2}{0}{-1}{1}
            }
            {
                \POSGsnapshot{2}{0}{-1}{1}
            }
            {
                \ShiftRight{1}{1}
                \POSGsnapshot{2}{0}{-1}{1}
            }
            {                  
                \ShiftRight{2}{2}
                {
                \tikzstyle{BeliefUpdateModelTwo}+=[style=InvisibleEdge]
                \POMDPghostsnapshot[2]
                }
                {
                \tikzstyle{BeliefNode}+=[style=EmptyNode]
                \POMDPend[2]
                }
            }
            }
        \end{pgfonlayer}
        }
        {
        \FadeAllEdges
        \FadeAllNodes
        \begin{pgfonlayer}{L3}
            {\ShiftBack{2}{2}

            {
                \ShiftLeft{3}{3}
                \POSGsnapshot{n}{0}{-2}{0}
            }
            {
                \ShiftLeft{2}{2}
                \POMDPghostsnapshot[n]
            }
            {
                \ShiftLeft{1}{1}
                \POSGsnapshot{n}{0}{-2}{0}
            }
            {
                \POSGsnapshot{n}{0}{-2}{0}
            }
            {
                \ShiftRight{1}{1}
                \POSGsnapshot{n}{0}{-2}{0}
            }
            {
                \ShiftRight{2}{2}
                {
                \tikzstyle{BeliefUpdateModelTwo}+=[style=InvisibleEdge]
                \POMDPghostsnapshot[n]
                }
                {
                \tikzstyle{BeliefNode}+=[style=EmptyNode]
                \POMDPend[n]
                }
            }
            }
        \end{pgfonlayer}
        }
    }
    \end{tikzpicture}
\end{scaletikzpicturetowidth}
}

%% file: Sections/3_Taxonomy.tex
\section{Approaches to the Core Modeling Tasks}\label{sec:model_taxonomy}

The second layer of our taxonomy involves a closer look at the characteristics of existing models that address one or more of the core driver modeling tasks: \emph{state estimation}, \emph{intention estimation}, \emph{trait estimation}, and \emph{motion prediction}.
  We devote most of our attention to the latter three.

Our first objective is to highlight---for each task---fundamental similarities and differences between the assumptions and methods employed by existing models.
  To this end, each of \cref{sec:state_estimation,sec:intention_estimation,sec:trait_estimation,sec:motion_prediction} introduces task-specific algorithmic axes along which models are compared and contrasted. 
  The discussion for each core task (except state estimation) is supplemented and summarized by a \emph{comparison} table that identifies where each surveyed model falls along the selected dimensions.
    Each term introduced in \textbf{bold text} corresponds to a column of the associated comparison table.

We also provide information about specific components and characteristics of existing models. This information is communicated via \emph{keyword} tables (one per task), which identify important task-specific keywords and list all surveyed models associated with each keyword. Keywords are separated into five categories.
  \emph{Architecture} keywords (e.g., dynamic Bayesian network, support vector machine, etc.) describe specific components or methodologies that a model incorporates. 
  \emph{Training} keywords (e.g., expectation maximization, genetic algorithms, etc.) describe how a model's parameter values are selected. 
  \emph{Theory} keywords (e.g., clustering, time series analysis) allude to the theoretical underpinnings of a model.
  \emph{Scope} keywords (e.g., intersection, highway merging) identify target applications for which a model is proposed, or on which it is evaluated.
  Finally, \emph{Evaluation} keywords (e.g., root-mean square error, precision over recall) identify specific metrics by which a model's performance is characterized.

To be clear, the comparison and keyword tables are not intended as mere summaries of the discussion in the body of the paper (indeed, they contain much information that does not appear in the body).
  Rather, these tables are intended as tools to help readers identify where reviewed models fit within specific algorithmic categories of the taxonomy. For example, one could use \cref{tab:motion_prediction} to identify models that represent predicted vehicle trajectories with multivariate Gaussian distributions. \Cref{tab:intent_estimation_keywords} could be used to find intention estimation models that use support vector machines (SVM), or that are applied to unsignalized intersections, etc. 


%% file: Sections/2_canonical_tasks_taxonomy.tex

\newcommand{\kw}[1]{#1}

\subsection{State Estimation}\label{sec:state_estimation}

The objective of \emph{state estimation} is to extract a coherent estimate of the physical environment state---including the current physical states $\state{1:\numcars}{t}$ of the surrounding vehicles---from a history of raw sensor information $\obs{\egoindex}{\initialtime:\currenttime}$.
  Though the state estimation task is explicitly addressed by relatively few (only \numstateestimation{}) of the reviewed publications, it is a foundational driver modeling task in the sense that all other modeling tasks are predicated on the information that it infers about $\state{1:\numcars}{t}$. Hence, we provide a very brief discussion and refer readers to \cref{tab:state_estimation_keywords}.

The algorithms underlying state estimation are usually some form of approximate Bayesian filter, including variants of the \kw{Kalman filter} and \kw{particle filter}. 
  Some advanced state estimation models---including those based on \kw{dynamic Bayesian networks} (DBN) and \kw{multiple model unscented Kalman filters} (MM-UKF), as well as the \kw{multi-perspective tracker}~\cite{Deo}---take advantage of the structure inherent in the driving environment to improve performance.
    \Cref{tab:state_estimation_keywords} identifies the state estimation keywords associated with models that address this task.


\subsection{Intention Estimation}\label{sec:intention_estimation}

The objective of \emph{intention estimation} is to infer what the drivers of surrounding vehicles intend to do in the immediate future.
This often involves computing a probability distribution over a finite set of high-level behavior modes---often corresponding to navigational goals (e.g., change lanes, overtake the lead vehicle)---that a driver might execute in the current situation.
  The intention estimation task is often addressed as part of a model's approach to the motion prediction task.

Intention estimation models can be compared in terms of their \emph{intention space}, their \emph{hypothesis representation}, and their \emph{inference paradigm}. \Cref{tab:intent_estimation} categorizes \numintentionestimation{} reviewed models along these axes. In addition, \cref{tab:intent_estimation_keywords} introduces a list of keywords that are associated with these intention estimation models. 

\newcommand{\intent}[1]{\text{\emph{#1}}}
\subsubsection{Intention Space}
\begin{figure}[t]
  \centering
  \includegraphics[width=\linewidth]{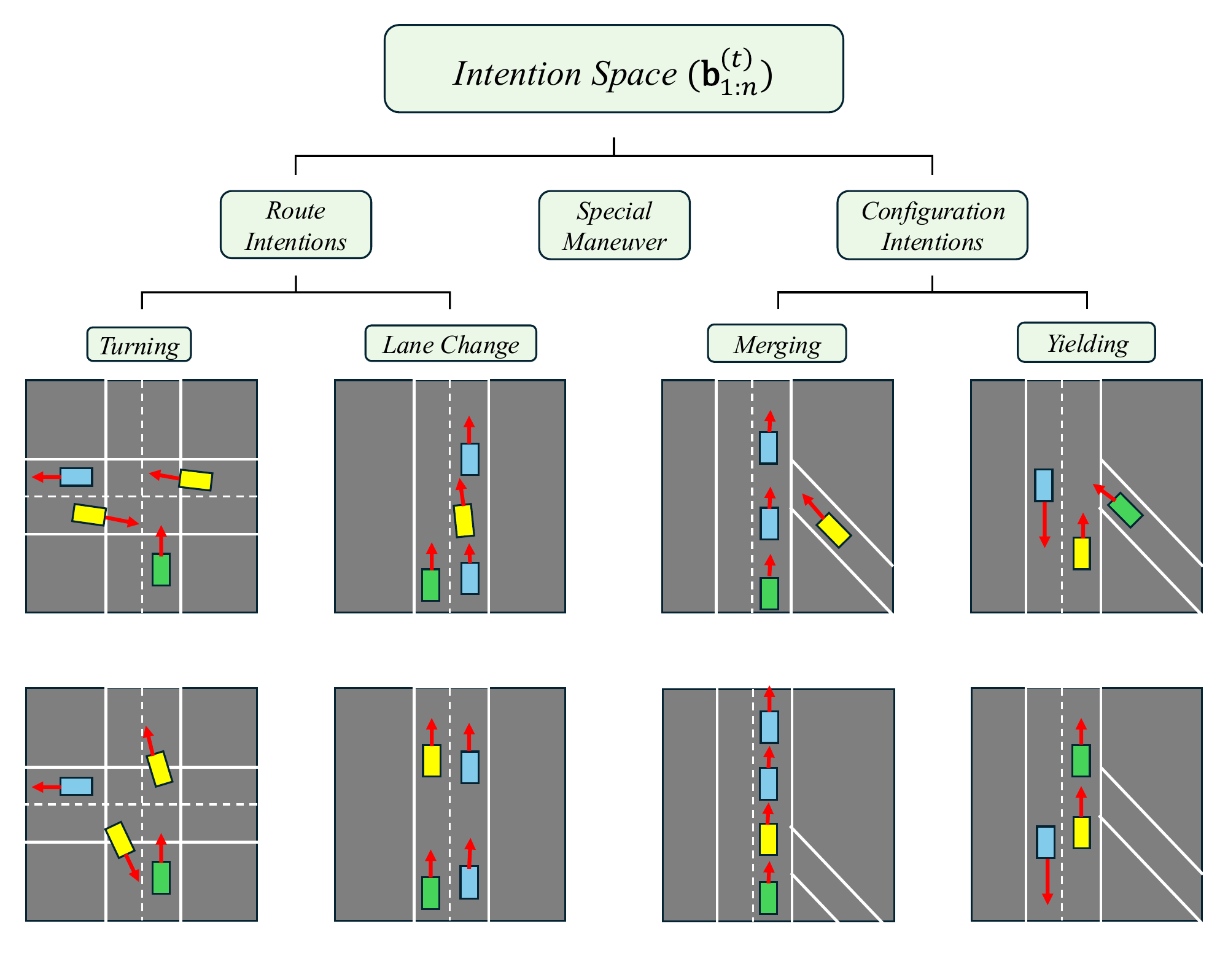}
  \caption{The intention space is the set of possible future behavior modes of the inference target. Consider the ego vehicle (green) performing intention estimation on a set of target vehicles (yellow) in the presence of some background vehicles (blue). The arrows indicate the directions of motion of the vehicles, and the traffic scene evolves from the top frame to the bottom frame. Route intentions describe how the target vehicles intend to navigate the road network. These include turning and lane change intentions. Configuration intentions describe the target vehicle's intended spatial relationships to other vehicles. These include merging intentions, which describe the vehicle gap the target car intends to enter in a merging scenario, and yielding intentions, which describe whether the target vehicle intends to yield to a merging vehicle. The intention space also encompasses possible intent for special maneuvers such as emergency pullovers.}
  \label{fig:intention_space}
\end{figure}

The \emph{intention space} refers to the set of possible behavior modes that may exist---according to the assumptions of a model---in a driver's internal state, $\internalstate{}{}$.
  The intention space is usually defined explicitly, although it can also be learned in an unsupervised manner~\cite{Guo2019}.
  We identify a non-exclusive, non-comprehensive list of behavior mode categories used by models in the literature. \Cref{fig:intention_space} shows our taxonomy of behavior modes.

\textbf{Route} behavior modes are defined in terms of the structure of the roadway network, and may consist of a single decision (e.g., $\intent{turn right}$) or a sequence of decisions (e.g., $\intent{turn right} \rightarrow \intent{go straight} \rightarrow \intent{turn right again}$) that a driver may intend to execute.
  \textbf{Lane-change} intentions are a fundamental case of route intentions.\footnote{A thorough review of lane-change intention inference models was published by \citeauthor{Xing2019} in 2019~\cite{Xing2019}.}

\textbf{Configuration} intentions are defined in terms of spatial relationships to other vehicles.
  For example, some intention estimation models reason about which gap between vehicles the target car intends to enter in a merging scenario \cite{Schulz,Schulz2018}. Other models consider the intentions of a car in the other lane (i.e., whether or not to yield and allow the merging vehicle to enter)~\cite{Dong2017Merging,Okuda2016,Okuda2017,Tran2019Merging}.
  Configuration behavior modes are sometimes described as homotopies~\cite{Schulz2017}, which correspond to the various ways vehicles might pass ahead of or behind each other. 

Some models consider modes of \textbf{longitudinal} driving behavior (e.g., $\intent{follow leader}$, $\intent{cruise}$) or modes of \textbf{lateral} driving behavior (e.g., $\intent{keep lane}$, $\intent{prepare to change lane}$, $\intent{change lane}$~\cite{Driggs-Campbell2015c}).
Examples of uncommon intention spaces include \textbf{intent to comply with traffic signals}~\cite{Aoudeb}, \textbf{possible emergency maneuvers}~\cite{Hillenbrand2006} and \textbf{intentional or unintentional maneuvers}~\cite{Lefevre2014b}.\footnote{See the 2019 article by \citeauthor{Tryhub2019} for a review of models for estimating whether a driver is attentive or distracted~\cite{Tryhub2019}.}

Behavior mode categories are often combined. For example, some models reason about routes through a road network and whether a driver intends to yield to conflicting traffic along any particular route \cite{Sun2019}. 
  The set of applicable behavior modes can vary depending on the specific context. While some models are tailored for a single operational context, others incorporate explicit \textbf{context-dependent} intention spaces with a scheme for restricting the set of applicable behavior modes based on observed features of the traffic environment \cite{Klingelschmitt2016,Klingelschmitt2016a,Hu2018a,Hu2018b,Bonnin2012a,Gindele,Galceran2017a}.
  
If a model's intention space is a good reflection of the actual behaviors exhibited by human drivers in a particular context, that model has a better chance of performing well at the intention estimation task. Model performance can also be impacted by the number of behavior modes considered, and how easy it is to distinguish between distinct behavior modes.

\subsubsection{Intention Hypothesis Representation}
A model's intention hypothesis $P(\internalstate{}{\currenttime})$ encodes uncertainty about the current intentions of drivers. 
Among reviewed models, the most common form of intention hypothesis is a \textbf{discrete probability distribution} over possible intentions of each driver under consideration.
  Some models compute a discrete \textbf{distribution over scenarios}, which corresponds to a (potentially sparse) joint distribution over intentions of multiple drivers within a given traffic scene~\cite{Hardy2013,Galceran2017a,Lawitzkyb}.
  A few surveyed models employ a \textbf{particle distribution}~\cite{Gonzalez2019,Brechtel2014}.
  Some use a \textbf{point estimate} hypothesis, which ignores uncertainty and simply assigns a probability of $1$ to a single (presumably the most likely) behavior mode.

In general, the form of the intention hypothesis implies a tradeoff between representational power and computational complexity.
  The appropriateness of any particular representation depends on the scope in which the model is used. For example, models applied to highly interactive situations (e.g., highway merging) are likely to benefit from a hypothesis representation that captures a joint distribution over intentions.

\newcommand{\ise}{f}
\subsubsection{Intention Inference Paradigms} \label{subsubsection: intention inference paradigms}

\begin{table*}[t]
    \caption{A comparison of the four main intention inference paradigms in terms of interpretability, representational capacity, computational efficiency, interaction awareness, and scalability to multiple agents. 
    }
    {\small
    \renewcommand{\arraystretch}{1.5}
    \begin{tabular*}{\linewidth}{@{}l @{\extracolsep{\fill}}ccccc}
        \toprule
        \makecell[l]{\textbf{Intention} \\ \textbf{Inference Paradigms}} &
        \makecell{\textbf{Inter-} \\ \textbf{pretability}} &
        \makecell{\textbf{Representational} \\ \textbf{Capacity}} &
        \makecell{\textbf{Computational} \\ \textbf{Efficiency}} &
        \makecell{\textbf{Interaction} \\ \textbf{Awareness}} &
        \makecell{\textbf{Scalability to} \\ \textbf{Multiple Agents}} \\
        \midrule
        \emph{Bayesian models} & high & medium & low & high & low \\ 
        \emph{Black-box models} & very low & very high & medium & high & high\\ 
        \emph{Prototype-based models} & very high & very low & very high & very low & very high \\ 
        \emph{Game-theoretic models} & high & medium & very low & very high & very low \\
        \bottomrule
        \hspace{1pt}
    \end{tabular*}
    }
    \label{tab:intention_inference_table}
\end{table*}

A model's \emph{inference paradigm} refers to the way the model actually computes the intention hypothesis (i.e., the way the model reasons about the processes that influence/determine each driver's intentions).
  Most reviewed models employ at least one of the following non-exclusive inference paradigms. 
  {\Cref{tab:intention_inference_table} evaluates these inference paradigms against a variety of metrics.

\textbf{Recursive estimation} algorithms 
operate by repeatedly updating the intention hypothesis at each time step based on the new information received, as in $P(\internalstate{}{t}) = \ise(P(\internalstate{}{t-1}), \obs{\egoindex}{t})$.
  They have the advantage of being able to ``remember'' useful information from arbitrarily far into the past, but may also have trouble ``forgetting'' that information when, for example, a driver's behavior suddenly changes to reflect a new intention.
  In contrast, \textbf{single-shot} estimators 
  compute a new hypothesis from scratch at each inference step, as in $P(\internalstate{}{t}) = \ise(\obs{\egoindex}{t-k:t})$, where $k \geq 0$ determines the length of the observation history that serves as the model's input. Single shot estimators do not store any information between successive inference iterations, and hence may be overly sensitive to noise in recent observations (though it is often possible to strike a good balance by choosing an appropriate value for $k$).

\textbf{Bayesian} models are based on Bayes' rule and the laws of conditional probability. 
  These models employ explicitly specified conditional probability distributions (which can be defined heuristically or by fitting to a dataset) based on $\obsmodel{\egoindex}$, $\policymodel{}$, $\transitionmodel{}$ and/or $\internaltransitionmodel{}$. 
  One example is the dynamic Bayesian network-based model proposed in~\cite{Gindele2015}.
  \textbf{Black box} models have many non-interpretable parameters whose values are usually set by minimizing some loss function over a training dataset. 
  For instance, recent studies have trained transformer neural network models to predict the route intentions of the surrounding vehicles based on observations of their physical states as well as the in-vehicle behavior of their drivers over a recent time period \cite{he2025novel, vellenga2025designing}.

Some Bayesian approaches may incorporate black box components (e.g., \cite{Hu2018b,Kumar2013}), but ``end-to-end'' black box models are not considered to operate within the Bayesian paradigm---even if their outputs are interpreted as probability distributions (e.g., \cite{Yoon2016}).
  Bayesian models are generally more interpretable than black box models, although the latter often have greater flexibility and representational capacity.

Many models operate by comparing observed motion history to a set of \textbf{prototype maneuvers} or \textbf{prototype policies}, and deducing which prototype is the ``closest'' match.
  A prototype maneuver can be represented by a single trajectory \cite{Woo2017,Sadigh2018,Hu2019}, a set of trajectories \cite{Kafer2010,Ward2015}, or a parametric or non-parametric distribution over trajectories \cite{Klingelschmitt2016,Tran2014,Guo2019Multi}.
  They can be based entirely on road geometry \cite{Houenou2013},
    extracted from a dataset of pre-recorded trajectories,
    or generated with a motion planning/prediction algorithm \cite{Galceran2017a,Sadigh2018,Liebner,Woo2017,Hu2019}.

Finally, \textbf{game-theoretic} models are ``interaction-aware'' in the sense that they explicitly consider the possible intentions of one or more other drivers when computing or refining the intention hypothesis for a given driver (i.e., $P(\internalstate{i}{t}) = \ise(\ldots, P(\internalstate{j}{t}))$).
  Some game-theoretic approaches incorporate interaction-aware models of $\internaltransitionmodel{}$ and $\policymodel{}$ \cite{Gonzalez2019,Schmerling2017}. Some use search or optimization to exclude scenarios with a high probability of conflicting intentions between drivers \cite{Woo2017,Lawitzkyb,Deo,Isele2019}. Others compute the Nash equilibrium of explicitly formulated games with payoff matrices \cite{Talebpour2015}.
    Models in this category often rely on motion prediction to inform or refine the output of intention estimation.
      Game-theoretic intention estimation models are well-suited to highly interactive scenarios (e.g., unsignalized intersections, highway merging), but can become intractable as the number of agents increases.

\subsection{Trait Estimation}\label{sec:trait_estimation}


\begin{figure*}[t]
  \centering
  \includegraphics[width=\linewidth]{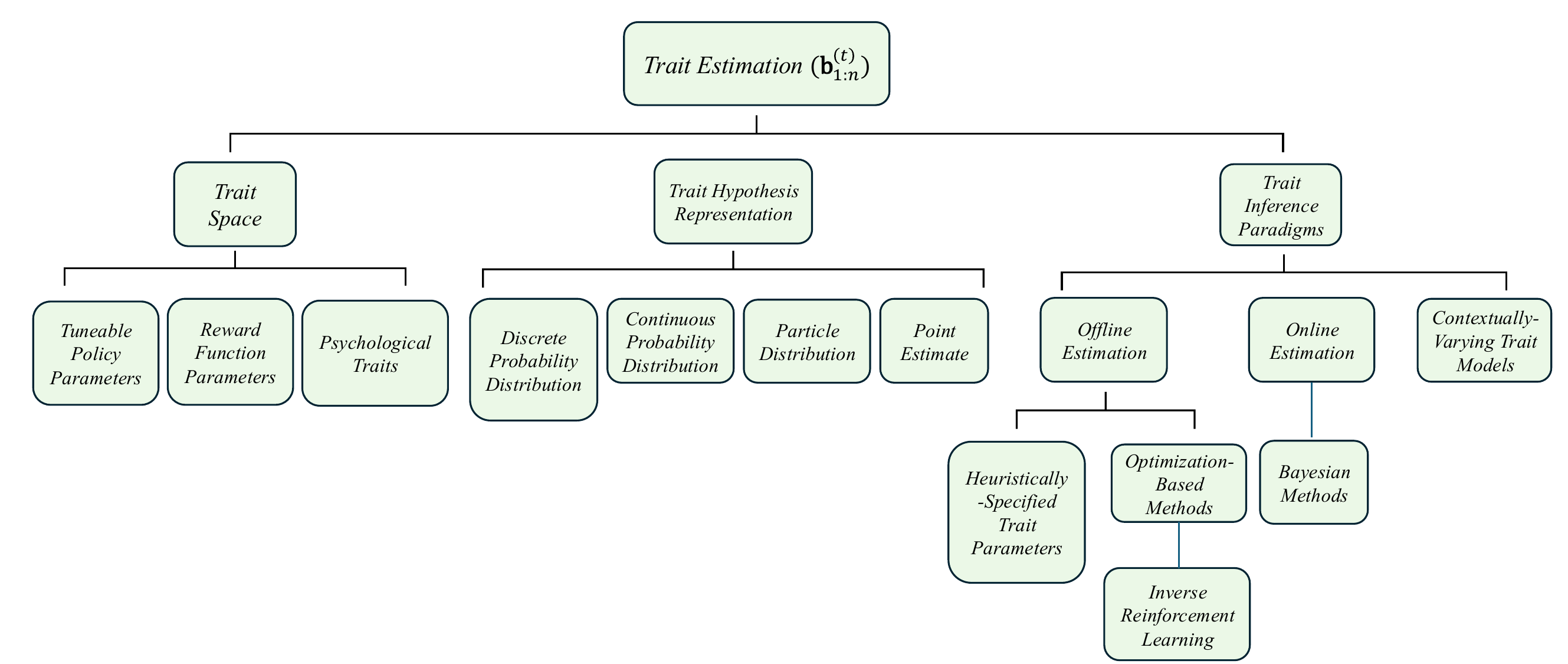}
  \caption{Trait estimation is the task of inferring the driver's skills, preferences, and driving styles. The trait of a driver can be represented as a policy model, a reward model, or a psychological model. Our estimate of the trait of a driver, i.e. our trait hypothesis, can be represented as a discrete probability distribution, a continuous probability distribution, a particle distribution, or a point estimate. The trait hypothesis can be computed offline based on expert knowledge or historical driving data. Bayesian methods allow us to update our trait hypotheses of the surrounding vehicles online based on real-time observations. If the vehicle needs to operate over a wide range of geographical areas and scenarios, contextually-varying trait models allow us to adapt our trait model to the current environment.}
  \label{fig:trait_estimation}
\end{figure*}

Whereas intentions denote what a driver is trying to do, traits encompass factors---e.g., skills, preferences, and ``style,'' as well as properties like fatigue, distractedness, etc.---that affect how the driver will do so.
  Driver traits vary across countries, cultures, and individuals~\cite{Wang2018Traits,Lajunen2004}.

Models that address the trait estimation task usually do so as part of their approach to the motion prediction task. In this sense, trait estimation can be thought of as ``training'' or ``calibrating'' the policy model that is to be used within a motion prediction model. 

We compare existing trait estimation models in terms of their \emph{trait spaces}, their \emph{trait hypotheses}, and their trait \emph{inference paradigms}. 
\Cref{fig:trait_estimation} illustrates the structure of our taxonomy of the trait estimation models.
\Cref{tab:trait_estimation} catalogues \numtraitestimation{} models along these dimensions, and \cref{tab:trait_estimation_keywords} introduces keywords associated with these models.


\newcommand{\tw}[1]{\emph{#1}}
\subsubsection{Trait Space}
A model's \emph{trait space} refers to the set of trait parameters about which the model reasons.
  Some of the most widely used driver models are simple parametric controllers with tuneable \textbf{policy parameters} that represent intuitive behavioral traits of drivers.
    For example, the Intelligent Driver Model (IDM) feedback control law has five parameters that govern longitudinal acceleration as a function of the relative distance and velocity to the lead vehicle: \emph{minimum desired gap}, \emph{desired time headway}, \emph{maximum feasible acceleration}, \emph{preferred deceleration}, \emph{maximum desired speed}~\cite{Treiber2000}.
    Some other examples of ``style'' parameters include \emph{aggressiveness}~\cite{Sunberg2017a} and \emph{politeness factor}~\cite{Kesting}.

Some models encode driver preferences in a parametric cost (or reward) function that drivers are assumed to be ``trying to optimize.''\footnote{The notion of a reward function can encompass both intentions and traits. For example, the discrete decision to change lanes can be viewed as a simple result of optimizing a reward function.}
  \textbf{Reward function parameters} are distinguished from policy parameters---though they often correspond to the same intuitive notions (e.g., preferred velocity, etc.)---because they parametrize a reward function rather than a closed-loop control policy.

A few models incorporate \textbf{attention parameters} to model whether (or to what extent) a driver is attentive to the driving task~\cite{Driggs-Campbell2015b}.
  In a similar vein are models that reason about \textbf{physiological traits} like ``reaction time''~\cite{Khodayari2012}.
    Such models are underrepresented in this survey (as we focus on higher-level behavior), but the interested reader may wish to consult the 2003 review by \citeauthor{Macadam2003}~\cite{Macadam2003} and the 2007 review by \citeauthor{Plochl}~\cite{Plochl}.

Finally, the \textbf{non-interpretable parameters} of black box (e.g., neural networks, Gaussian mixture models) policy models can be considered an implicit representation of driver traits.

\subsubsection{Trait Hypothesis}
In almost all cases, the \emph{trait hypothesis} $P(\internalstate{i}{\currenttime})$ is represented by a deterministic \textbf{point estimate} rather than a distribution.
  There are, however, a few notable exceptions:
  Several models employ \textbf{particle distributions} to represent a belief $P(\internalstate{i}{\currenttime})$ over policy parameters for individual agents~\cite{Sunberg2017a,Hoermann2017,Buyer2019,Bhattacharyya2020}.
  \citeauthor{Sadigh2018} maintain a \textbf{discrete distribution} over possible trait ``clusters'' to which a particular driver's reward function parameters might belong~\cite{Sadigh2018}.
  A few models employ \textbf{continuous distributions}, including a Gaussian distribution over policy parameters~\cite{Monteil2015} and a log-concave distribution over reward function parameters~\cite{Sadigh2017}. 


\subsubsection{Trait Inference Paradigms}
Finally, we consider the major inference paradigms that characterize how existing trait estimation models actually compute the trait hypothesis.

Trait estimation can be performed \textbf{offline} or \textbf{online}.
  In the offline paradigm, estimated trait parameters are computed prior to deploying the model. The selected parameter values usually remain fixed during operation, meaning that they only reflect the population of drivers whose behavior was observed in the training data.\footnote{In some ADAS applications, offline trait estimation is used to create individualized ``driver profiles''~\cite{Armand}.} 
  In the online trait estimation paradigm, models reason in real time about the traits of currently observed---perhaps previously unobserved---drivers. 
    Online estimation thus allows greater flexibility than the offline paradigm, but with the additional constraints of real-time operation.
  Some models combine the two paradigms by computing a prior parameter distribution offline, then tuning it online.
    Such online tuning procedures often rely on \textbf{Bayesian} methods.

Many trait estimation models employ an \textbf{optimization} algorithm to fit the values of driver trait parameters to a dataset.
  One particular class of optimization methods is \textbf{inverse reinforcement learning} (IRL), also known as inverse optimal control, which is used to infer the parameters of a reward function from observed behavior.
    Most optimization-based methods are offline, but a notable counterexample is the work of \citeauthor{Galceran2017a}, who use online maximum likelihood estimation to simultaneously regress policy parameters and infer driver intention \cite{Galceran2017a}. 
      The ``Training'' keywords section of \cref{tab:trait_estimation_keywords} identifies various algorithms used for optimization-based trait estimation.
    
Trait parameters are often set \textbf{heuristically} (i.e., manually).
  Indeed, many commonly used parametric models (e.g., \cite{Treiber2000}) come with ``recommended'' parameter settings.
Specifying parameter values manually is a way to incorporate expert domain knowledge into models. Optimization-based methods allow for a closer and more nuanced fit to actual recorded driving behavior, but also run the risk of overfitting to training data.

Some approaches allow parameter values to vary based on the region of the state space or the current behavior mode. 
  Such \textbf{contextually varying} trait models essentially define adaptive controllers, where adaptation laws can be stochastic or deterministic, and can be learned from data or specified heuristically.
  For example, \citeauthor{Liebner} use K-Means clustering to compute a set of geographically varying velocity profiles, which are then used to define a motion model whose parameters change as a function of road position and behavior mode~\cite{Liebner}.



\subsection{Motion Prediction}\label{sec:motion_prediction}
The objective of \emph{motion prediction} is to predict the future trajectories of the vehicles in a given situation, starting from the current time.
This is the canonical task of driver behavior modeling, in the sense that it involves reasoning about the variables ($\state{1:\numcars}{\currenttime+1:\finaltime}$) that have the most direct influence on the ego vehicle's motion planning.
Motion prediction models can be compared in terms of their \emph{state transition functions}, their \emph{scene-level} and their \emph{agent-level} motion hypotheses, and their \emph{prediction paradigms}. \Cref{tab:motion_prediction} categorizes \nummotionprediction{} reviewed models based on these considerations. \Cref{tab:motion_prediction_keywords} introduces keywords associated with these motion prediction models.

\subsubsection{State-Transition models}\label{sec:state_transition_models}

\begin{figure}[t]
  \centering
  \includegraphics[width=\linewidth]{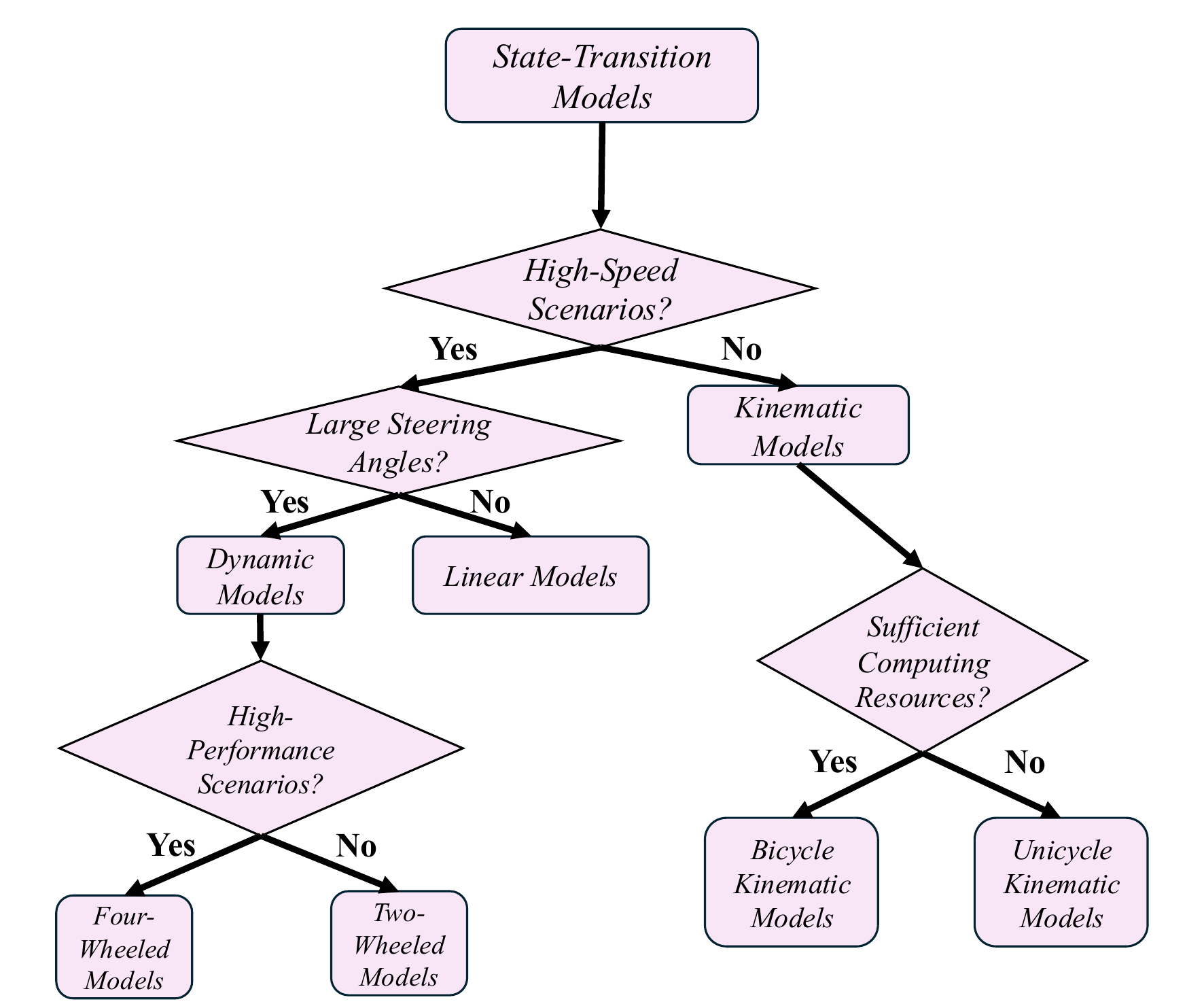}
  \caption{Decision tree for selecting state-transition models based on the modeling problem. Common state-transition models include dynamic models, kinematic models, and linear models. Dynamic models analyze the forces generated at the tire-road interface and are suitable for modeling scenarios involving strong tire forces, such as high-speed and large steering angle scenarios. Kinematic models assume zero slippage at the tire-road interface and are generally used in low-speed scenarios where tire forces are generally weak and negligible. Linear models compute affine approximations of the state transitions and are most applicable for highway-speed scenarios that do not involve turns, wherein the position of the vehicle evolves in a roughly linear fashion.}
  \label{fig:state_transition_models}
\end{figure}



Within our POSG framework, the state transition function $\transitionmodel{}$ encodes assumptions about how the physical state of a vehicle evolves over time as a driver executes control inputs.
  The choice of state transition model can influence the degree to which predicted motion is physically realizable.
  Vehicle models used in the driver modeling literature can be classified into several partially overlapping categories.\footnote{
    Some models either do not incorporate a state-transition model or simply fail to describe the specific model they use. The corresponding rows of \cref{tab:motion_prediction} are left blank.}
    \Cref{fig:state_transition_models} offers model recommendations based on the characteristics of the modeling problem.

Many state-transition models are physics- and/or geometry-based, but some are purely \textbf{learned}, in the sense that the observed correlation between consecutive states $\state{}{t}$ and $\state{}{t+1}$ results entirely from training on large datasets of trajectories.

Dynamic models operate by solving for the forces generated at the tire-road interface, then integrating these forces over time to propagate the vehicle state forward. 
\textbf{Four-wheel} or ``full car'' models explicitly consider all four tires, and often account for factors like nonlinear tire friction and stiffness, longitudinal and lateral load transfer, and suspension design. Such high-fidelity models are usually overkill for driver modeling applications, although exceptions include modeling driver behavior at the extremes of a vehicle's performance envelope~\cite{Lina}.
``Two-wheel'' \textbf{bicycle dynamic} models follow the same principles, but they lump the front wheels and rear wheels into a single wheel per axle. This reduction leads to a slight decrease in model fidelity in exchange for reduced computational complexity and fewer model parameters (e.g., stiffness and friction coefficients).

\textbf{Bicycle kinematic} models have the same two-wheel geometry as bicycle dynamic models, but they assume that the tires experience zero slippage. The no-slip assumption means that motion is computed purely from the geometry of the vehicle model---without reasoning about forces at the tire-road interface. Kinematic models have fewer parameters and are computationally less expensive than dynamic models, but can exhibit significant modeling errors in situations (e.g., at high speeds and large steering angles) where the no-slip assumption would require impossibly high tire forces.

``Single wheel'' \textbf{unicycle} models treat the vehicle as a point mass with a single point of road contact. 
It is generally assumed that the vehicle moves in the direction of its heading angle (no-slip), although certain models allow for lateral side-slip.

\textbf{Linear} state-transition models can be first order (i.e., output is position, input is velocity), second order (i.e., output is position, input is acceleration), and so forth. Longitudinal and lateral motion are usually decoupled, often with different order equations of motion (e.g., second order longitudinal dynamics with first order lateral dynamics).
    Linear models are usually not suitable for sitations where vehicles turn, although this can be partially alleviated by the common technique of adopting a curvilinear Frenet (lane-centric) coordinate system.
    The modeling error induced by the assumption of linear dynamics depends on the application. At highway speeds, for example, this assumption can be quite reasonable so long as lateral acceleration remains low.


A few state-transition models are based on parametric \textbf{splines} such as cubic splines, quintic splines, Chebychev polynomials, and B\'ezier curves. In such cases, the relevant spline interpolation equations constitute the state-transition function.
\textbf{Discrete} state-transition functions (i.e., the state space is discrete) are sometimes employed in applications where low-level vehicle dynamics can be abstracted away.
\textbf{Probabilistic} state-transition models capture uncertainty over the future state as a function of the current state and action. 


\subsubsection{Motion Hypothesis}\label{sec:motion_hypothesis}

The motion hypothesis refers to the way a model encodes uncertainty about $\state{1:\numcars}{\currenttime+1:\finaltime}$, the future trajectories of agents in the scene from the next time step $\currenttime+1$ to some prediction horizon $\finaltime$.
  We find it useful to discuss the various forms of motion hypothesis in terms of \emph{agent-level} representation and \emph{scene-level} representation.


On the agent-level, deterministic motion hypothesis representations from the literature include \textbf{single trajectories} (i.e., a single sequence of states per target agent), \textbf{sequences of bounding boxes}, and \textbf{splines}. 
Probabilistic agent-level motion hypotheses (i.e. distributions over trajectories) include unimodal \textbf{Gaussian} distributions, \textbf{Gaussian mixture} distributions, and \textbf{particle sets}.
  Some approaches use probabilistic \textbf{occupancy distributions}, created by binning a continuous space into finite cells and modeling the probability that any one cell is occupied at a given time.~\footnote{Occupancy distributions are called occupancy grids when the binning is rectilinear.}
  Agent-level motion hypotheses can also be represented in terms of reachable sets. A \textbf{forward reachable set} encodes the full set of states that a vehicle is be able to reach over a specified window. A \textbf{backward reachable set} encodes the set of joint states in which one vehicle would be able to force a collision with another. Reachability analysis is useful e.g., for worst-case analysis and the development of controllers that would be robust even to adversarial human behavior.
  Some models use \textbf{empirical reachable sets}, which balance the robustness of reachable sets and the expressiveness of probabilistic hypotheses.

At the scene level, \textbf{multi-scenario} motion prediction models reason about the different possible scenarios that may follow from an initial traffic scene, where each scenario is usually (though not necessarily) characterized by a unique combination of predicted behavior modes for each participant in the traffic scene.
Other models reason only about a \textbf{single scenario}, ignoring multimodal uncertainty at the scene level.
  Some models reason only about a \textbf{partial scenario}, meaning they predict the motion of only a subset of vehicles in the traffic scene, usually under a single scenario.

The motion hypothesis can also be represented as a \textbf{belief tree}.
  Belief trees generalize discrete distributions over scenarios because each individual node of the belief tree might represent a partial scenario.
    In most cases, the belief tree category applies to models that represent the motion hypothesis (perhaps implicitly) as part of behavior planning.

\subsubsection{Motion Prediction Paradigms}\label{sec:motion_prediction_paradigm}

\begin{figure*}[t]
  \centering
  \includegraphics[width=\linewidth]{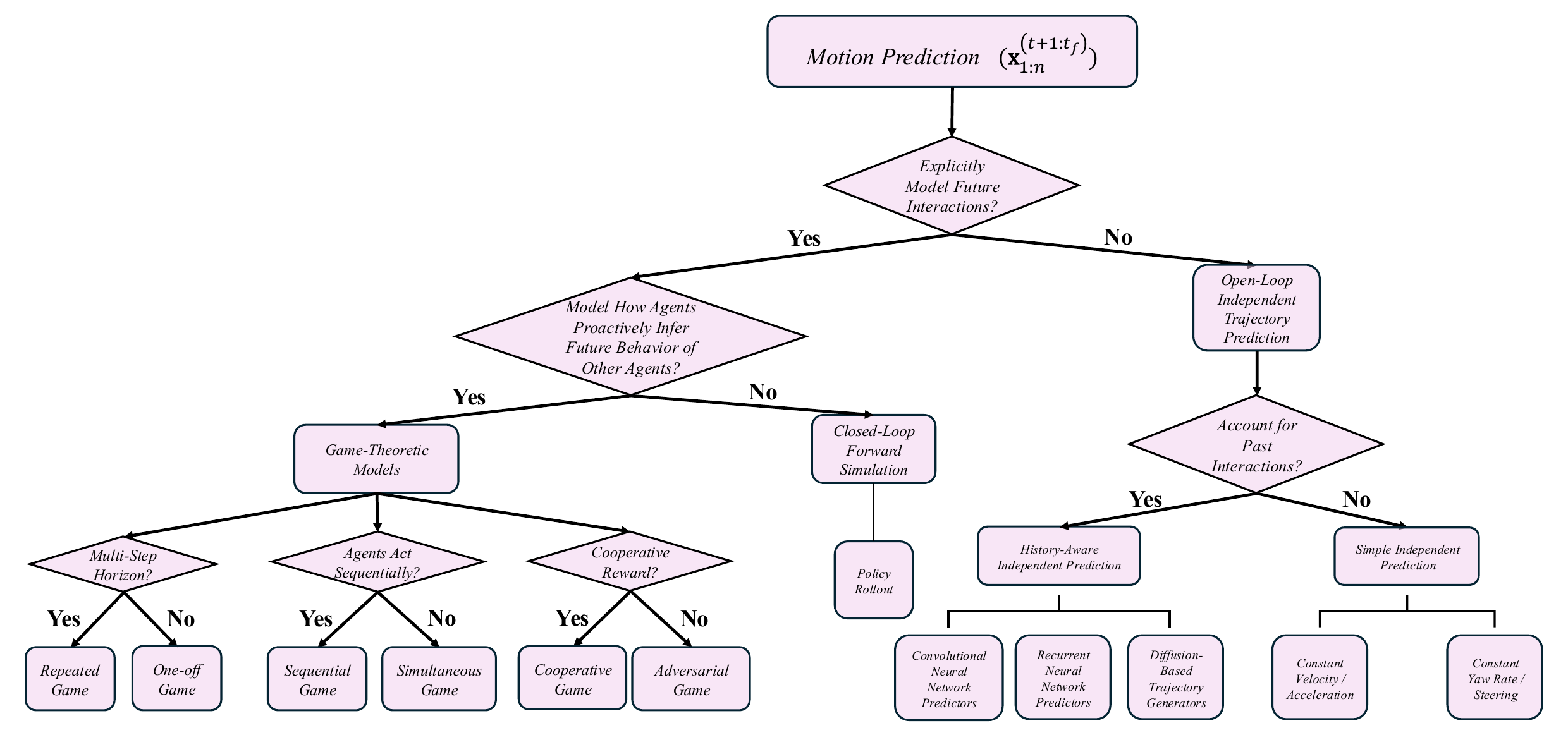}
  \caption{Motion prediction is the task of forecasting the future physical state trajectories of the surrounding vehicles. Motion prediction paradigms include interaction-aware prediction and open-loop independent trajectory prediction. Interaction-aware prediction paradigms explicitly model the future interactions among the surrounding vehicles. These include closed-loop forward simulation methods, which model how each vehicle reacts to the behavior of other vehicles through a rollout policy, and game-theoretic motion prediction paradigms, which model how each vehicle proactively predicts the future motion of other vehicles during decision-making. Open-loop independent trajectory prediction paradigms forecast the future motion of each vehicle independently. These include models that assume constant velocity, acceleration, yaw rate, or steering angles. More advanced independent prediction models account for the historical interactions among the vehicles.}
  \label{fig:motion_prediction_paradigms}
\end{figure*}

\begin{table*}[t]
    \caption{A comparison of the four main motion prediction paradigms in terms of long-horizon accuracy, computational efficiency, interaction awareness, and scalability to multiple agents. 
    }
    {\small
    \renewcommand{\arraystretch}{1.5}
    \begin{tabular*}{\linewidth}{@{}l @{\extracolsep{\fill}}cccc}
        \toprule
        \makecell[l]{\textbf{Motion} \\ \textbf{Prediction Paradigms}} &
        \makecell{\textbf{Long-horizon} \\ \textbf{accuracy}} &
        \makecell{\textbf{Computa-} \\ \textbf{tional Efficiency}} &
        \makecell{\textbf{Interaction} \\ \textbf{Awareness}} &
        \makecell{\textbf{Scalability to} \\ \textbf{Multiple Agents}} \\
        \midrule
        \emph{Closed-loop forward simulation} & high & low & high & low \\ 
        \emph{Simple independent prediction} & very low & very high & very low & very high\\ 
        \emph{History-aware independent prediction} & low & high & medium & high \\ 
        \emph{Game-theoretic models} & very high & very low & very high & very low \\
        \bottomrule
        \hspace{1pt}
    \end{tabular*}
    }
    \label{tab:motion_prediction_table}
\end{table*}

A model's motion prediction paradigm describes how the model actually computes the motion hypothesis.
  Existing approaches can be loosely grouped according to three key paradigms: Closed-loop forward simulation, open-loop independent trajectory prediction, and game theoretic prediction.
  {\Cref{fig:motion_prediction_paradigms} illustrates the structure of our taxonomy. \Cref{tab:motion_prediction_table} evaluates the major motion prediction paradigms based on a variety of metrics.

In the \textbf{forward simulation} paradigm, a motion hypothesis is computed by \emph{rolling out} a closed-loop control policy $\policymodel{}$ for each target vehicle.
    The model computes a control action for each agent at each time step based on the observations received up to and including that time step, then propagates the entire scene forward in time.
    This process is repeated until the prediction horizon $\finaltime$ is reached.
    Algorithm \ref{alg:forward_simulation} is a generic version of forward simulation prediction.

\begin{algorithm}
  \caption{Motion Prediction via Forward Simulation}
  \label{alg:forward_simulation}
  \begin{algorithmic}
    \For{$\tau \in \currenttime, \ldots, \finaltime-1$}
      \For{$i \in 1, \ldots, \numcars$}
        \State $\obs{i}{\tau} \gets \obsmodel{i}(\state{1:n}{\tau})$ \Comment{receive observation}
        \State $\internalstate{i}{\tau} \gets \internaltransitionmodel{i}(\internalstate{i}{\tau-1},\obs{i}{\tau})$ \Comment{update internal state}
        \State $\ctrlaction{i}{\tau} \gets \policymodel{i}(\internalstate{i}{\tau})$ \Comment{select action}
        \State $\state{i}{\tau+1} \gets \transitionmodel{i}(\state{i}{\tau},\ctrlaction{i}{\tau})$ \Comment{step forward}
      \EndFor
    \EndFor
  \end{algorithmic}
\end{algorithm}

Forward simulation can be performed with a deterministic state representation or a probabilistic state representation (e.g., a Gaussian state estimate~\cite{Bahram2016}).
  Some models reason about multimodal scene-level uncertainty by performing multiple (parallel) rollouts associated with different scenarios.

Motion prediction algorithms based on forward simulation can be described as nominally ``interaction aware.''
  The level of ``interaction-awareness'' depends on $\obsmodel{i}$, which encodes the model's assumptions about what agent $i$ observes at each time step, as well $\internaltransitionmodel{i}$ and $\policymodel{i}$, which encode what the agent does with that information. In most cases, $\internaltransitionmodel{i}$ simply updates the simulated agent's ``mental model'' of the surrounding environment (especially the states of other agents).\footnote{Most motion prediction models assume that the intentions and traits of each driver remain fixed throughout the prediction window.}
    A deeper notion of ``interaction awareness'' will be discussed in the context of game theoretic prediction paradigms.


The closed-loop action policies used in forward simulation-based models can take many forms.
    Simple examples include rule-based heuristic control laws, e.g., \cite{Treiber2000,Gipps1981,Kesting,Dagli2003,Krauss}.
      Many policy models are extensions of the IDM \cite{Treiber2000}.
      More sophisticated examples include closed-loop policies based on neural networks, dynamic Bayesian networks, and random forests.

Many models operate under the \textbf{independent prediction} paradigm, meaning that they predict a full trajectory independently for each agent in the scene.
  These approaches are more-or-less ``interaction-unaware'' because they are open-loop; though they may account for interaction between vehicles at the current time $\currenttime$, they do not explicitly reason about interaction over the prediction window from $\currenttime+1$ to $\finaltime$ (i.e., they do not reason about the drivers' observations $\obs{1:\numcars}{\currenttime+1:\finaltime}$ over that time interval).

The simplest motion prediction models in this class are based on various combinations of constant velocity, constant acceleration, constant yaw rate, and constant steering angle.
  Such models are truly interaction-unaware, because they rely exclusively on information about the target vehicle's current physical state.

On the other hand, some advanced independent prediction models could be described as ``implicitly'' interaction-aware because they account for much of the interaction between vehicles up to and/or including the current time step $\currenttime$.
      For example, \citeauthor{Deoa} employ convolutional social pooling to encode information about the scene history, which is then passed to a set of recurrent neural network models that generate independent predictions for all vehicles in the scene \cite{Deoa}. 
      More recently, diffusion models have been applied to realistic independent motion prediction, especially in the context of safety validation and robust planning. For instance, guided diffusion models have demonstrated the capability to independently generate potentially risky future trajectories of the background vehicles based on the scene history encoding \cite{chen2023advdiffuser}. 
      To enable the generation of possible background vehicle trajectories with specific characteristics of interest, partial diffusion processes have been utilized to provide additional control during trajectory generation \cite{chang2025safe}. 
      While these approaches do not explicitly consider interaction over the prediction window, they infer future interactions based on scene histories and are clearly more interaction-aware than, e.g., a constant velocity model.

One key advantage of independent prediction models is that they can be parallelized and made to run very fast. Moreover, some achieve excellent performance over short time horizons.
  Because these models don't explicitly account for interaction, however, their predictive power tends to quickly degrade as the prediction horizon extends further into the future.

Finally, some motion prediction models operate within a \textbf{game-theoretic} motion prediction paradigm, 
  meaning that they explicitly condition the predicted future motion of some agents on the predicted future motion of other agents in the scene. In other words, agents are modeled as ``looking ahead'' to reason about each others' behavior.
    This notion of looking ahead makes game-theoretic prediction models more deeply ``interaction aware'' than forward simulation models based on reactive closed-loop control.
    Models in this category are not always explicitly formulated as games, but it is helpful to examine them through the lens of formally defined games.



Game theoretic models vary in duration---each agent plays multiple times in a repeated game, but just once in a one-off game. 
  \citeauthor{Fisac2019} solve a two-player repeated game as part of a hierarchical game formulation for highway merging and overtaking \cite{Fisac2019}. \citeauthor{Sadigh2018} formulate a generic human-robot interaction problem as a repeated game~\cite{Sadigh2018}. \citeauthor{Isele2019} search over a game tree to solve a repeated game merging problem \cite{Isele2019}.
    Most other game-theoretic models reason about long horizons but avoid the complexity of a repeated game by defining a one-off game in which the actions are full trajectories.
  
Game theoretic models also vary in information structure (i.e., order of play)---agents play all at once in a simultaneous game, but take turns in a sequential (Stackelberg) game~\cite{HongYoo2012,HongYoo2013,Schester2019,Fisac2019}.
  The latter category notably includes continuous space two-player ``best-response'' formulations where one driver is modeled as having full access to the planned paths of the other vehicle~\cite{Sadigh2018, Schmerling2017, Leung}. 
  Most reviewed models are (implicitly) based on a simultaneous game structure where there is no information advantage.

Game theoretic models can also vary in reward structure---agents work together in a cooperative game, but against each other in an adversarial (zero-sum) game.
  Examples of cooperative games include formulations based on joint trajectory optimization (e.g., \cite{Schulz2017}).
  Most other reviewed models sit somewhere along the spectrum between fully cooperative and fully adversarial games.



Game theoretic models are well-suited for motion prediction in highly interactive situations (e.g., highway merging, unsignalized intersection navigation, etc.). Well-posed game formulations can be solved exactly (e.g., \cite{Pruekprasert2019,HongYoo2012,HongYoo2013}) by computing Nash equilibria \cite{Nash1951}, but exact methods tend to scale poorly to problems with many agents or large state- and action-spaces.
  Hence, many approaches rely on approximate solution techniques to reduce computational complexity. 
  Some models employ recursive reasoning (e.g., level-$k$ reasoning, cognitive hierarchy \cite{camerer2011behavioral}), where trajectory plans are recursively computed for each agent based on the most recent predicted plans for other agents~\cite{Jain2018,Garzon2019}.
  Others use game-theoretic recursive training procedures (for trait estimation), and then apply the trained models in a forward simulation paradigm for prediction~\cite{Oyler2016a}.
    Very long horizon games can be solved in a receding horizon fashion \cite{Sadigh2018}. One approach involves hierarchical decomposition of a game model into a ``short horizon'' game and a ``long horizon'' game \cite{Fisac2019}. Some models use schemes for ignoring interaction between ``independent'' groups of agents \cite{Kuhnt,Jain2018}.

A number of motion prediction frameworks use information gained from driver intention predictions. Such \textbf{intention-aware} motion prediction is partly empowered by advancements in transformer neural network models, which, as discussed in \cref{subsubsection: intention inference paradigms}, improves the effectiveness of driver intention inference. While they generally follow an open-loop independent prediction paradigm, these models operate under two distinct frameworks. \textbf{Dual transformer models} apply multi-head attention layers to both encode recent state trajectories as intention embeddings and decode the intention embeddings into future trajectory predictions \cite{jiang2023intention, gao2023dual}. \textbf{Diffusion-based models} similarly encode recent observations as the context embedding, but instead of explicit trajectory prediction, they use diffusion models to infer the probability distribution over the possible target positions and planned trajectories of the surrounding agents \cite{liu2025intention}.

%% file: Sections/5_Conclusion.tex
\section{Conclusion}\label{sec:conclusion}

In this review article, we have introduced a taxonomy for driver models. We have unified driver modeling tasks such as state estimation, intent estimation, trait estimation, and motion prediction into one framework by casting them as inference problems in a Partially Observable Stochastic Game. While previous reviews of the driver modeling literature focused on the motion prediction problem, we include intent and trait estimation as upstream components that feed into motion prediction.

The objective of intent estimation is to infer what the drivers of surrounding vehicles intend to do in the immediate future. Whereas intentions denote what a driver is trying to do, traits encompass factors that affect how a driver will do so. The objective of motion prediction is to predict the future trajectories of all vehicles in a given situation, starting from the current time. In the POSG framework, intention and trait estimation infer the internal states of the drivers, whereas motion prediction is the problem of inferring the future physical states of other vehicles.

Within each task, we classify models according to the space of variables, hypothesis representations, and inference paradigm. 
    Our classification of the models into the tables presented in this review allows easy access to a model according to the requirements of a downstream task. 
        Further, we classify models according to keywords that help make sense of the vast literature according to the architectures used, training algorithms, theoretical underpinnings, scope and evaluation metrics to assess the models. 
            The aim of this article is to provide a resource to help researchers navigate the complex landscape of driver behavior modeling research.

There are several important future directions for driver modeling. 
Efficient representations are needed for simultaneously modeling both long term and short term driver behavior. 
Promising ideas in this space include options-based reinforcement learning~\cite{du2025hierarchical} and hierarchical model predictive control~\cite{peng2025bilevel}. 
Currently, models are generally trained and tested in the same context. 
However, driver models are required that can generalize across conditions such as weather, lighting, and urban versus rural settings. 
Although there has been some progress on game-theoretic modeling, scalability to the number of vehicles remains an issue. 
Driver models that can account for human response to automated vehicle actions are necessary~\cite{dang2025dynamic}. 
The lack of standardized datasets and environments for rigorous comparison of driver models represents another challenge. 
Driver models should be able to generate realistic human behavior for automated vehicles to train against. 
Promising ideas include generative models and domain adaptation. 
Finally, foundation models provide an opportunity to include common sense knowledge in driver modeling. 
Approaches may include querying vision-language models in the loop~\cite{sinha2024real} and using foundation models for providing rewards to train policies using inverse reinforcement learning.

%% file: Sections/6_Appendices.tex

\section{}\label{sec:appendix}
\nopagebreak[4]



\renewcommand{\floatsep}{0pt}
\renewcommand{\dblfloatsep}{0pt}




\renewcommand{\floatpagefraction}{1.0}

\newcommand{\STAB}[1]{\begin{tabular}{@{}c@{}}\rotatebox[origin=c]{90}{\textbf{#1}}\end{tabular}}

\setlength{\tabcolsep}{3pt}
\renewcommand{\arraystretch}{0.9}
\newcolumntype{J}[1]{>{\raggedright\everypar{\hangindent2em \hangafter1}}p{#1}}
\newcolumntype{K}[1]{>{\tiny\everypar{\hangindent0.5em \hangafter1}}p{#1}}

\clearpage
\addtolength{\belowcaptionskip}{-0.25cm}
\addtolength{\abovecaptionskip}{-0.25cm}
\begin{table*}[hbtp!]
  \centering
  \caption{\textbf{Tasks} addressed by each model: State Estimation (SE), Intention Estimation (IE), Trait Estimation (TE), Motion Prediction (MP), Anomaly Detection (AD), Risk Estimation (RE), Traffic Simulation (TS), Behavior Imitation (Im).  Behavior Planning (BP) denotes that the model is introduced in the context of a behavior planning framework. }
  \label{tab:tasks}
  \renewcommand{\arraystretch}{0.98}
  {\scriptsize
  \setlength{\tabcolsep}{3.6pt}
  \begin{tabular}[t]{@{}l | c c c c c c c c c@{}}
      \toprule
      \multicolumn{1}{l}{\textbf{Ref}} & \multicolumn{9}{c}{\textbf{Tasks Addressed}} \\
      \midrule
      \input{appendices/tables/models-task_view1.tex} \\
      \bottomrule
  \end{tabular}%
  \hspace{.5cm}
  \begin{tabular}[t]{@{}l | c c c c c c c c c@{}}
    \toprule
    \multicolumn{1}{l}{\textbf{Ref}} & \multicolumn{9}{c}{\textbf{Tasks Addressed}} \\
    \midrule
    \input{appendices/tables/models-task_view2.tex} \\
    \bottomrule
  \end{tabular}%
  \hspace{.5cm}
  \begin{tabular}[t]{@{}l | c c c c c c c c c@{}}
    \toprule
    \multicolumn{1}{l}{\textbf{Ref}} & \multicolumn{9}{c}{\textbf{Tasks Addressed}} \\
    \midrule
    \input{appendices/tables/models-task_view3.tex} \\
    \bottomrule
  \end{tabular}
  }
  \hspace{1pt}
\end{table*}

\addtolength{\belowcaptionskip}{0.25cm}
\addtolength{\abovecaptionskip}{0.25cm}

\begin{table*}[hbtp!]
  \centering
  \captionsetup{width=10cm}
  \caption{\textbf{State Estimation keywords} and associated references.}
  \label{tab:state_estimation_keywords}
  {\scriptsize
  \begin{tabular}{@{}c J{4.8cm}| l @{}}
      \toprule
      & \multicolumn{1}{c}{\textbf{Algorithm Keyword}} & \multicolumn{1}{c}{\textbf{References}} \\ 
      \midrule
      \input{appendices/tables/tags-state_estimation_keyword_table_view.tex} \\
      \bottomrule
  \end{tabular}
  }
  \hspace{1pt}
\end{table*}


\clearpage

\addtolength{\belowcaptionskip}{-0.25cm}
\addtolength{\abovecaptionskip}{-0.25cm}
\begin{table*}[hbtp!]
  \centering
  \caption{\textbf{Intention Estimation keywords} and associated references.}
  \label{tab:intent_estimation_keywords}
  \renewcommand{\arraystretch}{1.0}
  {\scriptsize
  \begin{tabular}[t]{@{}c J{4.8cm}|K{3.4cm}@{}}
      \toprule
      & \multicolumn{1}{c}{\textbf{Algorithm Keyword}} & \multicolumn{1}{c}{\textbf{References}} \\ 
      \midrule
      \input{appendices/tables/tags-intention_estimation_keyword_table_view1.tex} \\
      \bottomrule
  \end{tabular}%
  \hspace{0.2cm}
  \begin{tabular}[t]{@{}c J{4.8cm}|K{3.4cm}@{}}
      \toprule
      & \multicolumn{1}{c}{\textbf{Algorithm Keyword}} & \multicolumn{1}{c}{\textbf{References}} \\ 
      \midrule
      \input{appendices/tables/tags-intention_estimation_keyword_table_view2.tex} \\
      \bottomrule
  \end{tabular}
  }
  \hspace{1pt}
\end{table*}

\begin{table*}[hbtp!]
  \centering
  \caption{\textbf{Intention Estimation} models classified according to
  \textbf{Intention Space}: Routes through road network (Ro), Joint Configurations (Co), Longitudinal modes (Lo), Lateral modes (La), Emergency Maneuvers (EM), Intentional vs. Unintentional (IU), Context Dependent (CD); 
  \textbf{Hypothesis Representation}: Point Estimate (S), Discrete Distribution (D), Discrete Distribution over scenarios (DS); 
  \textbf{Estimation Paradigm}: Recursive (Re), Single-Shot (SS), Bayesian (Ba), Black Box (BB), Maneuver Prototype (MP), Game Theoretic (GT).}
  \renewcommand{\arraystretch}{1.0}
  {\scriptsize
  \setlength{\tabcolsep}{3.1pt}
  \begin{tabular}[t]{@{}l|c c c c c c c|c c c c|c c c c c c@{}}
      \toprule
      \textbf{Ref} & \multicolumn{7}{c}{\textbf{Intention Space}} & \multicolumn{4}{c}{\textbf{Hypoth.}} & \multicolumn{6}{c}{\textbf{Paradigm}} \\ 
      \midrule
      \input{appendices/tables/models-intention_estimation_view1.tex} \\
      \bottomrule
  \end{tabular}%
  \hspace{0.5cm}
  \begin{tabular}[t]{@{}c|c c c c c c c|c c c c|c c c c c c@{}}
      \toprule
      \textbf{Ref} & \multicolumn{7}{c}{\textbf{Intention Space}} & \multicolumn{4}{c}{\textbf{Hypoth.}} & \multicolumn{6}{c}{\textbf{Paradigm}} \\ 
      \midrule
      \input{appendices/tables/models-intention_estimation_view2.tex} \\
      \bottomrule
  \end{tabular}
  }
  \hspace{1pt}
  \label{tab:intent_estimation}
\end{table*}

\clearpage

\begin{table*}[hbtp!]
  \centering
  \caption{\textbf{Trait Estimation keywords} and associated references.}
  \label{tab:trait_estimation_keywords}
  \renewcommand{\arraystretch}{1.0}
  {\scriptsize
  \begin{tabular}[t]{@{}c J{4.8cm}|K{3.4cm}@{}}
      \toprule
      & \multicolumn{1}{c}{\textbf{Algorithm Keyword}} & \multicolumn{1}{c}{\textbf{References}} \\ 
      \midrule
      \input{appendices/tables/tags-trait_estimation_keyword_table_view1.tex} \\
      \bottomrule
  \end{tabular}%
  \hspace{0.4cm}
  \begin{tabular}[t]{@{}c J{4.8cm}|K{3.4cm}@{}}
      \toprule
      & \multicolumn{1}{c}{\textbf{Algorithm Keyword}} & \multicolumn{1}{c}{\textbf{References}} \\ 
      \midrule
      \input{appendices/tables/tags-trait_estimation_keyword_table_view2.tex} \\
      \bottomrule
  \end{tabular}
  }
  \hspace{1pt}
\end{table*}

\begin{table*}[hbtp!]
  \centering
  \caption{\textbf{Trait Estimation} models classified according to:
      \textbf{Trait Space:} Control policy parameters (Pa), Reward model parameters (Re), Non-interpretable control policy parameters (NP), Physiological trait parameters (Ph), Attention parameters (At)
      \textbf{Hypothesis Representation:} Point Estimate (S), Continuous Distribution (C), Particle set (P)
      \textbf{Paradigm:} Online (On), Offline (Off), Heuristic (H), Optimization (Op), Bayesian (B), Inverse Reinforcement Learning (IRL), Contextually Varying (CV). }
  \label{tab:trait_estimation}
  \renewcommand{\arraystretch}{1.0}
  {\scriptsize
  \setlength{\tabcolsep}{3.9pt}
  \begin{tabular}[t]{@{}l|c c c c c|c c c|c c|c c c c c@{}}
      \toprule
      \textbf{Ref} & \multicolumn{5}{c}{\textbf{Trait Space}} & \multicolumn{3}{c}{\textbf{Hypoth.}} & \multicolumn{2}{c}{\textbf{Paradigm}} & \multicolumn{5}{c}{\textbf{Model Class}} \\ 
      \midrule
      \input{appendices/tables/models-trait_estimation_view1.tex} \\
      \bottomrule
  \end{tabular}%
  \hspace{0.5cm}
  \begin{tabular}[t]{@{}l|c c c c c|c c c|c c|c c c c c@{}}
      \toprule
      \textbf{Ref} & \multicolumn{5}{c}{\textbf{Trait Space}} & \multicolumn{3}{c}{\textbf{Hypoth.}} & \multicolumn{2}{c}{\textbf{Paradigm}} & \multicolumn{5}{c}{\textbf{Model Class}} \\ 
      \midrule
      \input{appendices/tables/models-trait_estimation_view2.tex} \\
      \bottomrule
  \end{tabular}
  }
  \hspace{1pt}
\end{table*}

\begin{table*}[hbtp!]
  \centering
  \caption{\textbf{Motion Prediction keywords} and associated references.}
  \label{tab:motion_prediction_keywords}
  \renewcommand{\arraystretch}{0.9}
  {\scriptsize
  \begin{tabular}[t]{@{}c J{4.8cm}|K{3.4cm}@{}}
      \toprule
      & \multicolumn{1}{c}{\textbf{Algorithm Keyword}} & \multicolumn{1}{c}{\textbf{References}} \\ 
      \midrule
      \input{appendices/tables/tags-motion_prediction_keyword_table_view1.tex} \\
      \bottomrule
  \end{tabular}%
  \hspace{0.4cm}
  \begin{tabular}[t]{@{}c J{4.8cm}|K{3.4cm}@{}}
      \toprule
      & \multicolumn{1}{c}{\textbf{Algorithm Keyword}} & \multicolumn{1}{c}{\textbf{References}} \\ 
      \midrule
      \input{appendices/tables/tags-motion_prediction_keyword_table_view2.tex} \\
      \bottomrule
  \end{tabular}
  }
  \hspace{1pt}
\end{table*}

\begin{table*}[hbtp!]
  \centering
  \caption{\textbf{Motion prediction} models classified according to:
        \textbf{Vehicle dynamics model:} (\emph{Veh. Model}):] Four Wheel (4W), Bicycle Dynamic (BD), Bicycle Kinematic (BK), Unicycle (U), Linear (L), Spline (S), Discrete(Dc), Probabilistic (P), Learned (X).
        \textbf{Scene-level uncertainty modeling:} Multi-Scenario (M), Single-Scenario (S), Partial Scenario (P), Belief Tree (Tr).
        \textbf{Agent-level uncertainty modeling:} Gaussian (G), Gaussian Mixture (GM), Particle Set (P), Single deterministic (S), Discrete Occupancy Distribution (O), Bounding Box (BB), Spline (Sp), Reachable Set (R), Backward Reachable Set (bR).
        \textbf{Prediction paradigm:} Forward Simulation (FS), Independent Prediction (IP), Game Theoretic (GT).}
  \label{tab:motion_prediction}
  \renewcommand{\arraystretch}{1.0}
  {
  \setlength{\tabcolsep}{1.0pt}
  \newcommand{\vlsep}{\hspace{4pt}}
  \scriptsize
  \begin{tabular}[t]{@{}l@{\vlsep}|@{\vlsep} c c c c c c c c c @{\vlsep}|@{\vlsep} c c c c @{\vlsep}|@{\vlsep} c c c c c c c c c@{\vlsep}|@{\vlsep} c c c@{}}
      \toprule
      \textbf{Ref} & \multicolumn{9}{c}{\textbf{Veh. Model}} & \multicolumn{4}{c}{\textbf{Scene}} & \multicolumn{9}{c}{\textbf{Agent}} & \multicolumn{3}{c}{\textbf{Paradigm}}\\ 
      \midrule
      \input{appendices/tables/models-motion_prediction_view1.tex} \\
      \bottomrule
  \end{tabular}%
  \hspace{0.5cm}
  \begin{tabular}[t]{@{}l@{\vlsep}|@{\vlsep} c c c c c c c c c @{\vlsep}|@{\vlsep} c c c c @{\vlsep}|@{\vlsep} c c c c c c c c c@{\vlsep}|@{\vlsep} c c c@{}}
      \toprule
      \textbf{Ref} & \multicolumn{9}{c}{\textbf{Veh. Model}} & \multicolumn{4}{c}{\textbf{Scene}} & \multicolumn{9}{c}{\textbf{Agent}} & \multicolumn{3}{c}{\textbf{Paradigm}}\\ 
      \midrule
      \input{appendices/tables/models-motion_prediction_view2.tex} \\
      \bottomrule
  \end{tabular}
  }
  \hspace{1pt}
\end{table*}

%% file: appendices/tables/models-task_view1.tex
\cite{Abbeela} & -- & -- & TE & MP & -- & -- & -- & -- & -- \\ 
\cite{Agamennoni2012b} & SE & IE & -- & MP & -- & -- & -- & -- & -- \\ 
\cite{Akita} & -- & IE & TE & MP & -- & -- & -- & -- & -- \\ 
\cite{Alizadeh2019} & -- & -- & -- & MP & -- & -- & -- & TS & BP \\ 
\cite{Altche2018} & -- & -- & -- & MP & -- & -- & -- & -- & -- \\ 
\cite{Althoff2009} & -- & -- & -- & MP & RE & -- & -- & -- & -- \\ 
\cite{Althoff2011a} & -- & -- & -- & MP & RE & -- & -- & -- & -- \\ 
\cite{Ammoun2009} & SE & -- & -- & MP & RE & -- & -- & -- & -- \\ 
\cite{Angkititrakul2009} & -- & IE & TE & MP & -- & -- & -- & -- & -- \\ 
\cite{Aoude2010a} & -- & IE & -- & -- & RE & -- & -- & -- & BP \\ 
\cite{Aoudea} & -- & IE & -- & -- & RE & -- & -- & -- & BP \\ 
\cite{Aoudeb} & -- & IE & -- & -- & -- & -- & -- & -- & -- \\ 
\cite{arbabi2022learning} & -- & -- & -- & MP & -- & -- & -- & TS & -- \\
\cite{Armand} & -- & -- & TE & MP & -- & -- & -- & -- & -- \\ 
\cite{Asljung2019} & -- & -- & -- & MP & RE & -- & -- & -- & -- \\ 
\cite{Bahram2016} & SE & IE & -- & MP & -- & -- & -- & -- & -- \\ 
\cite{Bansal2018} & -- & -- & -- & MP & -- & -- & -- & -- & BP \\
\cite{bao2023personalized} & -- & -- & TE & -- & -- & -- & -- & -- & BP \\
\cite{Barbier2017} & -- & IE & -- & -- & -- & -- & -- & -- & -- \\ 
\cite{Barth2008} & SE & -- & -- & MP & -- & -- & -- & -- & -- \\ 
\cite{Batza} & -- & -- & -- & MP & RE & -- & -- & -- & BP \\ 
\cite{Bender2015} & SE & IE & -- & -- & -- & -- & -- & -- & -- \\ 
\cite{Berndt2008} & -- & IE & -- & -- & -- & -- & -- & -- & -- \\ 
\cite{Bhattacharyya2020} & -- & -- & TE & MP & -- & -- & -- & TS & -- \\ 
\cite{Bonnin2012a} & -- & IE & -- & -- & -- & -- & -- & -- & -- \\ 
\cite{Bouton2019} & -- & IE & -- & MP & -- & -- & -- & -- & BP \\ 
\cite{Brannstrom2010a} & -- & -- & -- & -- & RE & -- & -- & -- & BP \\ 
\cite{Brechtel2014} & -- & IE & -- & MP & -- & -- & -- & -- & BP \\ 
\cite{Buyer2019} & -- & IE & TE & -- & -- & -- & -- & -- & -- \\ 
\cite{Chen2010} & -- & -- & TE & MP & -- & -- & -- & TS & -- \\
\cite{chen2023trajectory} & -- & -- & -- & MP & -- & -- & -- & -- & -- \\
\cite{chenescirl} & -- & -- & -- & MP & -- & -- & -- & -- & BP \\
\cite{cultrera2024addressing} & SE & -- & -- & -- & -- & -- & -- & TS & BP \\
\cite{Dagli2003} & -- & IE & -- & -- & -- & -- & -- & -- & -- \\ 
\cite{Dang2018} & -- & IE & -- & MP & -- & -- & -- & -- & -- \\ 
\cite{Demcenko2009a} & -- & -- & -- & MP & -- & -- & -- & -- & -- \\ 
\cite{Deng2019} & -- & IE & -- & -- & -- & -- & -- & -- & -- \\ 
\cite{Deo} & SE & IE & -- & MP & -- & -- & -- & -- & -- \\ 
\cite{Deoa} & -- & IE & -- & MP & -- & -- & -- & -- & -- \\ 
\cite{Deob} & -- & IE & -- & MP & -- & -- & -- & -- & -- \\ 
\cite{DiazAlvarcz2019} & -- & -- & -- & MP & -- & -- & -- & -- & -- \\ 
\cite{Diehl2019} & -- & -- & -- & MP & -- & -- & -- & -- & -- \\ 
\cite{Dong2017Merging} & -- & IE & -- & -- & -- & -- & -- & -- & BP \\ 
\cite{Doshi} & -- & IE & -- & -- & -- & -- & -- & -- & -- \\ 
\cite{Driggs-Campbell2015b} & -- & IE & TE & MP & -- & -- & -- & -- & -- \\ 
\cite{Driggs-Campbell2015c} & -- & IE & -- & -- & -- & -- & -- & -- & -- \\ 
\cite{Driggs-Campbell2017b} & -- & -- & -- & MP & -- & -- & IM & -- & BP \\ 
\cite{Driggs-Campbell2017c} & -- & IE & -- & MP & -- & -- & -- & -- & -- \\ 
\cite{Eggert} & -- & -- & TE & MP & -- & -- & -- & -- & -- \\ 
\cite{Feng2019} & -- & IE & -- & MP & -- & -- & -- & -- & -- \\ 
\cite{Ferguson} & -- & IE & -- & MP & -- & -- & -- & -- & BP \\ 
\cite{Fisac2019} & -- & -- & -- & MP & -- & -- & -- & -- & BP \\ 
\cite{Galceran2017a} & -- & IE & TE & MP & -- & AD & -- & -- & BP \\ 
\cite{gao2023dual} & -- & IE & -- & MP & -- & -- & -- & -- & -- \\ 
\cite{Garzon2019} & -- & -- & -- & MP & -- & -- & -- & -- & BP \\ 
\cite{Gebert2019} & -- & IE & -- & -- & -- & -- & -- & -- & -- \\ 
\cite{Gillmeier2019} & -- & IE & -- & MP & -- & -- & -- & -- & -- \\ 
\cite{Gindele} & SE & IE & -- & MP & -- & -- & -- & -- & -- \\ 
\cite{Gindele2015} & SE & IE & -- & MP & -- & -- & -- & -- & -- \\ 
\cite{Gipps1981} & -- & -- & -- & MP & -- & -- & -- & TS & -- \\ 
\cite{Gonzalez2019} & -- & IE & -- & MP & -- & -- & -- & -- & BP \\ 
\cite{Graf2019} & -- & -- & -- & MP & -- & -- & -- & -- & BP \\ 
\cite{Gray2013a} & -- & -- & -- & MP & -- & -- & -- & -- & BP \\ 
\cite{Guo2019} & -- & -- & -- & MP & -- & -- & -- & -- & -- \\ 
\cite{Guo2019Multi} & -- & IE & -- & MP & -- & -- & -- & -- & -- \\ 
\cite{Han2019} & -- & IE & TE & -- & -- & -- & -- & -- & -- \\ 
\cite{Han2019Short} & -- & -- & -- & MP & -- & -- & -- & -- & -- \\
\cite{haselberger2024situation} & -- & -- & TE & -- & -- & -- & -- & -- & -- \\
\cite{Hardy2013} & SE & IE & -- & MP & -- & -- & -- & -- & BP \\ 
\cite{Hart2019} & -- & -- & -- & MP & -- & -- & -- & -- & BP \\ 
\cite{Hayawaka2019} & SE & -- & -- & -- & -- & -- & -- & -- & -- \\ 
\cite{Hillenbrand2006} & -- & IE & -- & MP & -- & -- & -- & -- & BP \\ 
\cite{Hoermann2017} & SE & -- & TE & MP & -- & -- & -- & -- & -- \\ 
\cite{HongYoo2012} & -- & -- & -- & MP & -- & -- & -- & -- & -- \\ 
\cite{HongYoo2013} & -- & -- & -- & MP & -- & -- & -- & TS & -- \\ 
\cite{Houenou2013} & SE & IE & -- & MP & -- & -- & -- & -- & -- \\ 

%% file: appendices/tables/models-task_view2.tex
\cite{Hu2018a} & -- & IE & -- & MP & -- & -- & -- & -- & -- \\ 
\cite{Hu2018b} & -- & IE & -- & MP & -- & -- & -- & -- & -- \\ 
\cite{Hu2019} & -- & IE & -- & MP & -- & -- & -- & -- & -- \\ 
\cite{Hu2019Generic} & -- & -- & -- & MP & -- & -- & -- & -- & -- \\ 
\cite{Hu2020} & -- & -- & -- & MP & -- & -- & -- & TS & BP \\
\cite{hu2024highway} & -- & IE & -- & MP & -- & -- & -- & -- & -- \\
\cite{huang2021driving} & -- & -- & -- & MP & -- & -- & -- & TS & -- \\
\cite{huang2024driver} & -- & IE & -- & -- & -- & -- & -- & TS & -- \\
\cite{Hubmann2019} & -- & -- & -- & MP & -- & -- & -- & -- & BP \\
\cite{igl2022symphony} & -- & -- & -- & MP & -- & -- & -- & TS & -- \\
\cite{Isele2019} & -- & IE & -- & MP & -- & -- & -- & -- & BP \\
\cite{Jain2016} & -- & IE & -- & -- & -- & -- & -- & -- & -- \\ 
\cite{Jain2018} & -- & -- & -- & MP & -- & -- & -- & -- & -- \\ 
\cite{Jones2019} & -- & -- & -- & MP & -- & -- & -- & -- & -- \\ 
\cite{Jugade2019} & -- & -- & -- & MP & -- & -- & -- & -- & BP \\ 
\cite{Kaempchen2004} & SE & -- & -- & -- & -- & -- & -- & -- & -- \\ 
\cite{Kafer2010} & -- & IE & -- & -- & -- & -- & -- & -- & -- \\ 
\cite{Kesting} & -- & -- & TE & MP & -- & -- & -- & TS & -- \\ 
\cite{Kesting2008} & -- & -- & TE & -- & -- & -- & -- & TS & -- \\ 
\cite{Kesting2010} & -- & -- & -- & MP & -- & -- & -- & TS & -- \\ 
\cite{Khodayari2012} & -- & -- & TE & MP & -- & -- & -- & -- & -- \\ 
\cite{Khosroshahi} & -- & IE & -- & -- & -- & -- & -- & -- & -- \\ 
\cite{Kim2017} & -- & -- & -- & MP & -- & -- & -- & -- & -- \\ 
\cite{Klingelschmitt} & -- & IE & -- & -- & -- & -- & -- & -- & -- \\ 
\cite{Klingelschmitt2015} & -- & IE & -- & -- & RE & -- & -- & -- & -- \\ 
\cite{Klingelschmitt2015a} & SE & -- & -- & MP & -- & -- & -- & -- & -- \\ 
\cite{Klingelschmitt2016} & SE & IE & -- & -- & -- & -- & -- & -- & -- \\ 
\cite{Klingelschmitt2016a} & -- & IE & -- & -- & -- & -- & -- & -- & -- \\ 
\cite{Krajewski2019} & -- & -- & -- & MP & -- & -- & -- & TS & -- \\ 
\cite{Kruger2019} & -- & IE & -- & -- & -- & -- & -- & -- & -- \\ 
\cite{Kuderer2015a} & -- & -- & TE & MP & -- & -- & IM & -- & -- \\ 
\cite{Kuefler2017} & -- & -- & -- & MP & -- & -- & -- & TS & -- \\ 
\cite{Kuhnt} & -- & IE & -- & -- & -- & -- & -- & -- & -- \\ 
\cite{Kumar2013} & SE & IE & -- & -- & -- & -- & -- & -- & -- \\
\cite{kung2024looking} & -- & IE & -- & -- & -- & -- & -- & -- & -- \\
\cite{lange2024scene} & -- & -- & -- & MP & -- & -- & -- & -- & -- \\
\cite{Laugier2011a} & SE & IE & -- & MP & RE & -- & -- & -- & -- \\ 
\cite{Lawitzkyb} & -- & IE & -- & MP & RE & -- & -- & -- & -- \\ 
\cite{Lee2017a} & -- & -- & -- & MP & -- & -- & -- & -- & -- \\ 
\cite{Lefevre} & -- & IE & -- & -- & RE & -- & -- & -- & -- \\ 
\cite{Lefevre2012} & -- & IE & -- & -- & RE & -- & -- & -- & -- \\ 
\cite{Lefevre2013} & SE & IE & -- & -- & RE & -- & -- & -- & -- \\ 
\cite{Lefevre2014b} & -- & IE & -- & MP & -- & -- & IM & -- & BP \\ 
\cite{Lefevre2014c} & -- & -- & -- & MP & -- & -- & -- & -- & -- \\ 
\cite{Lefevre2015a} & -- & IE & -- & MP & -- & -- & IM & -- & BP \\ 
\cite{Lefevreb} & -- & IE & -- & -- & -- & -- & -- & -- & -- \\ 
\cite{Lenz2017} & -- & -- & -- & MP & -- & -- & -- & -- & -- \\ 
\cite{Leung} & -- & IE & -- & MP & -- & -- & -- & -- & BP \\ 
\cite{Levine2012a} & -- & -- & TE & MP & -- & -- & -- & -- & -- \\ 
\cite{Li2017} & -- & -- & -- & MP & -- & -- & -- & TS & BP \\ 
\cite{Li2019a} & -- & IE & -- & -- & -- & -- & -- & -- & -- \\ 
\cite{Li2019c} & -- & IE & -- & MP & -- & -- & -- & -- & BP \\ 
\cite{Li2019d} & -- & -- & -- & MP & -- & -- & -- & -- & -- \\ 
\cite{Li2019IV} & -- & IE & -- & MP & -- & -- & -- & -- & -- \\ 
\cite{Li2019Transferable} & -- & IE & -- & MP & -- & -- & -- & -- & -- \\
\cite{li2024interactive} & -- & IE & TE & MP & -- & -- & -- & -- & -- \\
\cite{li2024hierarchical} & -- & IE & -- & -- & -- & -- & -- & -- & -- \\
\cite{li2024deep} & -- & IE & -- & -- & -- & -- & -- & -- & BP \\
\cite{liao2024human} & -- & -- & -- & MP & -- & -- & -- & -- & -- \\
\cite{Liebner} & -- & IE & TE & MP & -- & -- & -- & -- & -- \\ 
\cite{Lin2019} & -- & IE & -- & -- & -- & -- & -- & -- & BP \\ 
\cite{Lina} & -- & -- & TE & MP & -- & -- & -- & -- & -- \\ 
\cite{Liu2001} & -- & IE & -- & -- & -- & -- & -- & -- & -- \\ 
\cite{Liu2019} & -- & IE & TE & MP & -- & -- & -- & -- & -- \\ 
\cite{liu2022multi} & -- & -- & -- & MP & -- & -- & -- & TS & -- \\
\cite{liu2021multimodal} & -- & -- & -- & MP & -- & -- & -- & TS & -- \\
\cite{liu2024augmenting} & -- & IE & -- & MP & -- & -- & -- & -- & BP \\
\cite{liu2022learning} & -- & -- & TE & -- & -- & -- & -- & -- & BP \\
\cite{Luo} & -- & -- & -- & MP & -- & -- & -- & -- & -- \\ 
\cite{Ma2019Wasserstein} & -- & -- & -- & MP & -- & -- & -- & -- & -- \\ 
\cite{Mandalia2005} & -- & IE & -- & -- & -- & -- & -- & -- & -- \\ 
\cite{Matousek2019} & -- & -- & -- & MP & -- & AD & -- & -- & -- \\ 
\cite{Messaoud2019Relational} & -- & -- & -- & MP & -- & -- & -- & -- & -- \\
\cite{mavrogiannis2022b} & -- & -- & -- & MP & -- & -- & -- & -- & -- \\
\cite{Messaoud2020} & -- & -- & -- & MP & -- & -- & -- & -- & -- \\ 
\cite{Meyer-Delius2008a} & -- & IE & -- & -- & -- & -- & -- & -- & -- \\ 
\cite{Monteil2015} & -- & -- & TE & -- & -- & -- & -- & -- & -- \\ 
\cite{Morales} & -- & -- & TE & MP & -- & -- & IM & -- & BP \\ 
\cite{Morris2011a} & -- & IE & -- & -- & -- & -- & -- & -- & -- \\ 
\cite{Morton2016} & -- & -- & TE & MP & -- & -- & -- & TS & -- \\ 

%% file: appendices/tables/models-task_view3.tex
\cite{mozaffari2023multimodal} & -- & -- & -- & MP & -- & -- & -- & -- & -- \\
\cite{Nava2019} & -- & -- & TE & MP & -- & -- & IM & -- & BP \\ 
\cite{Nishigaki2019} & -- & -- & TE & -- & -- & -- & -- & -- & -- \\ 
\cite{Okuda2016} & -- & IE & -- & -- & -- & -- & -- & -- & -- \\ 
\cite{Okuda2017} & -- & IE & -- & MP & -- & -- & -- & -- & BP \\ 
\cite{Oliver} & -- & IE & -- & -- & -- & -- & -- & -- & -- \\ 
\cite{Ortiz2011a} & -- & IE & -- & -- & -- & -- & -- & -- & -- \\ 
\cite{Oyler2016a} & -- & -- & TE & MP & -- & -- & -- & TS & -- \\ 
\cite{Panwai2005} & -- & -- & -- & MP & -- & -- & -- & -- & -- \\ 
\cite{Petrich2014} & SE & IE & -- & MP & -- & -- & -- & -- & -- \\ 
\cite{Phillips2017} & -- & IE & -- & -- & -- & -- & -- & -- & -- \\ 
\cite{Platho} & -- & -- & -- & MP & -- & -- & -- & -- & -- \\ 
\cite{Polychronopoulos2007b} & SE & -- & -- & MP & -- & -- & -- & -- & -- \\ 
\cite{Pruekprasert2019} & -- & -- & -- & MP & -- & -- & -- & -- & -- \\
\cite{rempe2022generating} & -- & -- & -- & MP & -- & -- & -- & TS & -- \\
\cite{Rodemerk} & -- & IE & -- & -- & -- & -- & -- & -- & -- \\ 
\cite{Roy2019} & -- & -- & -- & MP & -- & -- & -- & -- & -- \\ 
\cite{sachdeva2022gapformer} & -- & -- & TE & -- & -- & -- & -- & -- & -- \\ 
\cite{Sadigh2017} & -- & -- & TE & -- & -- & -- & -- & -- & -- \\ 
\cite{Sadigh2018} & -- & IE & TE & MP & -- & -- & -- & -- & BP \\ 
\cite{Salvuccib} & -- & IE & -- & -- & -- & -- & -- & -- & -- \\ 
\cite{Schester2019} & -- & -- & -- & MP & -- & -- & -- & -- & -- \\ 
\cite{Schlechtriemen} & -- & -- & -- & MP & -- & -- & -- & -- & -- \\ 
\cite{Schlechtriemen2014} & -- & IE & -- & -- & RE & -- & -- & -- & -- \\ 
\cite{Schlechtriemen2015} & -- & IE & -- & MP & -- & -- & -- & -- & -- \\ 
\cite{Schmerling2017} & -- & IE & -- & MP & -- & -- & -- & -- & BP \\ 
\cite{Schreier} & -- & IE & -- & MP & RE & -- & -- & -- & -- \\ 
\cite{Schreier2016} & -- & IE & -- & MP & RE & AD & -- & -- & -- \\ 
\cite{schrum2024maveric} & -- & -- & TE & -- & -- & -- & -- & -- & -- \\ 
\cite{Schubert} & SE & -- & -- & -- & -- & -- & -- & -- & -- \\ 
\cite{Schulz} & SE & IE & TE & MP & -- & -- & -- & -- & -- \\ 
\cite{Schulz2017} & -- & IE & TE & MP & -- & -- & -- & -- & -- \\ 
\cite{Schulz2018} & SE & IE & TE & MP & -- & -- & -- & -- & -- \\ 
\cite{Schulz2019} & SE & IE & TE & MP & -- & -- & -- & -- & -- \\ 
\cite{Sekizawa2007} & -- & IE & TE & -- & -- & -- & -- & -- & -- \\ 
\cite{Shen2012} & -- & -- & -- & -- & -- & -- & -- & TS & -- \\ 
\cite{Shi2019} & -- & -- & -- & -- & -- & -- & -- & -- & BP \\ 
\cite{Shia2014a} & -- & -- & -- & MP & -- & -- & -- & -- & BP \\ 
\cite{Shih2019} & -- & -- & -- & MP & -- & -- & -- & -- & -- \\ 
\cite{Shimosaka2014a} & -- & -- & TE & MP & -- & -- & -- & -- & -- \\ 
\cite{Shimosaka2015a} & -- & -- & TE & MP & -- & -- & -- & -- & -- \\ 
\cite{Sivaraman2014} & SE & -- & -- & MP & -- & -- & -- & -- & BP \\ 
\cite{Streubel2014} & -- & IE & -- & -- & -- & -- & -- & -- & -- \\ 
\cite{Sun2018} & -- & IE & TE & MP & -- & -- & -- & -- & -- \\ 
\cite{Sun2019} & -- & IE & -- & MP & -- & -- & -- & -- & -- \\ 
\cite{Sun2019IV} & -- & -- & TE & MP & -- & -- & IM & -- & BP \\ 
\cite{Sunberg2017a} & -- & -- & TE & MP & -- & -- & -- & -- & BP \\
\cite{sunberg2022improving} & -- & -- & TE & MP & -- & -- & -- & -- & BP \\
\cite{suo2021trafficsim} & -- & -- & -- & MP & -- & -- & -- & TS & -- \\
\cite{Talebpour2015} & -- & IE & TE & -- & -- & -- & -- & -- & -- \\ 
\cite{Toledo-Moreo2009} & SE & IE & -- & -- & -- & -- & -- & -- & -- \\ 
\cite{Tram2018} & -- & -- & -- & MP & -- & -- & -- & -- & BP \\ 
\cite{Tram2019} & -- & -- & -- & MP & -- & -- & -- & -- & BP \\ 
\cite{Tran2014} & -- & IE & -- & MP & -- & -- & -- & -- & -- \\ 
\cite{Tran2019Merging} & -- & IE & -- & MP & -- & -- & -- & -- & BP \\ 
\cite{Treiber2000} & -- & -- & TE & MP & -- & -- & -- & TS & -- \\ 
\cite{Tryhub2019} & -- & -- & TE & -- & -- & -- & -- & -- & -- \\ 
\cite{Ulbrich2015} & SE & -- & TE & MP & -- & -- & -- & -- & BP \\ 
\cite{Ward2015} & SE & IE & -- & -- & -- & -- & -- & -- & -- \\ 
\cite{Wheelera} & -- & -- & TE & MP & -- & -- & -- & TS & -- \\ 
\cite{Wiest2012} & -- & -- & -- & MP & -- & -- & -- & -- & -- \\ 
\cite{Wiest2013} & -- & -- & -- & MP & -- & -- & -- & -- & -- \\ 
\cite{Wissing2017} & -- & IE & -- & MP & -- & -- & -- & -- & -- \\ 
\cite{Woo2017} & SE & IE & -- & MP & -- & -- & -- & -- & -- \\ 
\cite{Worle2019} & -- & -- & TE & -- & -- & -- & -- & -- & -- \\ 
\cite{Worrall2008} & -- & -- & -- & MP & -- & -- & -- & -- & -- \\ 
\cite{Yan2019} & -- & IE & -- & -- & -- & -- & -- & -- & -- \\ 
\cite{Yan2020} & -- & IE & -- & -- & -- & -- & -- & -- & -- \\ 
\cite{yang2024multi} & -- & -- & -- & MP & -- & -- & -- & -- & -- \\ 
\cite{Yoon2016} & -- & IE & -- & MP & -- & -- & -- & -- & -- \\ 
\cite{Yuan2018} & SE & IE & -- & -- & -- & -- & -- & -- & -- \\
\cite{yuan2024temporal} & -- & IE & -- & MP & -- & -- & -- & -- & -- \\
\cite{Zhan2018a} & -- & IE & TE & -- & -- & -- & -- & -- & -- \\ 
\cite{Zhang2009} & -- & IE & -- & MP & RE & -- & -- & -- & -- \\
\cite{zhang2022systematic} & -- & IE & -- & MP & -- & -- & TS & -- & -- \\ 
\cite{zhao2023end} & -- & IE & -- & -- & -- & -- & -- & -- & -- \\ 
\cite{zhou2024i2t} & -- & IE & -- & MP & -- & -- & -- & -- & -- \\ 
\cite{zhou2024game} & -- & -- & -- & MP & -- & -- & -- & -- & BP \\ 

%% file: appendices/tables/tags-state_estimation_keyword_table_view.tex
\multirow{9}{*}{\STAB{Architecture}}
 & Bayesian Occupancy Filter  & \cite{Laugier2011a} \\ 
 & Dynamic Bayesian Network  & \cite{Agamennoni2012b,Gindele,Gindele2015,Ulbrich2015} \\ 
 & Extended Kalman Filter  & \cite{Petrich2014} \\ 
 & Kalman Filter  & \cite{Bahram2016,Bender2015,Houenou2013,Klingelschmitt2016,Ward2015,Woo2017,Yuan2018} \\ 
 & Markov State Space Model  & \cite{Lefevre2013} \\ 
 & Moving Average Filter  & \cite{Klingelschmitt2015a,Woo2017} \\ 
 & Multi-Perspective Tracker  & \cite{Deo} \\ 
 & Multiple Model Unscented Kalman Filter  & \cite{Schulz} \\ 
 & Particle Filter  & \cite{Hoermann2017,Kumar2013,Schulz2018} \\ 

%% file: appendices/tables/tags-intention_estimation_keyword_table_view1.tex
\multirow{54}{*}{\STAB{Architecture}}
 & Adaptive Cruise Control Policy  & \cite{Dong2017Merging,Tran2019Merging} \\ 
 & Bayesian Changepoint Estimation  & \cite{Galceran2017a} \\ 
 & Bayesian Filter  & \cite{Hardy2013,Kumar2013,Aoude2010a,Aoudeb,Aoudea} \\ 
 & Bayesian Network  & \cite{Bahram2016,Lefevreb,Liebner,Schreier,Gillmeier2019,Tran2019Merging,Okuda2016,Okuda2017,Schreier2016} \\ 
 & Conditional Probability Table  & \cite{Phillips2017} \\ 
 & Conditional Variational AutoEncoder  & \cite{Schmerling2017,Leung,Hu2018b,Feng2019} \\ 
 & Context Aware Scene Representation  & \cite{Rodemerk,Klingelschmitt} \\ 
 & Context Dependent  & \cite{Bonnin2012a,Driggs-Campbell2015b,Ortiz2011a} \\ 
 & Convolutional Neural Network  & \cite{Gebert2019} \\ 
 & Convolutional Social Pooling  & \cite{Deoa} \\ 
 & Counterfactual Reasoning  & \cite{Isele2019} \\ 
 & Coupled Hidden Markov Model  & \cite{Oliver} \\ 
 & Dirichlet Process  & \cite{Guo2019Multi} \\ 
 & Dynamic Bayesian Network  & \cite{Dagli2003,Gindele,Lefevre2012,Lefevre,Schulz2018,Schulz,Gindele2015,Agamennoni2012b,Kuhnt,Klingelschmitt,Brechtel2014,Zhang2009,Liu2019,Sun2019,Gonzalez2019,Schulz2019} \\ 
 & Dynamic Time Warping  & \cite{Hu2019} \\ 
 & Gated Recurrent Unit Network  & \cite{Yan2019} \\ 
 & Gaussian Mixture Model  & \cite{Zhan2018a,Feng2019,Gonzalez2019} \\ 
 & Gaussian Process  & \cite{Klingelschmitt2016,Tran2014,Kruger2019,Guo2019Multi} \\ 
 & Gaussian Radial Basis Kernel Function  & \cite{Aoude2010a,Aoudea} \\ 
 & Gibbs Sampling  & \cite{Klingelschmitt2016a,Guo2019Multi} \\ 
 & Hidden Markov Model  & \cite{Liu2001,Meyer-Delius2008a,Yuan2018,Streubel2014,Lefevre2015a,Lefevre2014b,Laugier2011a,Oliver,Deo,Zhan2018a,Aoudeb,Deng2019,Li2019IV,Schlechtriemen2014,Berndt2008} \\ 
 & Importance Weighting  & \cite{Li2019a} \\ 
 & Indicator Functions  & \cite{Rodemerk} \\ 
 & Intelligent Driver Model  & \cite{Isele2019} \\ 
 & Interacting Multiple-Model Kalman Filter  & \cite{Schulz2017,Toledo-Moreo2009} \\ 
 & K-Nearest Neighbors  & \cite{Ward2015} \\ 
 & Kalman Filter  & \cite{Gonzalez2019} \\ 
 & Least Common Subsequence  & \cite{Kafer2010} \\ 
 & Long Short-Term Memory Network  & \cite{Deoa,Deob,Dang2018,Jain2016,Phillips2017,Khosroshahi,Gebert2019,Yan2019,Han2019} \\ 
 & Marginal Composition  & \cite{Klingelschmitt2016a} \\ 
 & Marginal Probability Distribution  & \cite{Phillips2017} \\ 
 & Markov State Space Model  & \cite{Liu2001,Lefevre2013} \\ 
 & Mind Tracking  & \cite{Salvuccib} \\ 
 & Minimizing Overall Braking Induced by Lane Changes  & \cite{Buyer2019} \\ 
 & Mixture Density Network  & \cite{Hu2018a,Zhan2018a} \\ 
 & Mixture of Experts  & \cite{Schlechtriemen2015} \\ 
 & Multi-Layer Perceptron  & \cite{Yoon2016,Phillips2017,Ortiz2011a} \\ 
 & Multiple Model Unscented Kalman Filter  & \cite{Schulz2018,Schulz,Schulz2019} \\ 
 & Neural Network  & \cite{Hu2018a,Khosroshahi,Deng2019,Kruger2019} \\ 
 & Pairwise Probability Coupling  & \cite{Klingelschmitt2016} \\ 
 & Particle Filter  & \cite{Lefevre2012,Lefevre,Sun2019} \\ 
 & Piecewise Auto-Regressive Model  & \cite{Angkititrakul2009} \\ 
 & Polynomial Classifier  & \cite{Kafer2010} \\ 
 & Probabilistic Graphical Model  & \cite{Dong2017Merging} \\ 
 & Quadratic Discriminant Analysis  & \cite{Bender2015} \\ 
 & Quantile Regression Forest  & \cite{Hu2018a,Wissing2017} \\ 
 & Random Forest  & \cite{Driggs-Campbell2015c,Barbier2017,Ward2015,Schlechtriemen2015,Deng2019} \\ 
 & Recurrent Neural Network  & \cite{Dang2018,Jain2016,Phillips2017} \\ 
 & Relevance Vector Machine  & \cite{Morris2011a,Doshi} \\ 
 & Rule-Based  & \cite{Buyer2019,Bouton2019,Isele2019} \\ 
 & Single Layer Perceptron  & \cite{Bonnin2012a} \\ 
 & Stochastic Switched Autoregressive Exogenous Model  & \cite{Sekizawa2007} \\ 
 & Support Vector Machine  & \cite{Hu2018a,Driggs-Campbell2017c,Driggs-Campbell2015c,Barbier2017,Klingelschmitt2016,Ward2015,Woo2017,Kumar2013,Akita,Angkititrakul2009,Morris2011a,Aoudeb,Aoude2010a,Aoudea,Doshi,Deng2019,Yan2019,Li2019Transferable,Mandalia2005} \\ 
 & Two-player game  & \cite{Talebpour2015} \\ 
\midrule
\multirow{18}{*}{\STAB{Training}}
 & Bayesian Information Criterion  & \cite{Lefevre2014b,Schlechtriemen2014} \\ 
 & Continuous Inverse Optimal Control  & \cite{Sadigh2018,Sun2018} \\ 
 & Evolutionary Strategies  & \cite{Li2019c} \\ 
 & Expectation Maximization  & \cite{Lefevre2014b,Liu2019,Sekizawa2007} \\ 
 & Functional Discretization  & \cite{Barbier2017} \\ 
 & Gaussian Discriminant Analysis  & \cite{Bender2015} \\ 
 & Gaussian Mixture Regression  & \cite{Lefevre2014b} \\ 
 & Genetic Algorithms  & \cite{Deng2019} \\ 
 & Heuristic  & \cite{Lawitzkyb,Isele2019} \\ 
 & Input Selection  & \cite{Deng2019} \\ 
 & Iterative Forward Backward Algorithm  & \cite{Liu2019} \\ 
 & K-means Clustering  & \cite{Liebner,Akita,Driggs-Campbell2015b} \\ 
 & Linear Quantile Regression  & \cite{Wissing2017} \\ 
 & Logistic Regression  & \cite{Driggs-Campbell2015c,Klingelschmitt2016,Klingelschmitt2016a,Klingelschmitt2015,Klingelschmitt,Tran2019Merging,Okuda2016,Okuda2017,Yan2020} \\ 
 & Maximum Likelihood Estimation  & \cite{Okuda2016,Kruger2019} \\ 
 & Method of Simulated Moments  & \cite{Talebpour2015} \\ 
 & Non-Parametric Bayesian Learning  & \cite{Guo2019Multi} \\ 
 & Prefiltering  & \cite{Deng2019} \\ 

%% file: appendices/tables/tags-intention_estimation_keyword_table_view2.tex
\multirow{17}{*}{\STAB{Theory}}
 & Clustering  & \cite{Sun2018,Sadigh2018,Guo2019Multi} \\ 
 & Dempster Schafer Theory  & \cite{Petrich2014} \\ 
 & Distribution Adaptation  & \cite{Li2019a} \\ 
 & Domain Adaptation  & \cite{Li2019a,Li2019Transferable} \\ 
 & Game Theory  & \cite{Talebpour2015} \\ 
 & Interaction Detection  & \cite{Kuhnt,Hu2019,Li2019IV} \\ 
 & Inverse Reinforcement Learning  & \cite{Sun2018,Zhan2018a} \\ 
 & Level K Reasoning  & \cite{Isele2019} \\ 
 & Naive Bayes  & \cite{Rodemerk,Li2019IV,Gillmeier2019,Schlechtriemen2014} \\ 
 & Nash Equilibrium  & \cite{Talebpour2015} \\ 
 & Partially Observable Markov Decision Process  & \cite{Gonzalez2019,Bouton2019,Lin2019} \\ 
 & Reinforcement Learning  & \cite{Bouton2019} \\ 
 & Signal Detection Theory  & \cite{Aoudeb} \\ 
 & Time-Series Analysis  & \cite{Galceran2017a} \\ 
 & Trajectory Similarity  & \cite{Bahram2016,Galceran2017a,Liebner,Houenou2013,Klingelschmitt2016,Ward2015,Tran2014,Woo2017,Petrich2014,Zhan2018a,Kafer2010,Hu2019,Salvuccib} \\ 
 & Transfer Learning  & \cite{Li2019a} \\ 
 & Tree-Search Planning  & \cite{Isele2019} \\ 
\midrule
\multirow{18}{*}{\STAB{Scope}}
 & Advanced Driver Assistance Systems  & \cite{Gebert2019} \\ 
 & Assistive Braking  & \cite{Hillenbrand2006} \\ 
 & Assistive Steering  & \cite{Lefevre2014b} \\ 
 & Car Following  & \cite{Akita,Angkititrakul2009,Lefevre2015a} \\ 
 & Collision Mitigation  & \cite{Hillenbrand2006} \\ 
 & Highway Driving  & \cite{Toledo-Moreo2009,Gonzalez2019,Yan2019,Han2019,Kruger2019,Salvuccib,Schlechtriemen2014,Berndt2008,Bahram2016,Buyer2019,Dang2018,Deng2019,Deo,Deoa,Deob,Doshi,Driggs-Campbell2015b,Driggs-Campbell2015c,Feng2019,Guo2019Multi,Liu2019,Meyer-Delius2008a,Wissing2017,Yoon2016,Li2019c} \\ 
 & Highway Merging  & \cite{Tran2019Merging,Okuda2016,Okuda2017,Dong2017Merging,Isele2019,Klingelschmitt2016a,Sun2018} \\ 
 & In-cabin Sensing  & \cite{Gebert2019} \\ 
 & Intersection  & \cite{Khosroshahi,Kafer2010,Bouton2019,Aoudea,Aoudeb,Bender2015,Driggs-Campbell2015b,Guo2019Multi,Klingelschmitt,Klingelschmitt2015,Klingelschmitt2016,Kuhnt,Lefevre,Lefevre2012,Lefevre2013,Lefevreb,Liebner,Petrich2014,Phillips2017,Rodemerk,Schreier,Streubel2014,Tran2014,Zhang2009,Lin2019} \\ 
 & Lane Changing  & \cite{Driggs-Campbell2017c,Dang2018,Buyer2019,Kruger2019,Li2019a,Morris2011a,Salvuccib,Schlechtriemen2015,Schreier,Talebpour2015,Toledo-Moreo2009,Wissing2017,Woo2017,Yan2019,Yuan2018,Mandalia2005,Schlechtriemen2014,Li2019c} \\ 
 & Lane Keeping  & \cite{Lefevre2014b} \\ 
 & Merging at Intersection  & \cite{Schulz2017} \\ 
 & Overtaking  & \cite{Yan2020,Bahram2016,Meyer-Delius2008a} \\ 
 & Traffic Weaving  & \cite{Schmerling2017} \\ 
 & Unsignalized Intersections  & \cite{Barbier2017,Ward2015,Lin2019} \\ 
 & Urban Driving  & \cite{Ferguson,Berndt2008,Aoude2010a,Schulz,Schulz2018,Schulz2019} \\ 
 & Vehicle to Vehicle communication  & \cite{Talebpour2015} \\ 
 & Yielding at Intersection  & \cite{Ward2015} \\ 
\midrule
\multirow{21}{*}{\STAB{Evaluation}}
 & Area Under the ROC Curve  & \cite{Schlechtriemen2014,Han2019,Yoon2016} \\ 
 & Balanced Accuracy  & \cite{Kruger2019,Bahram2016} \\ 
 & Balanced Precision & \cite{Schlechtriemen2014,Bahram2016} \\ 
 & Brier Metric & \cite{Sun2018} \\ 
 & Classification Accuracy & \cite{Mandalia2005,Deng2019,Gebert2019,Han2019,Khosroshahi,Klingelschmitt,Kruger2019,Kuhnt,Li2019a,Li2019IV,Li2019Transferable,Okuda2016,Yan2019,Schulz2018,Schulz2017,Liu2019,Galceran2017a,Dang2018} \\ 
 & Classification Precision & \cite{Li2019IV,Yan2019,Galceran2017a} \\ 
 & Confusion Matrix  & \cite{Berndt2008,Khosroshahi,Schulz2017,Laugier2011a} \\ 
 & F1 Score  & \cite{Schlechtriemen2014,Gebert2019,Han2019,Kruger2019,Li2019IV,Yan2019,Hu2018a,Dang2018} \\ 
 & False Negative Rate  & \cite{Lefevre,Lefevre2012,Lefevre2013,Dang2018} \\ 
 & False Positive Rate  & \cite{Aoudeb,Deng2019,Klingelschmitt2015,Lefevre,Lefevre2012,Lefevre2013,Lefevre2014b,Dang2018} \\ 
 & Max Time to Detection  & \cite{Gebert2019,Kafer2010,Lefevre2014b} \\ 
 & Mean Lateral Offset before Detection  & \cite{Houenou2013} \\ 
 & Mean Time to Detection  & \cite{Gebert2019,Toledo-Moreo2009,Yoon2016,Houenou2013} \\ 
 & Median Time to Detection  & \cite{Lefevre2014b} \\ 
 & Min Time to Detection  & \cite{Liu2019,Lefevre2014b} \\ 
 & Negative Log Likelihood  & \cite{Kruger2019} \\ 
 & Precision over Recall  & \cite{Hu2018a} \\ 
 & Receiver Operating Characteristic curve  & \cite{Salvuccib,Schlechtriemen2014,Aoudeb,Doshi,Han2019,Klingelschmitt,Morris2011a,Ortiz2011a,Klingelschmitt2015,Hu2018a,Bahram2016,Yoon2016} \\ 
 & Root Mean Square Error  & \cite{Yan2019} \\ 
 & Standard Deviation of Time to Detection  & \cite{Yoon2016} \\ 
 & True Positive Rate  & \cite{Aoudeb,Deng2019} \\ 

%% file: appendices/tables/models-intention_estimation_view1.tex
\cite{Agamennoni2012b} & -- & -- & -- & -- & -- & -- & -- & -- & -- & -- & -- & Re & -- & Ba & -- & -- & -- \\ 
\cite{Akita} & -- & -- & Lo & -- & -- & -- & -- & -- & S & -- & -- & -- & SS & -- & BB & -- & -- \\ 
\cite{Angkititrakul2009} & -- & -- & Lo & -- & -- & -- & -- & -- & S & -- & -- & -- & SS & -- & BB & -- & -- \\ 
\cite{Aoude2010a} & Ro & -- & -- & -- & -- & -- & -- & D & S & -- & -- & Re & SS & -- & -- & -- & -- \\ 
\cite{Aoudea} & Ro & -- & -- & -- & -- & -- & -- & D & S & -- & -- & Re & SS & -- & -- & -- & -- \\ 
\cite{Aoudeb} & -- & -- & -- & -- & -- & -- & -- & D & S & -- & -- & Re & SS & -- & -- & -- & -- \\ 
\cite{Bahram2016} & -- & -- & -- & La & -- & -- & -- & D & -- & -- & -- & -- & SS & Ba & -- & MP & -- \\ 
\cite{Barbier2017} & Ro & -- & Lo & -- & -- & -- & -- & -- & S & -- & -- & -- & SS & -- & BB & -- & -- \\ 
\cite{Bender2015} & Ro & -- & -- & -- & -- & -- & -- & D & -- & -- & -- & -- & SS & -- & BB & -- & -- \\ 
\cite{Berndt2008} & Ro & -- & -- & La & -- & -- & -- & D & -- & -- & -- & Re & -- & Ba & -- & -- & -- \\ 
\cite{Bonnin2012a} & -- & -- & -- & La & -- & -- & CD & D & -- & -- & -- & -- & SS & -- & BB & -- & -- \\ 
\cite{Bouton2019} & -- & -- & Lo & -- & -- & -- & -- & D & -- & -- & -- & Re & -- & Ba & -- & -- & -- \\ 
\cite{Brechtel2014} & Ro & -- & -- & -- & -- & -- & -- & -- & -- & P & -- & Re & -- & Ba & -- & -- & -- \\ 
\cite{Buyer2019} & -- & -- & -- & La & -- & -- & -- & -- & S & -- & -- & -- & SS & -- & -- & -- & -- \\ 
\cite{Dagli2003} & -- & -- & -- & -- & -- & -- & -- & D & -- & -- & -- & Re & -- & Ba & -- & -- & -- \\ 
\cite{Dang2018} & -- & -- & -- & La & -- & -- & -- & -- & S & -- & -- & Re & -- & -- & BB & -- & -- \\ 
\cite{Deng2019} & -- & -- & -- & La & -- & -- & -- & D & -- & -- & -- & Re & SS & Ba & BB & -- & -- \\ 
\cite{Deo} & -- & Co & Lo & La & -- & -- & -- & D & -- & -- & -- & Re & -- & Ba & -- & -- & GT \\ 
\cite{Deoa} & -- & -- & -- & -- & -- & -- & -- & -- & -- & -- & -- & Re & -- & -- & BB & -- & -- \\ 
\cite{Deob} & -- & -- & -- & -- & -- & -- & -- & -- & -- & -- & -- & Re & -- & -- & BB & -- & -- \\ 
\cite{Dong2017Merging} & -- & Co & -- & -- & -- & -- & -- & -- & S & -- & -- & Re & -- & Ba & -- & -- & -- \\ 
\cite{Doshi} & -- & -- & -- & La & -- & -- & -- & D & -- & -- & -- & -- & SS & Ba & BB & -- & -- \\ 
\cite{Driggs-Campbell2015b} & -- & -- & -- & -- & -- & -- & -- & -- & S & -- & -- & -- & SS & -- & -- & -- & -- \\ 
\cite{Driggs-Campbell2015c} & -- & -- & -- & La & -- & -- & -- & -- & S & -- & -- & -- & SS & -- & BB & -- & -- \\ 
\cite{Driggs-Campbell2017c} & -- & -- & -- & -- & -- & -- & -- & -- & S & -- & -- & -- & SS & -- & BB & -- & -- \\ 
\cite{Feng2019} & -- & -- & -- & La & -- & -- & -- & D & -- & -- & -- & -- & SS & -- & BB & -- & -- \\ 
\cite{Ferguson} & Ro & -- & -- & -- & -- & -- & -- & D & -- & -- & -- & -- & SS & -- & -- & MP & -- \\ 
\cite{Galceran2017a} & -- & -- & -- & -- & -- & -- & CD & D & -- & -- & DS & Re & SS & Ba & -- & MP & -- \\ 
\cite{Gebert2019} & Ro & -- & -- & La & -- & -- & -- & D & -- & -- & -- & Re & -- & -- & BB & -- & -- \\ 
\cite{Gillmeier2019} & -- & -- & -- & -- & EM & -- & -- & D & -- & -- & -- & -- & SS & Ba & -- & -- & -- \\ 
\cite{Gindele} & -- & -- & -- & -- & -- & -- & CD & D & -- & -- & -- & Re & -- & Ba & -- & -- & -- \\ 
\cite{Gindele2015} & Ro & -- & -- & -- & -- & -- & -- & D & -- & -- & -- & Re & -- & Ba & -- & -- & -- \\ 
\cite{Gonzalez2019} & -- & -- & -- & La & -- & -- & -- & D & -- & P & -- & Re & -- & Ba & -- & -- & GT \\ 
\cite{Guo2019Multi} & -- & -- & -- & -- & -- & -- & -- & D & -- & -- & -- & -- & -- & -- & BB & MP & -- \\ 
\cite{Han2019} & -- & -- & -- & La & -- & -- & -- & D & -- & -- & -- & Re & -- & -- & BB & -- & -- \\ 
\cite{Hardy2013} & Ro & -- & -- & -- & -- & -- & -- & -- & -- & -- & DS & Re & -- & Ba & -- & -- & -- \\ 
\cite{Hillenbrand2006} & -- & -- & -- & -- & EM & -- & -- & -- & -- & -- & -- & -- & -- & -- & -- & -- & GT \\ 
\cite{Houenou2013} & -- & -- & -- & La & -- & -- & -- & -- & S & -- & -- & -- & SS & -- & -- & MP & -- \\ 
\cite{Hu2018a} & -- & Co & -- & -- & -- & -- & CD & D & -- & -- & -- & -- & SS & -- & BB & -- & -- \\ 
\cite{Hu2018b} & -- & Co & -- & -- & -- & -- & CD & D & -- & -- & -- & -- & SS & Ba & BB & -- & -- \\ 
\cite{Hu2019} & -- & -- & -- & -- & -- & -- & -- & D & -- & -- & -- & -- & -- & -- & -- & MP & -- \\ 
\cite{Isele2019} & -- & Co & -- & -- & -- & -- & -- & D & -- & -- & -- & Re & -- & Ba & -- & -- & GT \\ 
\cite{Jain2016} & Ro & -- & -- & La & -- & -- & -- & D & -- & -- & -- & Re & -- & -- & BB & -- & -- \\ 
\cite{Kafer2010} & Ro & -- & -- & -- & -- & -- & -- & D & -- & -- & DS & -- & SS & Ba & BB & MP & -- \\ 
\cite{Khosroshahi} & Ro & -- & -- & -- & -- & -- & -- & -- & S & -- & -- & Re & -- & -- & BB & -- & -- \\ 
\cite{Klingelschmitt} & Ro & -- & -- & -- & -- & -- & -- & D & -- & -- & -- & Re & -- & Ba & BB & -- & -- \\ 
\cite{Klingelschmitt2015} & Ro & -- & -- & -- & -- & -- & -- & D & -- & -- & -- & -- & SS & -- & BB & -- & -- \\ 
\cite{Klingelschmitt2016} & -- & -- & -- & -- & -- & -- & CD & D & -- & -- & -- & -- & SS & Ba & BB & MP & -- \\ 
\cite{Klingelschmitt2016a} & -- & -- & -- & -- & -- & -- & CD & D & -- & -- & -- & -- & SS & Ba & BB & -- & -- \\ 
\cite{Kruger2019} & -- & -- & -- & La & -- & -- & -- & D & -- & -- & -- & -- & SS & -- & BB & -- & -- \\ 
\cite{Kuhnt} & Ro & -- & -- & -- & -- & -- & -- & D & -- & -- & -- & Re & -- & Ba & BB & -- & -- \\ 
\cite{Kumar2013} & -- & -- & -- & La & -- & -- & -- & D & -- & -- & -- & -- & SS & Ba & BB & -- & -- \\ 
\cite{Laugier2011a} & Ro & Co & -- & -- & -- & -- & -- & D & -- & -- & -- & Re & -- & Ba & -- & -- & -- \\ 
\cite{Lawitzkyb} & -- & -- & -- & -- & -- & -- & -- & -- & -- & -- & DS & -- & SS & Ba & -- & -- & GT \\ 

%% file: appendices/tables/models-intention_estimation_view2.tex
\cite{Lefevre} & Ro & -- & -- & -- & -- & -- & -- & D & -- & -- & -- & Re & -- & Ba & -- & -- & GT \\ 
\cite{Lefevre2012} & Ro & -- & -- & -- & -- & -- & -- & D & -- & -- & -- & Re & -- & Ba & -- & -- & GT \\ 
\cite{Lefevre2013} & -- & -- & Lo & La & -- & -- & -- & D & -- & -- & -- & Re & -- & Ba & -- & -- & GT \\ 
\cite{Lefevre2014b} & -- & -- & -- & La & -- & IU & -- & D & -- & -- & -- & Re & -- & Ba & -- & -- & -- \\ 
\cite{Lefevre2015a} & -- & -- & -- & -- & -- & -- & -- & D & -- & -- & -- & Re & -- & Ba & -- & -- & -- \\ 
\cite{Lefevreb} & Ro & -- & -- & -- & -- & -- & -- & D & -- & -- & -- & -- & SS & Ba & -- & -- & -- \\ 
\cite{Leung} & -- & -- & -- & -- & -- & -- & -- & D & -- & -- & -- & Re & -- & -- & BB & -- & GT \\ 
\cite{Li2019a} & -- & -- & -- & La & -- & -- & -- & D & -- & -- & -- & -- & SS & -- & BB & -- & -- \\ 
\cite{Li2019c} & -- & -- & -- & La & -- & -- & -- & -- & S & -- & -- & -- & SS & -- & -- & -- & GT \\ 
\cite{Li2019IV} & -- & Co & -- & La & -- & -- & -- & D & -- & -- & -- & -- & SS & -- & BB & -- & -- \\ 
\cite{Li2019Transferable} & -- & -- & -- & -- & -- & -- & -- & -- & -- & -- & -- & Re & -- & -- & BB & -- & -- \\ 
\cite{Liebner} & Ro & -- & Lo & -- & -- & -- & -- & D & -- & -- & -- & -- & SS & Ba & -- & MP & -- \\ 
\cite{Lin2019} & Ro & -- & -- & -- & -- & -- & -- & D & -- & -- & -- & Re & -- & Ba & -- & MP & -- \\ 
\cite{Liu2001} & -- & -- & -- & -- & -- & -- & -- & D & -- & -- & -- & Re & -- & Ba & -- & -- & -- \\ 
\cite{Liu2019} & -- & -- & -- & La & -- & -- & -- & D & -- & -- & -- & -- & SS & Ba & -- & -- & -- \\ 
\cite{Mandalia2005} & -- & -- & -- & La & -- & -- & -- & -- & S & -- & -- & -- & SS & -- & BB & -- & -- \\ 
\cite{Meyer-Delius2008a} & -- & Co & -- & -- & -- & -- & -- & D & -- & -- & -- & Re & -- & Ba & -- & -- & -- \\ 
\cite{Morris2011a} & -- & -- & -- & La & -- & -- & -- & D & -- & -- & -- & -- & SS & Ba & BB & -- & -- \\ 
\cite{Okuda2016} & -- & Co & -- & -- & -- & -- & -- & D & -- & -- & -- & -- & SS & Ba & -- & -- & -- \\ 
\cite{Okuda2017} & -- & Co & -- & -- & -- & -- & -- & D & -- & -- & -- & -- & SS & Ba & -- & -- & -- \\ 
\cite{Oliver} & -- & -- & -- & -- & -- & -- & -- & D & -- & -- & -- & Re & -- & Ba & -- & -- & -- \\ 
\cite{Ortiz2011a} & -- & -- & Lo & -- & -- & -- & -- & -- & S & -- & -- & -- & SS & -- & BB & -- & -- \\ 
\cite{Petrich2014} & Ro & -- & -- & -- & -- & -- & -- & D & -- & -- & -- & -- & SS & Ba & -- & MP & -- \\ 
\cite{Phillips2017} & Ro & -- & -- & -- & -- & -- & -- & D & -- & -- & -- & Re & SS & Ba & BB & -- & -- \\ 
\cite{Rodemerk} & Ro & -- & -- & -- & -- & -- & -- & D & -- & -- & -- & -- & SS & -- & -- & -- & -- \\ 
\cite{Sadigh2018} & -- & -- & -- & -- & -- & -- & -- & D & -- & -- & -- & Re & -- & -- & -- & MP & -- \\ 
\cite{Salvuccib} & -- & -- & -- & La & -- & -- & -- & -- & S & -- & -- & -- & -- & -- & -- & MP & -- \\ 
\cite{Schlechtriemen2014} & -- & -- & -- & La & -- & -- & -- & D & -- & -- & -- & Re & SS & Ba & BB & -- & -- \\ 
\cite{Schlechtriemen2015} & -- & -- & -- & La & -- & -- & -- & D & -- & -- & -- & -- & SS & -- & BB & -- & -- \\ 
\cite{Schmerling2017} & -- & -- & -- & -- & -- & -- & -- & D & -- & -- & -- & Re & -- & -- & BB & -- & GT \\ 
\cite{Schreier} & -- & -- & -- & -- & -- & -- & -- & D & -- & -- & -- & -- & SS & Ba & -- & -- & -- \\ 
\cite{Schreier2016} & Ro & -- & -- & La & -- & -- & -- & D & -- & -- & -- & -- & SS & Ba & -- & -- & -- \\ 
\cite{Schulz} & Ro & Co & -- & -- & -- & -- & -- & D & -- & -- & -- & Re & -- & Ba & -- & -- & -- \\ 
\cite{Schulz2017} & -- & -- & -- & -- & -- & -- & -- & D & -- & -- & -- & Re & -- & Ba & -- & -- & -- \\ 
\cite{Schulz2018} & Ro & Co & -- & -- & -- & -- & -- & D & -- & -- & -- & Re & -- & Ba & -- & -- & -- \\ 
\cite{Schulz2019} & Ro & -- & -- & -- & -- & -- & -- & D & -- & -- & -- & Re & -- & Ba & -- & -- & -- \\ 
\cite{Sekizawa2007} & -- & Co & Lo & La & -- & -- & -- & D & -- & -- & -- & Re & -- & -- & BB & -- & -- \\ 
\cite{Streubel2014} & Ro & -- & -- & -- & -- & -- & -- & D & -- & -- & -- & Re & -- & Ba & -- & -- & -- \\ 
\cite{Sun2018} & -- & Co & -- & -- & -- & -- & -- & -- & -- & -- & DS & -- & SS & -- & BB & -- & -- \\ 
\cite{Sun2019} & Ro & Co & -- & -- & -- & -- & -- & D & -- & -- & -- & Re & -- & Ba & -- & -- & -- \\ 
\cite{Talebpour2015} & -- & -- & -- & La & -- & -- & -- & D & -- & -- & -- & -- & SS & -- & -- & -- & GT \\ 
\cite{Toledo-Moreo2009} & -- & -- & -- & La & -- & -- & -- & D & -- & -- & -- & Re & -- & Ba & -- & -- & -- \\ 
\cite{Tran2014} & Ro & -- & Lo & -- & -- & -- & -- & D & -- & -- & -- & -- & SS & -- & BB & MP & -- \\ 
\cite{Tran2019Merging} & -- & Co & -- & -- & -- & -- & -- & D & -- & -- & -- & -- & SS & Ba & -- & -- & -- \\ 
\cite{Ward2015} & -- & -- & Lo & -- & -- & -- & -- & D & S & -- & -- & -- & SS & -- & BB & MP & -- \\ 
\cite{Wissing2017} & -- & -- & -- & La & -- & -- & -- & -- & S & -- & -- & -- & SS & -- & BB & -- & -- \\ 
\cite{Woo2017} & -- & -- & -- & La & -- & -- & -- & -- & S & -- & -- & -- & SS & -- & BB & MP & GT \\ 
\cite{Yan2019} & -- & -- & -- & La & -- & -- & -- & -- & S & -- & -- & Re & -- & -- & BB & -- & -- \\ 
\cite{Yan2020} & -- & Co & -- & -- & -- & -- & -- & D & -- & -- & -- & -- & SS & -- & BB & -- & -- \\ 
\cite{Yoon2016} & -- & -- & -- & La & -- & -- & -- & D & -- & -- & -- & -- & SS & -- & BB & -- & -- \\ 
\cite{Yuan2018} & -- & -- & -- & La & -- & -- & -- & D & -- & -- & -- & Re & -- & Ba & -- & -- & -- \\ 
\cite{Zhan2018a} & -- & -- & -- & -- & -- & -- & -- & -- & -- & -- & -- & -- & SS & -- & -- & MP & GT \\ 
\cite{Zhang2009} & Ro & -- & -- & -- & -- & -- & -- & D & -- & -- & -- & Re & -- & Ba & -- & -- & -- \\ 

%% file: appendices/tables/tags-trait_estimation_keyword_table_view1.tex
\multirow{25}{*}{\STAB{Architecture}}
 & Attention Parameters  & \cite{Nishigaki2019,Tryhub2019,Driggs-Campbell2015b} \\ 
 & Band Pass Filter  & \cite{Nishigaki2019} \\ 
 & Bayesian Changepoint Estimation  & \cite{Galceran2017a} \\ 
 & Context Dependent  & \cite{Wheelera,Driggs-Campbell2015b,Shimosaka2015a,Eggert,Liebner} \\ 
 & Convolutional Neural Network  & \cite{Tryhub2019} \\ 
 & Dynamic Bayesian Network  & \cite{Liu2019} \\ 
 & Extended Kalman Filter  & \cite{Monteil2015} \\ 
 & Foresighted Driver Model  & \cite{Eggert} \\ 
 & Gaussian Process  & \cite{Armand} \\ 
 & Greedy Selection  & \cite{Wheelera} \\ 
 & Intelligent Driver Model  & \cite{Monteil2015,Buyer2019,Hoermann2017,Kesting,Kesting2008,Liebner,Morton2016,Schulz,Schulz2018,Sunberg2017a,Ulbrich2015,Chen2010,Eggert,Bhattacharyya2020,Treiber2000} \\ 
 & Minimizing Overall Braking Induced by Lane Changes  & \cite{Buyer2019,Kesting,Sunberg2017a} \\ 
 & MITSIM Driver Model  & \cite{Chen2010} \\ 
 & Multi Lane Intelligent Driver Model  & \cite{Buyer2019} \\ 
 & Neural Network  & \cite{Schulz2019} \\ 
 & Particle Filter  & \cite{Sunberg2017a,Hoermann2017,Buyer2019,Bhattacharyya2020} \\ 
 & Piecewise Auto-Regressive Model  & \cite{Lina,Akita,Angkititrakul2009} \\ 
 & Random Forest  & \cite{Worle2019} \\ 
 & Reaction time  & \cite{Khodayari2012} \\ 
 & Reward Parameters  & \cite{Kuderer2015a,Levine2012a,Sadigh2017,Sadigh2018,Schulz2017,Shimosaka2014a,Shimosaka2015a,Sun2018,Zhan2018a,Abbeela} \\ 
 & Stochastic Switched Autoregressive Exogenous Model  & \cite{Sekizawa2007} \\ 
 & Support Vector Machine  & \cite{Angkititrakul2009} \\ 
 & Time Delay Neural Network  & \cite{Nava2019} \\ 
 & Two-player game  & \cite{Talebpour2015} \\ 
 & Velocity Difference Model  & \cite{Kesting2008} \\ 

%% file: appendices/tables/tags-trait_estimation_keyword_table_view2.tex
\multirow{14}{*}{\STAB{Training}}
 & Active Preference-Based Learning  & \cite{Sadigh2017} \\ 
 & Bayesian Information Criterion  & \cite{Wheelera} \\ 
 & Continuous Inverse Optimal Control  & \cite{Sadigh2018,Levine2012a,Sun2018,Sun2019IV} \\ 
 & Expectation Maximization  & \cite{Wheelera,Liu2019,Sekizawa2007} \\ 
 & Gaussian Mixture Regression  & \cite{Angkititrakul2009} \\ 
 & Genetic Algorithms  & \cite{Kesting2008,Chen2010,Han2019} \\ 
 & Gradient-Based Optimization  & \cite{Schulz2019} \\ 
 & Heuristic  & \cite{Han2019,Talebpour2015,Treiber2000} \\ 
 & Iterative Forward Backward Algorithm  & \cite{Liu2019} \\ 
 & K-fold Cross-Validation  & \cite{Wheelera} \\ 
 & Maximum Likelihood Estimation  & \cite{Wheelera} \\ 
 & Method of Simulated Moments  & \cite{Talebpour2015} \\ 
 & Nonlinear Optimization  & \cite{Kesting2008} \\ 
 & Ridge Regression  & \cite{Wheelera} \\ 
\midrule
\multirow{7}{*}{\STAB{Theory}}
 & Clustering  & \cite{Sadigh2018} \\ 
 & Inverse Reinforcement Learning  & \cite{Kuderer2015a,Abbeela,Shimosaka2015a,Shimosaka2014a,Sun2018,Zhan2018a,Morales,Sun2019IV} \\ 
 & Level K Reasoning  & \cite{Oyler2016a} \\ 
 & People as Sensors  & \cite{Sun2019IV} \\ 
 & Reinforcement Learning  & \cite{Oyler2016a} \\ 
 & Time-Series Analysis  & \cite{Galceran2017a} \\ 
 & Trajectory Similarity  & \cite{Galceran2017a} \\ 
\midrule
\multirow{11}{*}{\STAB{Scope}}
 & Adaptive Cruise Control  & \cite{Nava2019} \\ 
 & Blind Intersection  & \cite{Morales} \\ 
 & Car Following  & \cite{Monteil2015,Akita,Angkititrakul2009,Bhattacharyya2020,Chen2010,Kesting2008,Khodayari2012,Morton2016,Nava2019,Wheelera} \\ 
 & Drifting  & \cite{Lina} \\ 
 & Highway Driving  & \cite{Han2019,Buyer2019,Driggs-Campbell2015b,Kesting,Liu2019,Nishigaki2019,Oyler2016a,Sunberg2017a,Ulbrich2015} \\ 
 & Highway Merging  & \cite{Sun2018,Eggert} \\ 
 & Intersection  & \cite{Armand,Driggs-Campbell2015b,Liebner,Sun2019IV} \\ 
 & Lane Changing  & \cite{Buyer2019,Kesting} \\ 
 & Merging at Intersection  & \cite{Schulz2017} \\ 
 & Unsignalized Intersections  & \cite{Armand} \\ 
 & Urban Driving  & \cite{Hoermann2017,Schulz,Schulz2018,Schulz2019} \\ 

%% file: appendices/tables/models-trait_estimation_view1.tex
\cite{Abbeela} & Re & -- & -- & -- & -- & S & -- & -- & -- & Off & -- & -- & -- & IRL & -- \\ 
\cite{Akita} & -- & Pa & -- & -- & -- & S & -- & -- & -- & Off & -- & -- & -- & -- & CV \\ 
\cite{Angkititrakul2009} & -- & Pa & NP & -- & -- & S & -- & -- & -- & Off & -- & -- & -- & -- & CV \\ 
\cite{Armand} & -- & -- & -- & -- & -- & S & -- & -- & -- & Off & -- & Op & -- & -- & -- \\ 
\cite{Bhattacharyya2020} & -- & Pa & -- & -- & -- & -- & -- & P & On & -- & -- & -- & B & -- & -- \\ 
\cite{Buyer2019} & -- & Pa & -- & -- & -- & -- & -- & P & On & -- & -- & -- & B & -- & -- \\ 
\cite{Chen2010} & -- & Pa & -- & -- & -- & S & -- & -- & -- & Off & -- & Op & -- & -- & -- \\ 
\cite{Driggs-Campbell2015b} & -- & -- & -- & -- & At & S & -- & -- & On & -- & H & -- & -- & -- & -- \\ 
\cite{Eggert} & -- & Pa & -- & -- & -- & -- & -- & -- & -- & Off & H & -- & -- & -- & CV \\ 
\cite{Galceran2017a} & -- & Pa & -- & -- & -- & S & -- & -- & On & Off & -- & Op & B & -- & -- \\ 
\cite{Han2019} & -- & Pa & -- & -- & -- & S & -- & -- & On & Off & -- & -- & -- & -- & -- \\ 
\cite{Hoermann2017} & -- & Pa & -- & -- & -- & -- & -- & P & On & -- & -- & -- & B & -- & -- \\ 
\cite{Kesting} & -- & Pa & -- & -- & -- & S & -- & -- & -- & Off & -- & Op & -- & -- & -- \\ 
\cite{Kesting2008} & -- & Pa & -- & -- & -- & S & -- & -- & -- & Off & -- & Op & -- & -- & -- \\ 
\cite{Khodayari2012} & -- & -- & -- & Ph & -- & S & -- & -- & -- & Off & -- & Op & -- & -- & -- \\ 
\cite{Kuderer2015a} & Re & -- & -- & -- & -- & S & -- & -- & -- & Off & -- & -- & -- & IRL & -- \\ 
\cite{Levine2012a} & Re & -- & -- & -- & -- & S & -- & -- & -- & Off & -- & -- & -- & IRL & -- \\ 
\cite{Liebner} & -- & Pa & -- & -- & -- & -- & -- & -- & -- & Off & -- & -- & -- & -- & CV \\ 
\cite{Lina} & -- & Pa & -- & -- & -- & S & -- & -- & -- & Off & -- & Op & -- & -- & -- \\ 
\cite{Liu2019} & -- & Pa & -- & -- & -- & -- & -- & -- & -- & Off & -- & -- & -- & -- & -- \\ 
\cite{Monteil2015} & -- & Pa & -- & -- & -- & -- & C & -- & On & -- & -- & -- & B & -- & -- \\ 
\cite{Morales} & Re & -- & -- & -- & -- & S & -- & -- & -- & Off & -- & -- & -- & IRL & -- \\ 
\cite{Morton2016} & -- & Pa & -- & -- & -- & S & -- & -- & -- & Off & -- & Op & -- & -- & -- \\ 

%% file: appendices/tables/models-trait_estimation_view2.tex
\cite{Nava2019} & -- & -- & -- & -- & -- & S & -- & -- & -- & Off & -- & Op & -- & -- & -- \\ 
\cite{Nishigaki2019} & -- & -- & -- & -- & At & S & -- & -- & On & -- & H & -- & -- & -- & -- \\ 
\cite{Oyler2016a} & -- & Pa & -- & -- & -- & S & -- & -- & -- & Off & -- & -- & -- & -- & -- \\ 
\cite{Sadigh2017} & Re & -- & -- & -- & -- & -- & C & -- & -- & Off & -- & -- & -- & IRL & -- \\ 
\cite{Sadigh2018} & Re & -- & -- & -- & -- & S & -- & -- & On & Off & -- & Op & B & IRL & -- \\ 
\cite{Schulz} & -- & Pa & -- & -- & -- & -- & -- & -- & On & Off & H & -- & -- & -- & CV \\ 
\cite{Schulz2017} & Re & -- & -- & -- & -- & S & -- & -- & -- & Off & H & -- & -- & -- & -- \\ 
\cite{Schulz2018} & -- & Pa & -- & -- & -- & -- & -- & -- & -- & Off & H & -- & -- & -- & CV \\ 
\cite{Schulz2019} & -- & -- & -- & -- & -- & S & -- & -- & -- & Off & -- & Op & -- & -- & -- \\ 
\cite{Sekizawa2007} & -- & Pa & -- & -- & -- & S & -- & -- & -- & Off & -- & Op & -- & -- & -- \\ 
\cite{Shimosaka2014a} & Re & -- & -- & -- & -- & S & -- & -- & -- & Off & -- & -- & -- & IRL & -- \\ 
\cite{Shimosaka2015a} & Re & -- & -- & -- & -- & S & -- & -- & -- & Off & -- & -- & -- & IRL & -- \\ 
\cite{Sun2018} & Re & -- & -- & -- & -- & S & -- & -- & -- & Off & -- & -- & -- & IRL & -- \\ 
\cite{Sun2019IV} & Re & -- & -- & -- & -- & S & -- & -- & -- & Off & -- & -- & -- & IRL & -- \\ 
\cite{Sunberg2017a} & -- & Pa & -- & -- & -- & -- & -- & P & On & -- & -- & -- & B & -- & -- \\ 
\cite{sunberg2022improving} & -- & Pa & -- & -- & -- & -- & -- & P & On & -- & -- & -- & B & -- & -- \\
\cite{Talebpour2015} & Re & -- & -- & -- & -- & S & -- & -- & -- & Off & -- & Op & -- & -- & -- \\ 
\cite{Treiber2000} & -- & Pa & -- & -- & -- & S & -- & -- & -- & Off & H & -- & -- & -- & -- \\ 
\cite{Tryhub2019} & -- & -- & -- & -- & At & S & -- & -- & On & -- & -- & Op & -- & -- & -- \\ 
\cite{Ulbrich2015} & -- & Pa & -- & -- & -- & S & -- & -- & -- & Off & H & -- & -- & -- & -- \\ 
\cite{Wheelera} & -- & Pa & NP & -- & -- & S & -- & -- & -- & Off & -- & Op & -- & -- & CV \\ 
\cite{Worle2019} & -- & Pa & -- & -- & -- & S & -- & -- & -- & Off & -- & Op & -- & -- & -- \\ 
\cite{Zhan2018a} & Re & -- & -- & -- & -- & S & -- & -- & -- & Off & -- & -- & -- & IRL & -- \\ 

%% file: appendices/tables/tags-motion_prediction_keyword_table_view1.tex
\multirow{72}{*}{\STAB{Architecture}}
 & Adaptive  & \cite{Schulz2018,Schulz,Petrich2014,Gray2013a,Hu2020,Polychronopoulos2007b} \\ 
 & Adaptive Cruise Control Policy  & \cite{Tran2019Merging} \\ 
 & Bezier Curves  & \cite{Krajewski2019} \\ 
 & Cognitive Hierarchy  & \cite{Garzon2019} \\ 
 & Conditional Expectation  & \cite{Lefevre2015a} \\ 
 & Conditional Variational AutoEncoder  & \cite{Hu2018b,Feng2019,Hu2019Generic,Li2019IV,Hu2019} \\ 
 & Constant Acceleration  & \cite{Lefevre2014c,Lefevre2014b,Deo,Kesting2010,Schlechtriemen,Platho,Polychronopoulos2007b,Schreier2016} \\ 
 & Constant Acceleration Constant Steering Angle  & \cite{Hillenbrand2006} \\ 
 & Constant Turn Rate Constant Tangential Acceleration  & \cite{Polychronopoulos2007b} \\ 
 & Constant Velocity  & \cite{Deo,Lefevre2014c,Lenz2017,Schlechtriemen,Ammoun2009,Lin2019,Sivaraman2014} \\ 
 & Constant Velocity Constant Turnrate  & \cite{Deo,Polychronopoulos2007b} \\ 
 & Constant Yaw Rate and Acceleration  & \cite{Houenou2013} \\ 
 & Context Dependent  & \cite{Platho} \\ 
 & Convolutional Neural Network  & \cite{Luo,Jain2018,Li2019d,Messaoud2020,Han2019Short} \\ 
 & Convolutional Social Pooling  & \cite{Messaoud2020} \\ 
 & Dirichlet Process  & \cite{Guo2019Multi} \\ 
 & Dynamic Bayesian Network  & \cite{Gindele,Gindele2015,Brechtel2014,Zhang2009,Sun2019} \\ 
 & Dynamic Forest  & \cite{Wheelera} \\ 
 & Empirical Reachable Set  & \cite{Driggs-Campbell2017c,Driggs-Campbell2017b} \\ 
 & Encoder-Decoder  & \cite{Deoa,Deob,Lee2017a,Feng2019,Messaoud2020,Hu2019,Messaoud2019Relational,Matousek2019} \\ 
 & Extended Kalman Filter  & \cite{Batza,Barth2008} \\ 
 & Foresighted Driver Model  & \cite{Eggert} \\ 
 & Forward Reachable Set  & \cite{Driggs-Campbell2017c,Driggs-Campbell2015b,Driggs-Campbell2017b,Shia2014a} \\ 
 & Gated Recurrent Unit Network  & \cite{Schulz2019} \\ 
 & Gaussian Mixture Model  & \cite{Deo,Lefevre2014c,Wheelera,Lenz2017,Schlechtriemen2015,Angkititrakul2009,Schlechtriemen,Gonzalez2019,Morton2016,Wiest2012} \\ 
 & Gaussian Process  & \cite{Tran2014,Laugier2011a,Armand,Liu2019,Guo2019Multi} \\ 
 & Gaussian Radial Basis Kernel Function  & \cite{Hillenbrand2006} \\ 
 & Generative Adversarial Network  & \cite{Li2019IV,Ma2019Wasserstein,Roy2019} \\ 
 & Gibbs Sampling  & \cite{Guo2019Multi} \\ 
 & Gipps Car Following Model  & \cite{Panwai2005,Li2019c} \\ 
 & Graph Neural Network  & \cite{Diehl2019} \\ 
 & Hidden Markov Model  & \cite{Lefevre2015a} \\ 
 & Hierarchical Mixture of Experts  & \cite{Wiest2013} \\ 
 & Intelligent Driver Model  & \cite{Sunberg2017a,Hoermann2017,Schulz2018,Schulz,Petrich2014,Kesting,Ulbrich2015,Lefevre2014c,Morton2016,Lenz2017,Kesting2010,Shen2012,Chen2010,Gonzalez2019,Hubmann2019,Treiber2000,Isele2019,Hu2020,Hart2019,Alizadeh2019,Guo2019} \\ 
 & Interaction Graph  & \cite{Li2019d,Jain2018,Diehl2019} \\ 
 & Iterative Semi-Network Form Game  & \cite{Garzon2019} \\ 
 & Kalman Prediction  & \cite{Ammoun2009,Graf2019,Batza} \\ 
 & Linear Gaussian  & \cite{Gray2013a,Wheelera} \\ 
 & Long Short-Term Memory Network  & \cite{Morton2016,Messaoud2020,Feng2019,Jones2019,Schulz2019,Roy2019,Han2019Short,Shih2019,Messaoud2019Relational,Li2019d,Lee2017a,Jain2018,Altche2018,Deoa,Deob,Kim2017} \\ 
 & Markov Chain  & \cite{Althoff2009,Althoff2011a,Asljung2019} \\ 
 & Minimizing Overall Braking Induced by Lane Changes  & \cite{Sunberg2017a,Kesting,Shen2012,Alizadeh2019} \\ 
 & MITSIM Driver Model  & \cite{Chen2010,Panwai2005} \\ 
 & Mixture of Experts  & \cite{Schlechtriemen2015} \\ 
 & Monte Carlo Simulation  & \cite{Althoff2011a,Sun2019,Asljung2019} \\ 
 & Monte Carlo Tree Search  & \cite{Gonzalez2019} \\ 
 & Multi-Fidelity  & \cite{Jain2018,Ulbrich2015} \\ 
 & Multi-Layer Perceptron  & \cite{Hu2019} \\ 
 & Neural Network  & \cite{Khodayari2012,Bansal2018,Demcenko2009a,Lefevre2014c,Hu2018b,Morton2016,Lenz2017,Schulz2019,Bouton2019,Messaoud2019Relational,Hu2020,Guo2019,Klingelschmitt2015a} \\ 
 & Optimum Velocity Model  & \cite{Dagli2003,Panwai2005} \\ 
 & Particle Filter  & \cite{Sun2019,Hoermann2017} \\ 
 & Perfect Information Game  & \cite{Pruekprasert2019} \\ 
 & Piecewise Auto-Regressive Model  & \cite{Akita,Angkititrakul2009,Lina} \\ 
 & Piecewise Uniform Distribution  & \cite{Morton2016} \\ 
 & Potential Field  & \cite{Bahram2016,Woo2017} \\ 
 & Proportional Derivative Feedback Control  & \cite{Tran2019Merging,Okuda2017} \\ 
 & Quantile Regression Forest  & \cite{Hu2018a} \\ 
 & Random Forest  & \cite{Gindele2015,Wheelera,Platho} \\ 
 & Recurrent Neural Network  & \cite{Schmerling2017,Kuefler2017,Leung,Morton2016,Lenz2017,Messaoud2020,Messaoud2019Relational,Shih2019,Bansal2018} \\ 
 & Relational Recurrent Neural Network  & \cite{Messaoud2019Relational} \\ 
 & Rule-Based  & \cite{Tran2019Merging,Schreier2016} \\ 
 & Simultaneous Game  & \cite{Schester2019} \\ 
 & Spline  & \cite{Gindele,Wiest2013,Kuderer2015a,Hardy2013,Gillmeier2019,Jugade2019,Houenou2013} \\ 
 & Stackelberg Game  & \cite{Fisac2019,HongYoo2012,HongYoo2013,Schester2019} \\ 
 & Static Gaussian  & \cite{Wheelera} \\ 
 & SUMO Model  & \cite{Lefevre2014c} \\ 
 & Switching  & \cite{Agamennoni2012b,Okuda2017,Tran2019Merging} \\ 
 & Tabular policy  & \cite{Shimosaka2015a,Shimosaka2014a,Oyler2016a,Li2017} \\ 
 & Time Delay Neural Network  & \cite{Nava2019} \\ 
 & Two-player game  & \cite{Fisac2019,Schester2019} \\ 
 & Variational  & \cite{Deo,Gonzalez2019} \\ 
 & Variational Autoencoder  & \cite{Krajewski2019,Li2019IV,Ma2019Wasserstein} \\ 
 & Wasserstein Auto Encoder  & \cite{Ma2019Wasserstein} \\ 

%% file: appendices/tables/tags-motion_prediction_keyword_table_view2.tex
\multirow{13}{*}{\STAB{Training}}
 & Continuous Inverse Optimal Control  & \cite{Hu2019Generic} \\ 
 & Evolutionary Strategies  & \cite{Li2019c} \\ 
 & Expectation Maximization  & \cite{Schlechtriemen} \\ 
 & Gaussian Mixture Regression  & \cite{Lefevre2014b,Schlechtriemen,Li2019Transferable} \\ 
 & Generative Adversarial Imitation Learning  & \cite{Kuefler2017} \\ 
 & Heuristic  & \cite{Galceran2017a,Schreier,Althoff2011a,Althoff2009,Gipps1981,Tram2018,Tram2019} \\ 
 & Inverse Model Predictive Control  & \cite{Guo2019} \\ 
 & Nelder-Mead Simplex  & \cite{Guo2019} \\ 
 & Nonlinear Optimization  & \cite{Lefevre2014c} \\ 
 & Polynomial Regression  & \cite{Jugade2019} \\ 
 & Q Learning  & \cite{Schester2019} \\ 
 & Semi-Supervised  & \cite{Krajewski2019} \\ 
 & Structural Risk Minimization  & \cite{Schlechtriemen} \\ 
\midrule
\multirow{22}{*}{\STAB{Theory}}
 & Apprenticeship Learning  & \cite{Abbeela} \\ 
 & Clustering  & \cite{Yoon2016} \\ 
 & Domain Adaptation  & \cite{Li2019Transferable} \\ 
 & Fuzzy Logic  & \cite{DiazAlvarcz2019} \\ 
 & Game Theory  & \cite{Pruekprasert2019,Schester2019} \\ 
 & Hierarchical Planning and Control  & \cite{Fisac2019} \\ 
 & Information Theory  & \cite{Ma2019Wasserstein} \\ 
 & Interaction Detection  & \cite{Hu2019,Worrall2008,Li2019IV} \\ 
 & Interpretability  & \cite{Hu2019} \\ 
 & Inverse Reinforcement Learning  & \cite{Abbeela} \\ 
 & Level K Reasoning  & \cite{Oyler2016a,Li2017,Garzon2019,Isele2019} \\ 
 & Model Predictive Control  & \cite{Sadigh2018,Schulz2017,Lina,Hu2019Generic,Sun2019IV,Tran2019Merging,Okuda2017} \\ 
 & Model Verification  & \cite{Bouton2019} \\ 
 & Nash Equilibrium  & \cite{Pruekprasert2019,Schester2019} \\ 
 & Partially Observable Markov Decision Process  & \cite{Gonzalez2019,Bouton2019,Hubmann2019,Brechtel2014} \\ 
 & People as Sensors  & \cite{Sun2019IV} \\ 
 & Reachability  & \cite{Althoff2009,Althoff2011a,Hubmann2019} \\ 
 & Recursive Reasoning  & \cite{Jain2018,Oyler2016a} \\ 
 & Reinforcement Learning  & \cite{Oyler2016a,Li2017,Schester2019,Bouton2019,Garzon2019,Hu2020} \\ 
 & Stochastic Reachability  & \cite{Althoff2009,Althoff2011a} \\ 
 & Trajectory Optimization  & \cite{Schulz2017,Levine2012a,Hardy2013,Sun2018,Lawitzkyb,Morales} \\ 
 & Tree-Search Planning  & \cite{Gonzalez2019,Isele2019} \\ 
\midrule
\multirow{19}{*}{\STAB{Scope}}
 & Adaptive Cruise Control  & \cite{Kesting2010,Nava2019} \\ 
 & Assistive Braking  & \cite{Hillenbrand2006} \\ 
 & Assistive Steering  & \cite{Gray2013a,Lefevre2014b,Shia2014a} \\ 
 & Car Following  & \cite{Gipps1981,Chen2010,Nava2019,Bhattacharyya2020,Jones2019,Treiber2000,Panwai2005,Guo2019,Akita,Altche2018,Angkititrakul2009,Khodayari2012,Lefevre2014c,Lefevre2015a,Morton2016,Wheelera} \\ 
 & Cooperative Maneuvering  & \cite{Batza} \\ 
 & Drifting  & \cite{Lina} \\ 
 & Highway  & \cite{Schlechtriemen2015,HongYoo2012,Gonzalez2019,Messaoud2020,Diehl2019,Li2019Transferable,Graf2019,Messaoud2019Relational,Alizadeh2019,Polychronopoulos2007b,Altche2018,Deo,Deoa,Deob,Driggs-Campbell2015b,Driggs-Campbell2017b,Feng2019,Guo2019Multi,Jain2018,Kim2017,Krajewski2019,Kuefler2017,Lenz2017,Li2017,Li2019d,Oyler2016a,Schlechtriemen,Shih2019,Sunberg2017a,Ulbrich2015,Yoon2016,Liu2019} \\ 
 & Highway Merging  & \cite{arbabi2022learning,Fisac2019,HongYoo2013,Li2019IV,Tran2019Merging,Okuda2017,Isele2019,Hu2020,Hart2019,Eggert,Schester2019,Sun2018,Garzon2019} \\ 
 & Intersection  & \cite{Armand,Sun2019IV,Bouton2019,Hubmann2019,Roy2019,Guo2019Multi,Klingelschmitt2015a,Petrich2014,Platho,Schreier,Tran2014,Zhang2009} \\ 
 & Intersection --- Blind  & \cite{Morales} \\ 
 & Intersection --- Merging  & \cite{Brechtel2014,Schulz2017} \\ 
 & Intersection --- Unsignalized  & \cite{Armand,Pruekprasert2019,Li2019IV,Roy2019,Lin2019} \\ 
 & Lane Bifurcation  & \cite{Leung,Schmerling2017} \\ 
 & Lane Changing  & \cite{Driggs-Campbell2017c,Kesting,Li2019c,Schlechtriemen2015,Schreier,Woo2017,Shen2012} \\ 
 & Lane Keeping  & \cite{Gray2013a,Lefevre2014b} \\ 
 & Overtaking  & \cite{Fisac2019,Graf2019,Polychronopoulos2007b} \\ 
 & Roundabout  & \cite{Ma2019Wasserstein,Hu2019Generic,Li2019IV} \\ 
 & Shared Control  & \cite{Jugade2019} \\ 
 & Urban Driving  & \cite{DiazAlvarcz2019,Bouton2019,Ferguson,Hoermann2017,Hubmann2019,Luo,Matousek2019,Schulz,Schulz2018,Schulz2019,Shimosaka2014a,Shimosaka2015a} \\ 
\midrule
\multirow{27}{*}{\STAB{Evaluation}}
 & Absolute Lateral Position Error  & \cite{Yoon2016,Feng2019} \\ 
 & Absolute Longitudinal Position Error  & \cite{Klingelschmitt2015a,Feng2019,Diehl2019,Bahram2016} \\ 
 & Absolute Longitudinal Velocity Error  & \cite{Guo2019,Platho,Klingelschmitt2015a,Bahram2016} \\ 
 & Collision Probability  & \cite{Althoff2011a,Althoff2009} \\ 
 & Collision Rate  & \cite{arbabi2022learning,Bhattacharyya2020,Schester2019,Pruekprasert2019,Oyler2016a,Kuefler2017,Lefevre2013,Lefevre2012,Lefevre,Hardy2013} \\ 
 & Final Euclidean Distance  & \cite{Li2019IV} \\ 
 & Hausdorff Distance  & \cite{Driggs-Campbell2017c,Driggs-Campbell2017b} \\ 
 & KL Divergence --- Acceleration  & \cite{Morton2016,Kuefler2017} \\ 
 & KL Divergence --- Inverse Time To Collision  & \cite{Morton2016,Kuefler2017} \\ 
 & KL Divergence --- Jerk  & \cite{Wheelera,Morton2016,Kuefler2017} \\ 
 & KL Divergence --- Speed  & \cite{Morton2016,Kuefler2017} \\ 
 & KL Divergence --- Turnrate  & \cite{Kuefler2017} \\ 
 & Mean Absolute Error  & \cite{Schlechtriemen2015,Schlechtriemen,Ma2019Wasserstein,Lefevre2014c,Kim2017,Khodayari2012,Deo} \\ 
 & Mean Cross Validated Log Likelihood  & \cite{Wheelera} \\ 
 & Mean Euclidean Distance  & \cite{Sun2018,Li2019IV} \\ 
 & Mean Log Likelihood  & \cite{Gindele2015,Gindele} \\ 
 & Mean Square Error  & \cite{Sun2019,Hu2019} \\ 
 & Median Absolute Error  & \cite{Deo} \\ 
 & Modified Hausdorff Distance  & \cite{Shimosaka2015a,Shimosaka2014a,Petrich2014,Morales} \\ 
 & Negative Log Likelihood  & \cite{Schulz2019,Schmerling2017,Leung,Lenz2017,Hu2019,Deob,Deoa} \\ 
 & Reachable Set Accuracy  & \cite{Shia2014a,Driggs-Campbell2017c,Driggs-Campbell2017b,Driggs-Campbell2015b} \\ 
 & Reachable Set Intrusion  & \cite{Driggs-Campbell2017c,Driggs-Campbell2017b} \\ 
 & Reachable Set Overlap  & \cite{Driggs-Campbell2017c,Driggs-Campbell2017b} \\ 
 & Reachable Set Precision  & \cite{Shia2014a,Driggs-Campbell2017c,Driggs-Campbell2017b,Driggs-Campbell2015b} \\ 
 & Root Mean Square Error  & \cite{Panwai2005,Polychronopoulos2007b,Altche2018,Bhattacharyya2020,Wissing2017,Shih2019,Schulz2019,Schulz2018,Schulz,Nava2019,Messaoud2020,Messaoud2019Relational,Ma2019Wasserstein,Li2019Transferable,Li2019IV,Li2019d,Lefevre2014c,Lee2017a,Krajewski2019,Klingelschmitt2015a,Khodayari2012,Jones2019,Jain2018,Hu2019Generic,Hu2018b,Houenou2013,Han2019Short,Gindele2015,Gindele,DiazAlvarcz2019,Deob,Deoa,Demcenko2009a,Dang2018,Chen2010} \\ 
 & Root Weighted Square Error  & \cite{arbabi2022learning,Wheelera,Morton2016,Lenz2017,Kuefler2017} \\ 
 & Time to Collision  & \cite{Ammoun2009} \\ 

%% file: appendices/tables/models-motion_prediction_view1.tex
\cite{Abbeela} & -- & -- & -- & -- & -- & -- & Dc & -- & -- & -- & -- & P & -- & -- & -- & -- & S & -- & -- & -- & -- & -- & FS & -- & -- \\ 
\cite{Agamennoni2012b} & -- & -- & BK & -- & -- & -- & -- & -- & -- & -- & -- & -- & -- & -- & -- & -- & -- & -- & -- & -- & -- & -- & -- & -- & -- \\ 
\cite{Akita} & -- & -- & -- & -- & -- & -- & -- & -- & -- & -- & -- & -- & -- & -- & -- & -- & -- & -- & -- & -- & -- & -- & -- & -- & -- \\ 
\cite{Alizadeh2019} & -- & -- & BK & -- & -- & -- & -- & -- & -- & -- & -- & -- & -- & -- & -- & -- & -- & -- & -- & -- & -- & -- & -- & -- & -- \\ 
\cite{Altche2018} & -- & -- & -- & -- & -- & -- & -- & -- & X & -- & -- & P & -- & -- & -- & -- & S & -- & -- & -- & -- & -- & -- & IP & -- \\ 
\cite{Althoff2009} & -- & -- & -- & -- & L & -- & -- & -- & -- & -- & S & -- & -- & -- & -- & -- & -- & O & -- & -- & R & -- & -- & IP & GT \\ 
\cite{Althoff2011a} & -- & -- & -- & -- & L & -- & -- & -- & -- & -- & S & -- & -- & -- & -- & P & -- & O & -- & -- & R & -- & -- & IP & GT \\ 
\cite{Ammoun2009} & -- & BD & -- & -- & -- & -- & -- & -- & -- & -- & S & -- & -- & G & -- & -- & -- & -- & -- & -- & -- & -- & -- & IP & -- \\ 
\cite{Angkititrakul2009} & -- & -- & -- & -- & L & -- & -- & -- & -- & -- & -- & P & -- & -- & -- & -- & S & -- & -- & -- & -- & -- & FS & -- & -- \\
\cite{arbabi2022learning} & -- & -- & BK & -- & -- & -- & -- & -- & -- & -- & S & -- & -- & -- & -- & -- & S & -- & -- & -- & -- & -- & FS & -- & -- \\
\cite{Armand} & -- & -- & -- & -- & -- & -- & -- & -- & -- & -- & -- & P & -- & -- & -- & -- & S & -- & -- & -- & -- & -- & -- & IP & -- \\ 
\cite{Asljung2019} & -- & -- & -- & -- & -- & -- & Dc & -- & -- & -- & S & -- & -- & -- & -- & P & -- & -- & -- & -- & -- & -- & -- & IP & -- \\ 
\cite{Bahram2016} & -- & -- & -- & -- & L & -- & -- & P & -- & -- & S & -- & -- & G & -- & -- & -- & -- & -- & -- & -- & -- & FS & -- & -- \\ 
\cite{Bansal2018} & -- & -- & -- & -- & -- & -- & -- & -- & X & -- & S & -- & -- & -- & -- & -- & -- & O & -- & -- & -- & -- & FS & -- & -- \\ 
\cite{Barth2008} & -- & -- & -- & U & -- & -- & -- & -- & -- & -- & -- & P & -- & -- & -- & -- & S & -- & -- & -- & -- & -- & -- & IP & -- \\ 
\cite{Batza} & -- & -- & -- & U & -- & -- & -- & -- & -- & -- & S & -- & -- & -- & -- & -- & S & -- & -- & -- & -- & -- & -- & IP & -- \\ 
\cite{Bhattacharyya2020} & -- & -- & -- & -- & L & -- & -- & -- & -- & -- & S & -- & -- & -- & -- & -- & S & -- & -- & -- & -- & -- & FS & -- & -- \\ 
\cite{Bouton2019} & -- & -- & -- & -- & -- & -- & Dc & -- & -- & -- & -- & -- & Tr & -- & -- & -- & -- & -- & -- & -- & -- & -- & -- & -- & GT \\ 
\cite{Brechtel2014} & -- & -- & -- & U & -- & -- & -- & -- & -- & -- & -- & -- & Tr & -- & -- & P & -- & -- & -- & -- & -- & -- & -- & -- & GT \\ 
\cite{Chen2010} & -- & -- & -- & -- & L & -- & -- & -- & -- & -- & S & -- & -- & -- & -- & -- & S & -- & -- & -- & -- & -- & FS & -- & -- \\ 
\cite{Dang2018} & -- & -- & -- & -- & -- & -- & -- & -- & -- & -- & -- & P & -- & -- & -- & -- & S & -- & -- & -- & -- & -- & -- & -- & -- \\ 
\cite{Demcenko2009a} & -- & -- & -- & -- & -- & -- & -- & -- & -- & -- & -- & P & -- & -- & -- & -- & S & -- & -- & -- & -- & -- & -- & -- & -- \\ 
\cite{Deo} & -- & -- & -- & -- & -- & -- & -- & P & X & -- & S & -- & -- & G & -- & -- & -- & -- & -- & -- & -- & -- & -- & IP & -- \\ 
\cite{Deoa} & -- & -- & -- & -- & -- & -- & -- & -- & X & -- & -- & P & -- & -- & GM & -- & -- & -- & -- & -- & -- & -- & -- & IP & -- \\ 
\cite{Deob} & -- & -- & -- & -- & -- & -- & -- & -- & X & -- & -- & P & -- & -- & GM & -- & -- & -- & -- & -- & -- & -- & -- & IP & -- \\ 
\cite{DiazAlvarcz2019} & -- & -- & -- & -- & -- & -- & -- & -- & -- & -- & -- & P & -- & -- & -- & -- & S & -- & -- & -- & -- & -- & FS & -- & -- \\ 
\cite{Diehl2019} & -- & -- & -- & -- & -- & -- & -- & -- & X & -- & S & -- & -- & -- & -- & -- & S & -- & -- & -- & -- & -- & -- & IP & -- \\ 
\cite{Driggs-Campbell2015b} & -- & -- & -- & -- & -- & -- & -- & -- & X & -- & -- & P & -- & -- & -- & -- & -- & -- & -- & -- & R & -- & -- & IP & -- \\ 
\cite{Driggs-Campbell2017b} & -- & -- & -- & -- & -- & -- & -- & -- & -- & -- & -- & P & -- & -- & -- & -- & -- & -- & -- & -- & R & -- & -- & IP & -- \\ 
\cite{Driggs-Campbell2017c} & -- & -- & -- & -- & -- & -- & -- & -- & X & -- & -- & P & -- & -- & -- & -- & -- & -- & -- & -- & R & -- & -- & IP & -- \\ 
\cite{Eggert} & -- & -- & -- & -- & -- & -- & -- & -- & -- & -- & -- & P & -- & -- & -- & -- & S & -- & -- & -- & -- & -- & FS & -- & -- \\ 
\cite{Feng2019} & -- & -- & -- & -- & -- & -- & -- & -- & X & M & -- & -- & -- & -- & -- & P & -- & -- & -- & -- & -- & -- & -- & IP & -- \\ 
\cite{Ferguson} & -- & -- & -- & U & -- & -- & -- & -- & -- & M & -- & -- & -- & -- & -- & -- & S & -- & -- & -- & -- & -- & -- & IP & -- \\ 
\cite{Fisac2019} & -- & -- & -- & -- & -- & -- & -- & -- & -- & -- & -- & -- & -- & -- & -- & -- & S & -- & -- & -- & -- & -- & -- & -- & GT \\ 
\cite{Galceran2017a} & -- & -- & -- & -- & -- & -- & -- & P & -- & M & -- & -- & -- & G & -- & -- & -- & -- & -- & -- & -- & -- & FS & -- & -- \\ 
\cite{Garzon2019} & -- & -- & -- & -- & -- & -- & Dc & -- & -- & M & -- & -- & -- & -- & -- & P & -- & -- & -- & -- & -- & -- & FS & -- & GT \\ 
\cite{Gillmeier2019} & -- & -- & BK & -- & -- & -- & -- & -- & -- & -- & S & -- & -- & -- & -- & -- & S & -- & -- & -- & -- & -- & -- & IP & -- \\ 
\cite{Gindele} & -- & -- & BK & -- & -- & -- & -- & -- & -- & -- & S & -- & -- & -- & -- & P & -- & -- & -- & -- & -- & -- & FS & -- & -- \\ 
\cite{Gindele2015} & -- & -- & -- & U & -- & -- & -- & -- & -- & -- & S & -- & -- & -- & -- & P & -- & -- & -- & -- & -- & -- & FS & -- & -- \\ 
\cite{Gipps1981} & -- & -- & -- & -- & L & -- & -- & -- & -- & -- & S & -- & -- & -- & -- & -- & S & -- & -- & -- & -- & -- & FS & -- & -- \\ 
\cite{Gonzalez2019} & -- & -- & -- & -- & L & -- & -- & -- & -- & M & -- & -- & Tr & -- & -- & -- & S & -- & -- & -- & -- & -- & -- & -- & GT \\ 
\cite{Graf2019} & -- & -- & -- & -- & L & -- & -- & -- & -- & -- & S & -- & -- & G & -- & -- & -- & -- & -- & -- & -- & -- & -- & IP & -- \\ 
\cite{Gray2013a} & -- & -- & -- & -- & L & -- & -- & -- & -- & -- & S & -- & -- & -- & -- & P & -- & -- & -- & -- & -- & -- & FS & -- & -- \\ 
\cite{Guo2019} & -- & -- & -- & -- & L & -- & -- & -- & -- & -- & S & -- & -- & -- & -- & -- & S & -- & -- & -- & -- & -- & -- & IP & -- \\ 
\cite{Guo2019Multi} & -- & -- & -- & -- & -- & -- & -- & -- & X & M & -- & -- & -- & -- & -- & -- & S & -- & -- & -- & -- & -- & FS & -- & -- \\ 
\cite{Han2019Short} & -- & -- & -- & -- & -- & -- & -- & -- & -- & -- & -- & P & -- & -- & -- & -- & S & -- & -- & -- & -- & -- & FS & -- & -- \\ 
\cite{Hardy2013} & -- & -- & -- & -- & -- & -- & -- & -- & -- & M & -- & -- & -- & G & -- & -- & -- & -- & -- & -- & -- & -- & -- & IP & -- \\ 
\cite{Hart2019} & -- & -- & -- & -- & -- & -- & -- & -- & -- & -- & S & -- & -- & -- & -- & -- & S & -- & -- & -- & -- & -- & FS & -- & -- \\ 
\cite{Hillenbrand2006} & -- & -- & -- & -- & -- & -- & -- & -- & -- & -- & -- & -- & -- & -- & -- & -- & S & -- & -- & -- & -- & -- & -- & -- & -- \\ 
\cite{Hoermann2017} & -- & -- & -- & -- & L & -- & -- & -- & -- & -- & S & -- & -- & -- & -- & P & -- & -- & -- & -- & -- & -- & FS & -- & -- \\ 
\cite{HongYoo2012} & -- & -- & -- & -- & L & -- & -- & -- & -- & -- & S & -- & -- & -- & -- & -- & S & -- & -- & -- & -- & -- & -- & -- & GT \\ 
\cite{HongYoo2013} & -- & -- & -- & -- & -- & -- & Dc & -- & -- & -- & S & -- & -- & -- & -- & -- & S & -- & -- & -- & -- & -- & -- & -- & GT \\ 
\cite{Houenou2013} & -- & -- & -- & -- & -- & -- & -- & -- & -- & -- & -- & -- & -- & -- & -- & P & -- & -- & -- & -- & -- & -- & -- & IP & -- \\ 
\cite{Hu2018a} & -- & -- & -- & -- & -- & -- & -- & -- & X & -- & -- & P & -- & -- & GM & -- & -- & -- & -- & -- & -- & -- & -- & -- & -- \\ 
\cite{Hu2018b} & -- & -- & -- & -- & -- & -- & -- & -- & X & -- & -- & P & -- & -- & -- & P & -- & -- & -- & -- & -- & -- & FS & -- & -- \\ 
\cite{Hu2019} & -- & -- & -- & -- & -- & -- & -- & -- & X & -- & S & -- & -- & -- & -- & P & -- & -- & -- & -- & -- & -- & -- & -- & GT \\ 
\cite{Hu2019Generic} & -- & -- & -- & -- & -- & -- & -- & -- & X & M & -- & -- & -- & -- & -- & P & -- & -- & -- & -- & -- & -- & -- & IP & GT \\ 
\cite{Hu2020} & -- & -- & -- & -- & L & -- & -- & -- & -- & -- & S & -- & -- & -- & -- & -- & S & -- & -- & -- & -- & -- & FS & -- & -- \\ 
\cite{Hubmann2019} & -- & -- & -- & -- & L & -- & -- & -- & -- & -- & -- & -- & Tr & -- & -- & -- & -- & -- & -- & -- & -- & -- & -- & -- & GT \\ 
\cite{igl2022symphony} & -- & -- & BK & -- & -- & -- & -- & -- & -- & -- & S & -- & -- & -- & -- & -- & S & -- & -- & -- & -- & -- & FS & -- & -- \\ 
\cite{Isele2019} & -- & -- & -- & -- & -- & -- & Dc & -- & -- & -- & -- & -- & Tr & -- & -- & -- & -- & -- & -- & -- & -- & -- & -- & -- & GT \\ 
\cite{Jain2018} & -- & -- & -- & -- & -- & -- & -- & -- & X & -- & S & -- & -- & G & -- & -- & -- & -- & -- & -- & -- & -- & -- & -- & GT \\ 
\cite{Jones2019} & -- & -- & -- & -- & L & -- & -- & -- & -- & -- & S & -- & -- & -- & -- & P & -- & -- & -- & -- & -- & -- & -- & IP & -- \\ 
\cite{Jugade2019} & -- & -- & -- & -- & -- & -- & -- & -- & -- & -- & S & -- & -- & -- & -- & -- & S & -- & -- & S & -- & -- & -- & IP & -- \\ 
\cite{Kesting} & -- & -- & -- & -- & L & -- & -- & -- & -- & -- & S & -- & -- & -- & -- & -- & S & -- & -- & -- & -- & -- & FS & -- & -- \\ 
\cite{Kesting2010} & -- & -- & -- & -- & L & -- & -- & -- & -- & -- & S & -- & -- & -- & -- & -- & S & -- & -- & -- & -- & -- & FS & -- & -- \\ 
\cite{Khodayari2012} & -- & -- & -- & -- & -- & -- & -- & -- & -- & -- & -- & P & -- & -- & -- & -- & S & -- & -- & -- & -- & -- & FS & -- & -- \\ 
\cite{Kim2017} & -- & -- & -- & -- & -- & -- & -- & -- & X & -- & S & -- & -- & -- & -- & -- & -- & O & -- & -- & -- & -- & -- & IP & -- \\ 
\cite{Klingelschmitt2015a} & -- & -- & -- & -- & -- & -- & Dc & -- & -- & -- & S & -- & -- & -- & -- & -- & S & -- & -- & -- & -- & -- & -- & IP & -- \\ 
\cite{Krajewski2019} & -- & -- & -- & -- & -- & -- & -- & -- & X & -- & -- & P & -- & -- & -- & P & -- & -- & -- & -- & -- & -- & -- & IP & -- \\ 
\cite{Kuderer2015a} & -- & -- & -- & -- & -- & S & -- & -- & -- & -- & -- & P & -- & -- & -- & -- & S & -- & -- & S & -- & -- & FS & -- & GT \\ 

%% file: appendices/tables/models-motion_prediction_view2.tex
\cite{Kuefler2017} & -- & -- & BK & -- & -- & -- & -- & -- & -- & -- & S & -- & -- & -- & -- & -- & S & -- & -- & -- & -- & -- & FS & -- & -- \\ 
\cite{Laugier2011a} & -- & -- & -- & -- & -- & -- & Dc & -- & X & -- & S & -- & -- & -- & -- & P & -- & -- & -- & -- & -- & -- & -- & IP & -- \\ 
\cite{Lawitzkyb} & -- & -- & -- & -- & -- & -- & -- & -- & -- & -- & S & -- & -- & -- & -- & P & -- & -- & -- & -- & -- & -- & -- & IP & -- \\ 
\cite{Lee2017a} & -- & -- & -- & -- & -- & -- & -- & -- & -- & -- & S & -- & -- & -- & -- & P & -- & -- & -- & -- & -- & -- & -- & IP & -- \\ 
\cite{Lefevre2014b} & -- & BD & -- & -- & -- & -- & -- & -- & -- & -- & S & -- & -- & -- & -- & -- & S & -- & -- & -- & -- & -- & FS & -- & -- \\ 
\cite{Lefevre2014c} & -- & -- & -- & -- & L & -- & -- & -- & -- & -- & -- & P & -- & -- & -- & -- & S & -- & -- & -- & -- & -- & FS & -- & -- \\ 
\cite{Lefevre2015a} & -- & -- & -- & -- & L & -- & -- & -- & -- & -- & S & -- & -- & -- & -- & -- & S & -- & -- & -- & -- & -- & FS & -- & -- \\ 
\cite{Lenz2017} & -- & -- & -- & -- & L & -- & -- & -- & -- & -- & S & -- & -- & -- & -- & -- & S & -- & -- & -- & -- & -- & FS & -- & -- \\ 
\cite{Leung} & -- & BD & -- & -- & -- & -- & -- & -- & -- & M & -- & P & -- & -- & -- & P & -- & -- & -- & -- & -- & bR & -- & -- & GT \\ 
\cite{Levine2012a} & -- & -- & -- & -- & -- & -- & -- & -- & -- & -- & -- & P & -- & -- & -- & -- & S & -- & -- & -- & -- & -- & FS & -- & GT \\ 
\cite{Li2017} & -- & -- & -- & -- & L & -- & -- & -- & -- & -- & S & -- & -- & -- & -- & -- & S & -- & -- & -- & -- & -- & FS & -- & -- \\ 
\cite{Li2019c} & -- & -- & -- & -- & -- & -- & Dc & -- & -- & -- & S & -- & -- & -- & -- & -- & S & -- & -- & -- & -- & -- & FS & -- & -- \\ 
\cite{Li2019d} & -- & -- & -- & -- & -- & -- & -- & -- & X & -- & S & -- & -- & -- & -- & -- & S & -- & -- & -- & -- & -- & -- & IP & -- \\ 
\cite{Li2019IV} & -- & -- & -- & -- & -- & -- & -- & -- & X & -- & S & -- & -- & -- & -- & P & -- & -- & -- & -- & -- & -- & -- & IP & -- \\ 
\cite{Li2019Transferable} & -- & -- & -- & -- & -- & -- & -- & -- & -- & -- & S & -- & -- & -- & -- & -- & S & -- & -- & -- & -- & -- & FS & -- & -- \\ 
\cite{Liebner} & -- & -- & -- & -- & -- & -- & -- & -- & -- & -- & -- & -- & -- & -- & -- & -- & -- & -- & -- & -- & -- & -- & -- & -- & -- \\ 
\cite{Lina} & 4W & -- & -- & -- & -- & -- & -- & -- & -- & -- & S & -- & -- & -- & -- & -- & S & -- & -- & -- & -- & -- & -- & IP & -- \\ 
\cite{Liu2019} & -- & -- & -- & -- & -- & -- & -- & P & -- & -- & -- & P & -- & G & -- & -- & -- & -- & -- & -- & -- & -- & -- & IP & -- \\ 
\cite{Luo} & -- & -- & -- & -- & -- & -- & -- & -- & X & -- & S & -- & -- & -- & -- & -- & -- & -- & BB & -- & -- & -- & -- & IP & -- \\ 
\cite{Ma2019Wasserstein} & -- & -- & BK & -- & -- & -- & -- & -- & -- & -- & S & -- & -- & -- & -- & P & -- & -- & -- & -- & -- & -- & -- & IP & -- \\ 
\cite{Matousek2019} & -- & -- & -- & -- & -- & -- & -- & -- & X & -- & S & -- & -- & -- & -- & -- & S & -- & -- & -- & -- & -- & -- & IP & -- \\ 
\cite{Messaoud2019Relational} & -- & -- & -- & -- & -- & -- & -- & -- & X & -- & S & -- & -- & -- & -- & -- & S & -- & -- & -- & -- & -- & -- & IP & -- \\ 
\cite{Messaoud2020} & -- & -- & -- & -- & -- & -- & -- & -- & X & -- & S & -- & -- & G & -- & -- & -- & -- & -- & -- & -- & -- & -- & IP & -- \\ 
\cite{Morales} & -- & -- & -- & -- & L & -- & -- & -- & -- & -- & -- & P & -- & -- & -- & -- & S & -- & -- & -- & -- & -- & -- & IP & -- \\ 
\cite{Morton2016} & -- & -- & -- & -- & L & -- & -- & -- & -- & -- & -- & P & -- & -- & -- & P & -- & -- & -- & -- & -- & -- & FS & -- & -- \\ 
\cite{Nava2019} & -- & -- & -- & -- & L & -- & -- & -- & -- & -- & -- & P & -- & -- & -- & -- & S & -- & -- & -- & -- & -- & FS & -- & -- \\ 
\cite{Okuda2017} & -- & -- & -- & -- & L & -- & -- & -- & -- & -- & S & -- & -- & -- & -- & -- & S & -- & -- & -- & -- & -- & -- & -- & GT \\ 
\cite{Oyler2016a} & -- & -- & -- & -- & -- & -- & Dc & -- & -- & -- & S & -- & -- & -- & -- & -- & S & -- & -- & -- & -- & -- & FS & -- & -- \\ 
\cite{Panwai2005} & -- & -- & -- & -- & L & -- & -- & -- & -- & -- & -- & P & -- & -- & -- & -- & S & -- & -- & -- & -- & -- & FS & -- & -- \\ 
\cite{Petrich2014} & -- & -- & BK & -- & -- & -- & -- & -- & -- & -- & S & -- & -- & -- & -- & P & -- & -- & -- & -- & -- & -- & -- & IP & -- \\ 
\cite{Platho} & -- & -- & -- & -- & -- & -- & -- & -- & -- & -- & S & -- & -- & -- & -- & -- & S & -- & -- & -- & -- & -- & -- & IP & -- \\ 
\cite{Polychronopoulos2007b} & -- & -- & -- & -- & L & -- & -- & -- & -- & -- & -- & -- & -- & -- & -- & -- & -- & -- & -- & -- & -- & -- & -- & -- & -- \\ 
\cite{Pruekprasert2019} & -- & -- & -- & -- & L & -- & -- & -- & -- & -- & S & -- & -- & -- & -- & -- & S & -- & -- & -- & -- & -- & -- & -- & GT \\ 
\cite{Roy2019} & -- & -- & -- & -- & -- & -- & -- & -- & X & -- & S & -- & -- & -- & -- & -- & S & -- & -- & -- & -- & -- & -- & IP & -- \\ 
\cite{Sadigh2018} & -- & -- & -- & U & -- & -- & -- & -- & -- & -- & S & -- & -- & -- & -- & -- & S & -- & -- & -- & -- & -- & -- & -- & GT \\ 
\cite{Schester2019} & -- & -- & -- & -- & -- & -- & Dc & -- & -- & -- & S & -- & -- & -- & -- & -- & S & -- & -- & -- & -- & -- & -- & -- & GT \\ 
\cite{Schlechtriemen} & -- & -- & -- & -- & L & -- & -- & -- & -- & -- & -- & P & -- & -- & GM & -- & -- & -- & -- & -- & -- & -- & -- & IP & -- \\ 
\cite{Schlechtriemen2015} & -- & -- & -- & -- & -- & -- & -- & -- & -- & -- & -- & P & -- & -- & GM & -- & -- & -- & -- & -- & -- & -- & -- & IP & -- \\ 
\cite{Schmerling2017} & -- & -- & -- & -- & L & -- & -- & -- & -- & M & -- & -- & -- & -- & -- & P & -- & -- & -- & -- & -- & -- & -- & -- & GT \\ 
\cite{Schreier} & -- & -- & -- & -- & -- & -- & -- & P & -- & -- & S & -- & -- & G & -- & -- & -- & -- & -- & -- & -- & -- & -- & IP & -- \\ 
\cite{Schreier2016} & -- & -- & -- & -- & L & -- & -- & -- & -- & -- & S & -- & -- & -- & -- & P & -- & -- & -- & -- & -- & -- & FS & IP & -- \\ 
\cite{Schulz} & -- & -- & -- & U & -- & -- & -- & -- & -- & M & -- & -- & -- & -- & -- & -- & S & -- & -- & -- & -- & -- & FS & -- & -- \\ 
\cite{Schulz2017} & -- & -- & -- & -- & -- & -- & -- & -- & -- & M & -- & -- & -- & -- & -- & -- & S & -- & -- & -- & -- & -- & -- & -- & GT \\ 
\cite{Schulz2018} & -- & -- & -- & U & -- & -- & -- & -- & -- & M & -- & -- & -- & -- & -- & -- & S & -- & -- & -- & -- & -- & FS & -- & -- \\ 
\cite{Schulz2019} & -- & -- & BK & -- & -- & -- & -- & -- & -- & M & -- & -- & -- & -- & -- & -- & S & -- & -- & -- & -- & -- & FS & -- & -- \\ 
\cite{Shia2014a} & -- & BD & -- & -- & -- & -- & -- & -- & -- & -- & S & -- & -- & -- & -- & P & -- & -- & -- & -- & R & -- & -- & -- & -- \\ 
\cite{Shih2019} & -- & -- & -- & -- & -- & -- & -- & -- & -- & -- & S & -- & -- & -- & -- & -- & S & -- & -- & -- & -- & -- & -- & IP & -- \\ 
\cite{Shimosaka2014a} & -- & -- & -- & -- & -- & -- & Dc & -- & -- & -- & S & -- & -- & -- & -- & -- & S & -- & -- & -- & -- & -- & FS & -- & -- \\ 
\cite{Shimosaka2015a} & -- & -- & -- & -- & -- & -- & Dc & -- & -- & -- & S & -- & -- & -- & -- & -- & S & -- & -- & -- & -- & -- & FS & -- & -- \\ 
\cite{Sivaraman2014} & -- & -- & -- & -- & -- & -- & -- & -- & -- & -- & -- & P & -- & -- & -- & -- & S & -- & -- & -- & -- & -- & -- & IP & -- \\ 
\cite{Sun2018} & -- & -- & -- & -- & -- & -- & -- & -- & -- & -- & -- & P & -- & -- & -- & -- & S & -- & -- & -- & -- & -- & FS & -- & GT \\ 
\cite{Sun2019} & -- & -- & -- & -- & -- & S & -- & P & -- & -- & S & -- & -- & -- & -- & P & -- & -- & -- & -- & -- & -- & FS & -- & -- \\ 
\cite{Sun2019IV} & -- & -- & -- & -- & -- & -- & -- & -- & -- & -- & S & -- & -- & -- & -- & -- & S & -- & -- & -- & -- & -- & -- & -- & GT \\ 
\cite{Sunberg2017a} & -- & -- & -- & -- & L & -- & -- & -- & -- & -- & -- & -- & Tr & -- & -- & -- & -- & -- & -- & -- & -- & -- & FS & -- & -- \\
\cite{sunberg2022improving} & -- & -- & -- & -- & L & -- & -- & -- & -- & -- & -- & -- & Tr & -- & -- & -- & -- & -- & -- & -- & -- & -- & FS & -- & -- \\
\cite{suo2021trafficsim} & -- & -- & -- & -- & -- & -- & -- & -- & X & -- & S & -- & -- & -- & -- & -- & S & -- & -- & -- & -- & -- & FS & -- & -- \\
\cite{Tram2018} & -- & -- & -- & -- & -- & -- & -- & -- & -- & -- & -- & -- & Tr & -- & -- & -- & -- & -- & -- & -- & -- & -- & FS & -- & -- \\ 
\cite{Tram2019} & -- & -- & -- & -- & -- & -- & -- & -- & -- & -- & -- & -- & Tr & -- & -- & -- & -- & -- & -- & -- & -- & -- & FS & -- & -- \\ 
\cite{Tran2014} & -- & -- & -- & -- & -- & -- & -- & -- & X & -- & S & -- & -- & -- & -- & P & -- & -- & -- & -- & -- & -- & -- & IP & -- \\ 
\cite{Tran2019Merging} & -- & -- & -- & -- & L & -- & -- & -- & -- & M & -- & -- & -- & -- & -- & -- & S & -- & -- & -- & -- & -- & -- & -- & GT \\ 
\cite{Treiber2000} & -- & -- & -- & -- & L & -- & -- & -- & -- & -- & S & -- & -- & -- & -- & -- & S & -- & -- & -- & -- & -- & FS & -- & -- \\ 
\cite{Ulbrich2015} & -- & -- & -- & -- & L & -- & -- & -- & -- & -- & -- & -- & Tr & -- & -- & -- & -- & -- & -- & -- & -- & -- & FS & -- & -- \\ 
\cite{Wheelera} & -- & -- & BK & -- & -- & -- & -- & -- & -- & -- & S & -- & -- & -- & -- & -- & S & -- & -- & -- & -- & -- & FS & -- & -- \\ 
\cite{Wiest2012} & -- & -- & -- & -- & -- & S & -- & -- & -- & -- & -- & P & -- & -- & GM & -- & -- & -- & -- & -- & -- & -- & -- & IP & -- \\ 
\cite{Wiest2013} & -- & -- & -- & -- & -- & S & -- & P & X & -- & -- & P & -- & G & -- & -- & -- & -- & -- & S & -- & -- & -- & IP & -- \\ 
\cite{Wissing2017} & -- & -- & -- & -- & -- & -- & -- & -- & -- & -- & -- & P & -- & -- & -- & -- & -- & -- & -- & -- & -- & -- & -- & -- & -- \\ 
\cite{Woo2017} & -- & -- & -- & -- & -- & -- & -- & -- & -- & -- & -- & -- & -- & -- & -- & -- & -- & -- & -- & -- & -- & -- & -- & -- & -- \\ 
\cite{Worrall2008} & -- & -- & -- & -- & -- & -- & Dc & -- & -- & -- & S & -- & -- & -- & -- & -- & -- & O & -- & -- & -- & -- & -- & IP & -- \\ 
\cite{Yoon2016} & -- & -- & -- & -- & L & -- & -- & -- & -- & -- & -- & P & -- & -- & -- & P & -- & -- & -- & -- & -- & -- & -- & IP & -- \\ 
\cite{Zhang2009} & -- & -- & -- & -- & -- & -- & Dc & P & -- & -- & -- & -- & -- & -- & -- & -- & -- & -- & -- & -- & -- & -- & -- & -- & -- \\ 